\documentclass{aa}
 
\usepackage{natbib} 
\usepackage[usenames]{color}
\bibpunct{(}{)}{;}{a}{}{,} 
\usepackage{amsmath} 
\usepackage{graphicx,txfonts} 
\usepackage{longtable} 
\usepackage{url} 
\usepackage{xspace}

\begin{document} 
\title{The CoRoT B-type binary HD50230: a prototypical hybrid pulsator with g-mode period and p-mode frequency spacings\thanks{The CoRoT space mission was developed and is operated by the French space agency CNES, with participation of ESA's RSSD and Science Programmes, Austria, Belgium, Brazil, Germany, and Spain. Based on observations made with the ESO telescopes at La Silla Observatory under the ESO Large Programme LP182.D-0356, and on observations made with the Mercator Telescope, operated on the island of La Palma by the Flemish Community, at the Spanish Observatorio del Roque de los Muchachos of the Instituto de Astrof\'isica de Canarias, and on observations obtained with the HERMES spectrograph, which is supported by the Fund for Scientific Research of Flanders (FWO), Belgium, the Research Council of K.U.\,Leuven, Belgium, the Fonds National Recherches Scientific (FNRS), Belgium, the Royal Observatory of Belgium, the Observatoire de Gen\`eve, Switzerland and the Th\"uringer Landessternwarte Tautenburg, Germany.}}
 \author{P.~Degroote\inst{\ref{inst:leuven},\ref{inst:KITP}}\thanks{Postdoctoral Fellow of the Fund for Scientific Research, Flanders}
 \and C.~Aerts\inst{\ref{inst:leuven},\ref{inst:KITP},\ref{inst:nijmegen}} 
\and E.~Michel\inst{\ref{inst:lesia}}
 \and M.~Briquet\inst{\ref{inst:leuven},\ref{inst:liege}}\thanks{F.R.S.-FNRS Postdoctoral Researcher, Belgium}
 \and P.~I.~P\'apics\inst{\ref{inst:leuven}}
 \and P.~Amado\inst{\ref{inst:inaf}}
 \and P.~Mathias\inst{\ref{inst:toul1},\ref{inst:toul2}}
 \and E.~Poretti\inst{\ref{inst:inaf}}
 \and M.~Rainer\inst{\ref{inst:inaf}}
 \and R.~Lombaert\inst{\ref{inst:leuven}}
 \and M.~Hillen\inst{\ref{inst:leuven}}
 \and T.~Morel\inst{\ref{inst:liege}}
\and M.~Auvergne\inst{\ref{inst:lesia}}
\and A.~Baglin\inst{\ref{inst:lesia}} 
\and F.~Baudin\inst{\ref{inst:ias}}
\and C.~Catala\inst{\ref{inst:lesia}} 
\and R.~Samadi\inst{\ref{inst:lesia}}
 }

\institute{Instituut voor Sterrenkunde, K.U.Leuven, Celestijnenlaan 200D, 3001 
Leuven, Belgium\label{inst:leuven} 
\and Kavli Institute for Theoretical Physics, University of California Santa Barbara, USA\label{inst:KITP}
\and Department of Astrophysics, IMAPP, Radboud University Nijmegen, PO Box 9010, 6500 GL Nijmegen, The Netherlands\label{inst:nijmegen} 
\and LESIA, Observatoire de Paris, CNRS UMR 8109, Universit\'e Pierre et Marie Curie, Universit\'e Denis Diderot, 5 place J. Janssen, 92105 Meudon, France\label{inst:lesia} 
\and Institut d'Astrophysique et de G\'eophysique Universit\'e de Li\`{e}ge, 
All\'{e}e du 6 Ao\^{u}t 17, 4000 Li\`{e}ge, Belgium\label{inst:liege} 
\and Institut d'Astrophysique Spatiale, CNRS/Universit\'e Paris XI UMR 8617, F-091405 Orsay, France\label{inst:ias}
\and INAF - Osservatorio Astronomico di Brera, via E. Bianchi 46, 23807 Merate (LC), Italy\label{inst:inaf}
\and Universit\'e de Toulouse; UPS-OMP; IRAP;  F-65000 Tarbes, France\label{inst:toul1}
\and CNRS; IRAP; 57, Avenue d'Azereix, BP 826, F-65008 Tarbes, France\label{inst:toul2} 
} 
 
\date{Received 30 November 2011; accepted 8 April 2012} 
\authorrunning{Degroote et al.} 
\titlerunning{The CoRoT B-type binary HD50230}
 
\abstract 
  {B-type stars are promising targets for asteroseismic modelling, since their
  frequency spectrum is relatively simple.}
  {We deduce and summarise observational constraints for the hybrid pulsator, HD\,50230, earlier reported to have deviations from a uniform period spacing of its gravity modes. The combination of spectra and a high-quality light
curve measured by the CoRoT satellite allow a combined approach to fix the
position of HD\,50230 in the HR diagram.}
  {To describe the observed pulsations, classical Fourier analysis was combined
with short-time Fourier transformations and frequency spacing analysis techniques. Visual spectra were
used to constrain the projected rotation rate of the star and the fundamental parameters of the target. In a first approximation, the combined
information was used to interpret multiplets and spacings to infer the true surface rotation
rate and a rough estimate of the inclination angle.}
  {We identify HD\,50230 as a spectroscopic binary and characterise the two components. We detect the simultaneous presence of high-order g modes and low-order p and g-modes in the CoRoT light curve, but were unable to link them to line profile variations in the spectroscopic time series. We extract the relevant information from the frequency spectrum, which can be used for seismic modelling, and explore possible interpretations of the pressure mode spectrum.}
{}

\maketitle

\section{Introduction}
Prior to the launch of the CoRoT space mission, HD\,50230 (spectral type B3, $V$ mag = 8.95) was a poorly known early type star. It is located in an overlapping region in the pulsational Hertzsprung-Russel diagram, where gravity modes and pressure modes are expected to occur. The target was therefore chosen to be observed as part of the study of massive stars within CoRoT's asteroseismology programme \citep{baglin2006,michel2006}, although ground-based all-sky monitoring programmes reveal no variability \citep[e.g., ASAS,][]{pojmanski2002}. The CoRoT light curve, as shown in Fig.\,1 in \citet{degroote2010a}, immediately shows that detection limitations lie at the basis of this non-detection. Indeed, these authors revealed a gravity mode spacing in the low-frequency region below 2\,d$^{-1}$ that was linked with the chemical gradient left by the receding core and suggests extra-mixing around the convective core.

In addition to the gravity modes, $\beta$\,Cep-like pressure modes (above 4\,d$^{-1}$) are clearly visible in the Fourier spectrum (Suppl.\,Fig.\,3 in \citet{degroote2010a}), but up to now they were not studied in detail. This simultaneous occurrence of different types of modes yields the potential of constraining the whole interior of the star, from the core (probed by the gravity modes), to the envelope (probed by the pressure modes).

The simultaneous presence of pressure and gravity modes was already observed in hot B-stars from ground-based data, but the detected gravity mode frequencies in these stars are not always predicted to be excited by the theoretical models. This concerns the stars $\nu$\,Eri \citep{handler2004,deridder2004}, 12\,Lac \citep{handler2006,desmet2009} and $\gamma$\,Peg \citep{handler2009}. In all these examples, however, the pressure modes are dominant over the gravity modes. Space-based data from Kepler \citep{balona2011} uncovered seven additional B-type hybrid stars.

In this paper, we aim at preparing an in-depth seismic study of HD\,50230, motivated by the detection of period spacings. To this end, we perform an analysis of the pressure modes visible in the CoRoT spectrum. Subsequently, we determine the fundamental parameters of the star via a set of spectra obtained with three different instruments within a time span of three years. The observed spectroscopic variability is then finally linked back to the results obtained from the CoRoT light curve.

\section{The CoRoT light curve}

\subsection{Frequency analysis}

The CoRoT observations of HD\,50230 were made during the first long run in the direction of the Galactic anticentre (LRa01), and lasted $\sim$\,137\,d. The observations started at CoRoT Julian Day 2846.95139 (or Julian Day 2454391.95139). The raw light curve contains nearly 400\,000 observations of which $\sim40\,000$ flagged data points are removed because of various instrumental and environmental causes \citep{auvergne2009}. The power in the frequency spectrum is concentrated in the low-frequency region between 0 and 15\,d$^{-1}$. At higher frequencies (15-100\,d$^{-1}$), the only power present is due to harmonics of the satellite's orbital frequency ($f_\mathrm{sat}\approx 13.97$\,d$^{-1}$),  which shows slight variations both in frequency and amplitude during the course of observations. Beyond 120\,d$^{-1}$, no significant power is present.

The long-term linear trend in the light curve is linked to the decline of the CCD's gain \citep{auvergne2009}. We treated it as a multiplicative instrumental effect and divided the light curve with a fitted linear polynomial. This way, we neglect a small deviation from the linear trend between day 70 and day 100, which is also visible in the light curves of the simultaneously observed A5 star HD\,49862 and A2 star HD\,49294, and is therefore probably of instrumental origin. This introduces low amplitude low frequencies (e.g. $f_{013}$ in Table\,\ref{tbl:freqanal}). For a description of the noise properties and the window function, we refer to \citet{degroote2009b}, where a detailed analysis for LRc01 is presented.

\begin{figure*}
 \includegraphics[width=2\columnwidth]{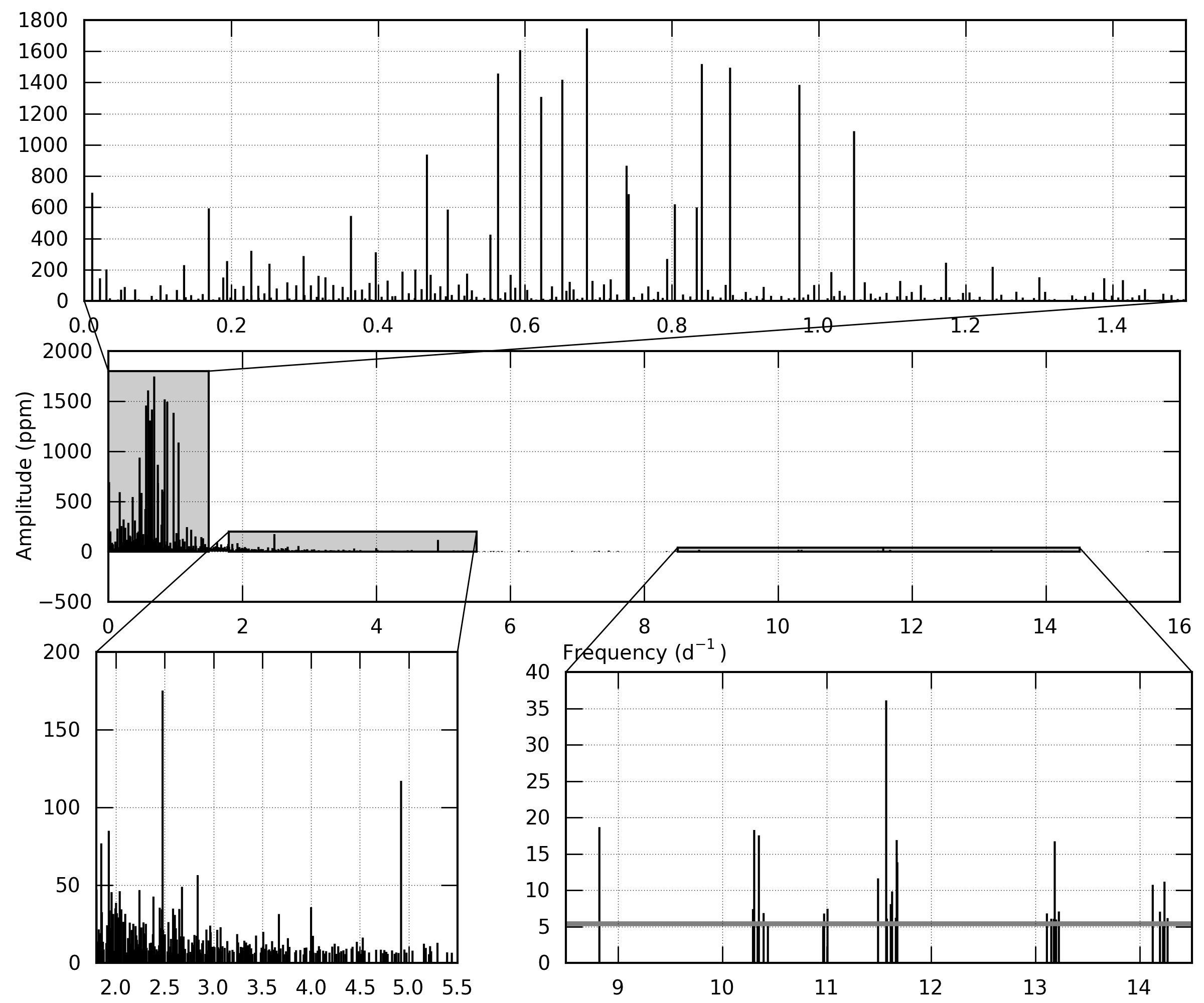}
\caption{Frequency diagram (units of d$^{-1}$ on the $x$-axis, and ppm on the $y$-axis) containing all frequencies from the iterative prewhitening analysis. The centre panel shows the whole range of detected frequencies. The top panel is a zoom-in on the gravity mode region. The bottom right panel is a zoom-in on the pressure mode region (grey line is the detection threshold). The bottom left panel shows the intermediate region with two modes clearly standing out amongst the other frequencies..}
\label{fig:bardiagram}
\end{figure*}

\begin{figure}
\includegraphics[width=\columnwidth]{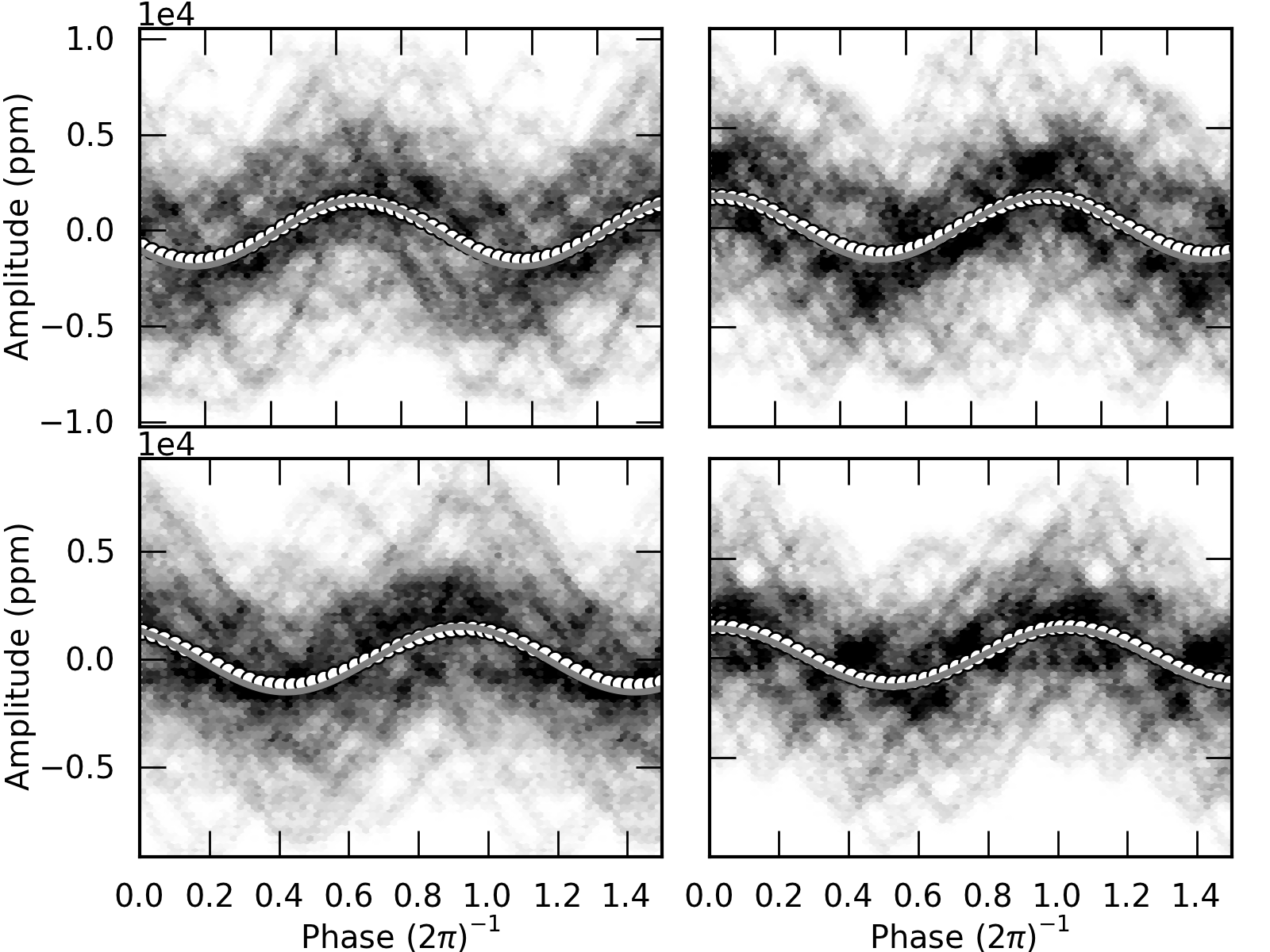}
\caption{Phase-folded light curves of the highest amplitude frequencies show that only one
sine component is needed to fit the
shape (\emph{from left to right, top to bottom:}
$f_{001}$, $f_{002}$, $f_{004}$,
and $f_{005}$). The black dots are observations, the
white circles are phase-binned averages, the thin grey
line is a sine-fit (the exponent of the y-axis scale is given at the top of the plot).}\label{fig:phaseplots}
\end{figure}

\begin{figure}
\includegraphics[width=\columnwidth]{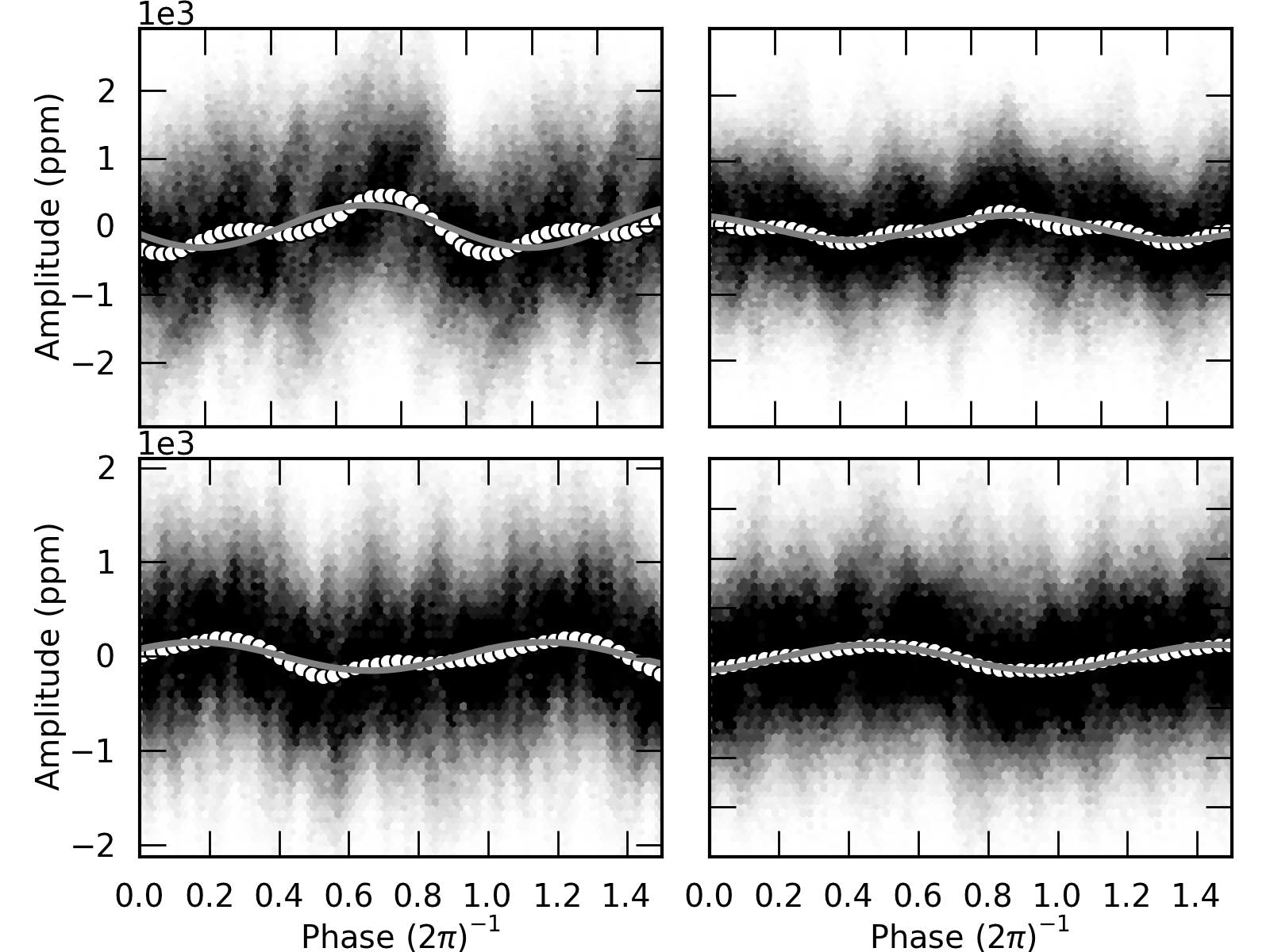}
\caption{Phase-folded light curves, using the frequencies (\emph{from
left to right, top to bottom:} $f_{020}$, $f_{033}$,
$f_{038}$, and $f_{047}$). A clear deviation from the
single-harmonic case in the form of a distorted phase shape
is visible. For an explanation of the symbols, see
Fig.\ref{fig:phaseplots}.}\label{fig:phaseplots2}
\end{figure}

We used an iterative prewhitening procedure, each time with a nonlinear update of the last 20 frequencies \citep[see ][]{degroote2010a} to extract the frequency content from the CoRoT light curve. The results are shown in Fig.\,\ref{fig:bardiagram} and listed in Table\,\ref{tbl:freqanal}. This table contains a list of all frequency identification numbers, the amplitudes, frequency values and phases (including errors), as well as the signal-to-noise ratio (S/N) and the BIC value. All amplitudes range between 5 and 1800 ppm, or, equivalently, between 6\,$\mu$mag and 1.8\,mmag. The general trend is that at higher frequencies, the amplitudes become lower, with a few exceptions (e.g. $f_{034}$ and $f_{050}$). If we limit the analysis to the first 20 frequencies, no signs of nonlinear mode behaviour are detected in the frequency spectrum of HD\,50230, neither in the form of harmonics nor higher-order combination frequencies (e.g. Fig.\,\ref{fig:phaseplots}). If we include the frequencies beyond the first 20, we do identify harmonics and even combination frequencies. However, since the chance of finding combinations increases with the number of detected frequencies, these may be coincidental, and, therefore, only harmonics were taken into consideration. The frequencies for which harmonics have been detected are identified in Table\,\ref{tbl:freqanal} in the last column, together with the relative strength present in the higher harmonics, compared to the main frequency. As an example, for $f_{012}$, we list $A_r(4)=0.38$, which means that four harmonics of $f_{012}$ are detected, and that the sum of the amplitudes of all these harmonics amounts to 38\% of the amplitude of $f_{012}$ itself. Frequencies with a high $A_r$ have a highly distorted phase shape (e.g. Fig\,\ref{fig:phaseplots2}).

\begin{figure}
\includegraphics[width=\columnwidth]{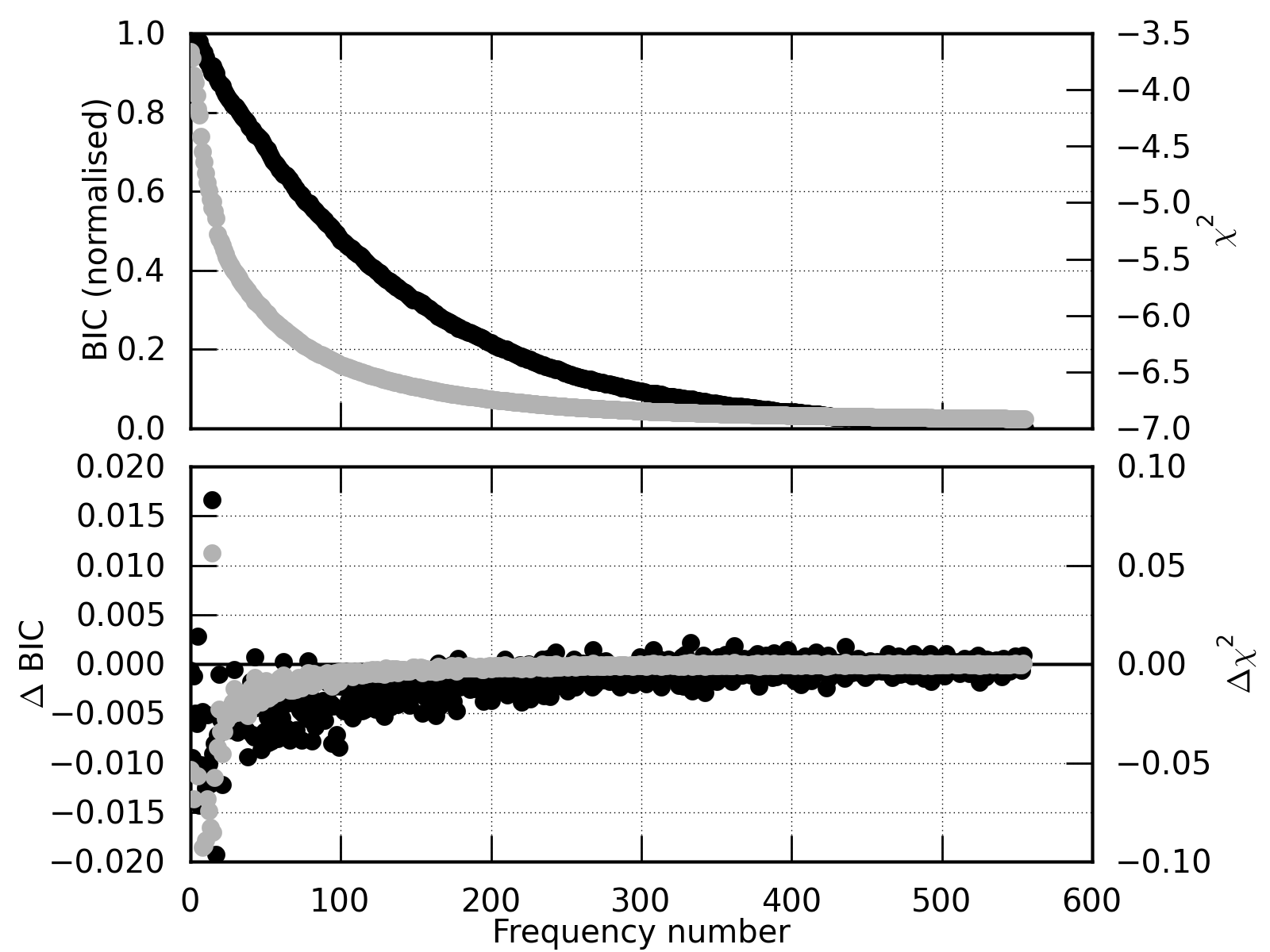}
\caption{Goodness-of-fit statistics of the sine-model. Both
the BIC (\emph{black}) and $\chi^2$ (\emph{grey}) show a general decreasing
trend (\emph{top panel}), suggesting that sines are a good way of modelling the
light curve. Despite the general decrease, however, not all frequencies are
adequately modelled by sines: some frequencies in particular induce a worse BIC
($\Delta {\rm BIC}>0$) (\emph{lower panel}).}\label{fig:bic}
\end{figure}

To test the validity of the sine model, the variance reduction was computed for an increasing number of single-sinusoids. Because our method of iteratively
prewhitening sinusoids necessarily implies an increasing variance reduction, we computed the $\chi^2$ (with constant error terms) and Bayesian information
criterion (BIC) as well \citep[see also][]{degroote2009a}. The latter criterion is only meaningful in a relative way, and weighs the introduction of extra parameters against the gain in variance reduction. The model with the lowest BIC value has the ideal number of parameters of all tested models. In general, the BIC and $\chi^2$ show a decreasing trend when more single-sinusoids are considered, which means that sinusoids are considered to be appropriate models to explain the observed variability (Fig.\,\ref{fig:bic}). When the BIC for each model is explicitly compared to the previous model, we conclude that in some cases adding three parameters to the model does \emph{not} outweigh the resulting variance reduction. Because the BIC puts more weight on the number of parameters used, it puts the threshold for accepting sinusoids from the complete frequency analysis at $\sim200$ frequencies (see column 9 in Table\,\ref{tbl:freqanal}), while the $\chi^2$ allows the use of 300 frequencies. Below these thresholds, the differences between the consecutive BIC and $\chi^2$ are no longer monotonically decreasing.

For the last nine frequencies of the 566 found from the linear frequency analysis, the BIC starts to rise, implying that the additional number of parameters are not worth the gain in variance reduction. This coincides with the use of a p-value of $p=0.1\%$ as a stop criterion under the assumption of an exponential distribution of the Fourier periodogram, and a Bonferroni correction for multiple trial frequencies. The stepwise nonlinearly updated version of the frequency analysis finds nine frequencies less than the linear analysis (using the $p$ value as stop criterion), while it still explains slightly more of the variance.

The high-amplitude gravity mode spectrum is discussed in detail by \citet{degroote2010a}. For completeness, we list the frequencies connected to the period spacing in Table\,\ref{tbl:period_spacing}. They are also marked in italics in Table\,\ref{tbl:freqanal}. In the following, we focus on the (low-amplitude) features in the frequency spectrum above 1.5\,d$^{-1}$.

\begin{table}
\caption{Periods and associated period spacings detected in the gravity mode spectrum.}\label{tbl:period_spacing}
 \begin{tabular}{ccc}
  \hline\hline
Frequency ID & Period (s) & $\Delta P$ (s) \\\hline
$f_{052}$ &    $69854\pm7$  &  $9241.379  \pm   21$\\
$f_{001}$ &    $79096\pm20$ &  $9640.178  \pm   20$\\
$f_{016}$ &    $88736\pm5$  &  $9522.432  \pm    9$\\
$f_{011}$ &    $98258\pm7$  &  $9186.452  \pm   10$\\
$f_{005}$ &   $107445\pm8$  &  $9562.853  \pm   12$\\
$f_{006}$ &   $117008\pm10$ &  $9233.798  \pm   15$\\
$f_{028}$ &   $126241\pm11$ &  $9430.037  \pm   33$\\
$f_{101}$ &   $126241\pm32$ &  -  \\\hline\hline
 \end{tabular}

\end{table}

\subsection{Interpretation of the p-mode frequency spectrum}

\begin{figure}
 \includegraphics[width=\columnwidth]{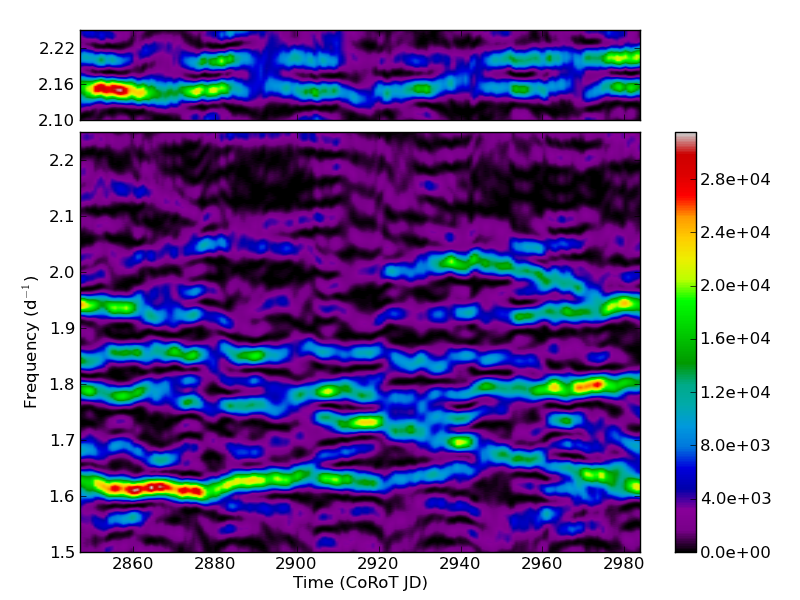}
\caption{Short-time Fourier transformation of the frequency region between 1.5 and 2.3\,d$^{-1}$ with a rectangular window of width 30\,d$^{-1}$. All signal outside of these regions is removed. Clear signal is present (colours denote squared amplitude in ppm$^2$), and the frequencies and amplitudes show hints of variability. As a reference, the top panel shows a copy of the upper part of plot, where two artificial constant sine curves are introduced.}\label{fig:stft_lowfreqs}
\end{figure}

The majority of the frequencies above 1.5\,d$^{-1}$ have amplitudes below 50 ppm, making it difficult to separate them. The origin of these closely packed frequencies (between $\sim$1.5 and $\sim$5\,d$^{-1}$, see, e.g., the lower left panel in Fig.\,\ref{fig:bardiagram}) is not obvious. They are not instrumental, since for quiet stars, the noise level of the CoRoT photometry is much lower. We do not expect them to be an artifact from the prewhitening procedure, partly because we used nonlinear optimisation, but mostly because the frequencies are located far away ($\gg1/T$ with $T$ the total time span of the observations) from the dominant frequencies. They could originate from many closely spaced time-variable modes, or background stellar noise. A short-time Fourier transformation (STFT) shows hints for the time-variability of the modes (Fig.\,\ref{fig:stft_lowfreqs}).

\begin{figure}
 \includegraphics[width=0.495\columnwidth]{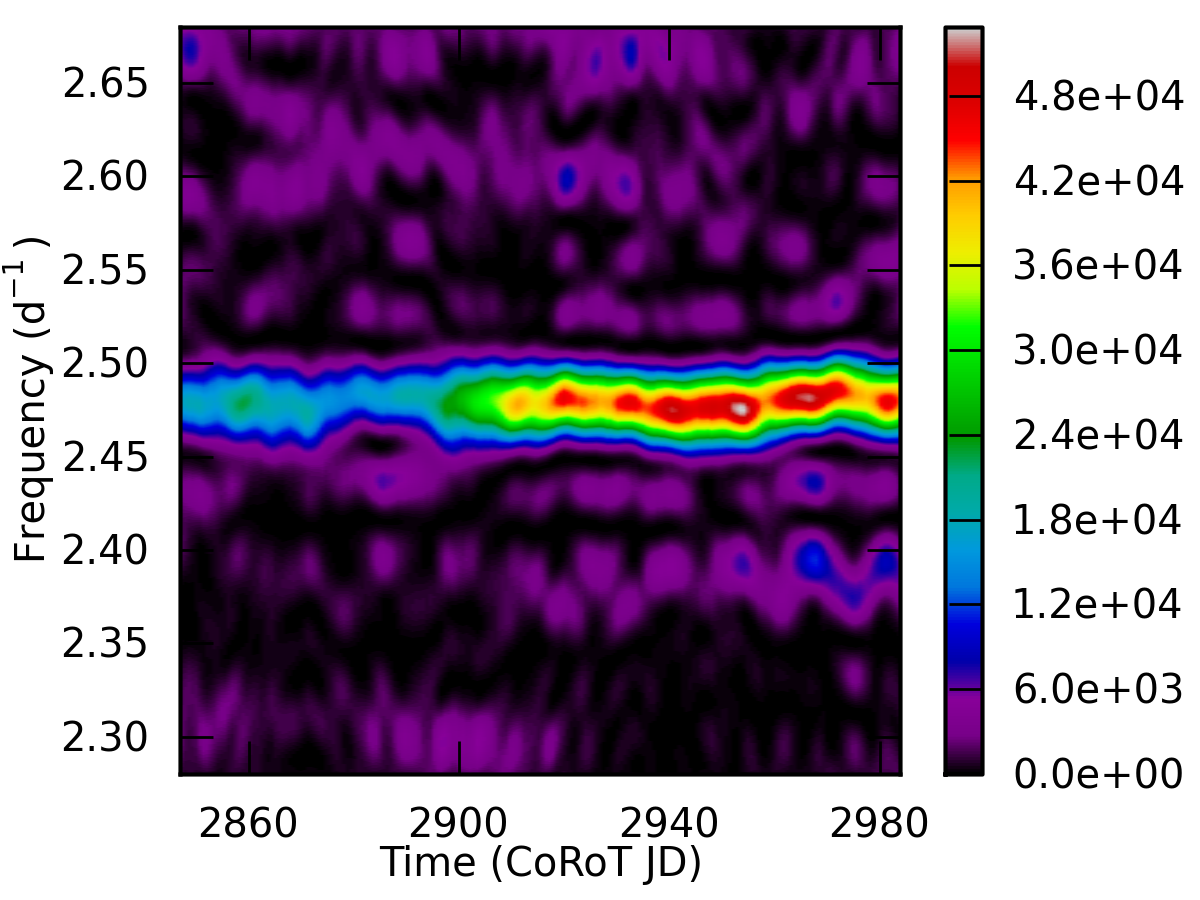}
 \includegraphics[width=0.495\columnwidth]{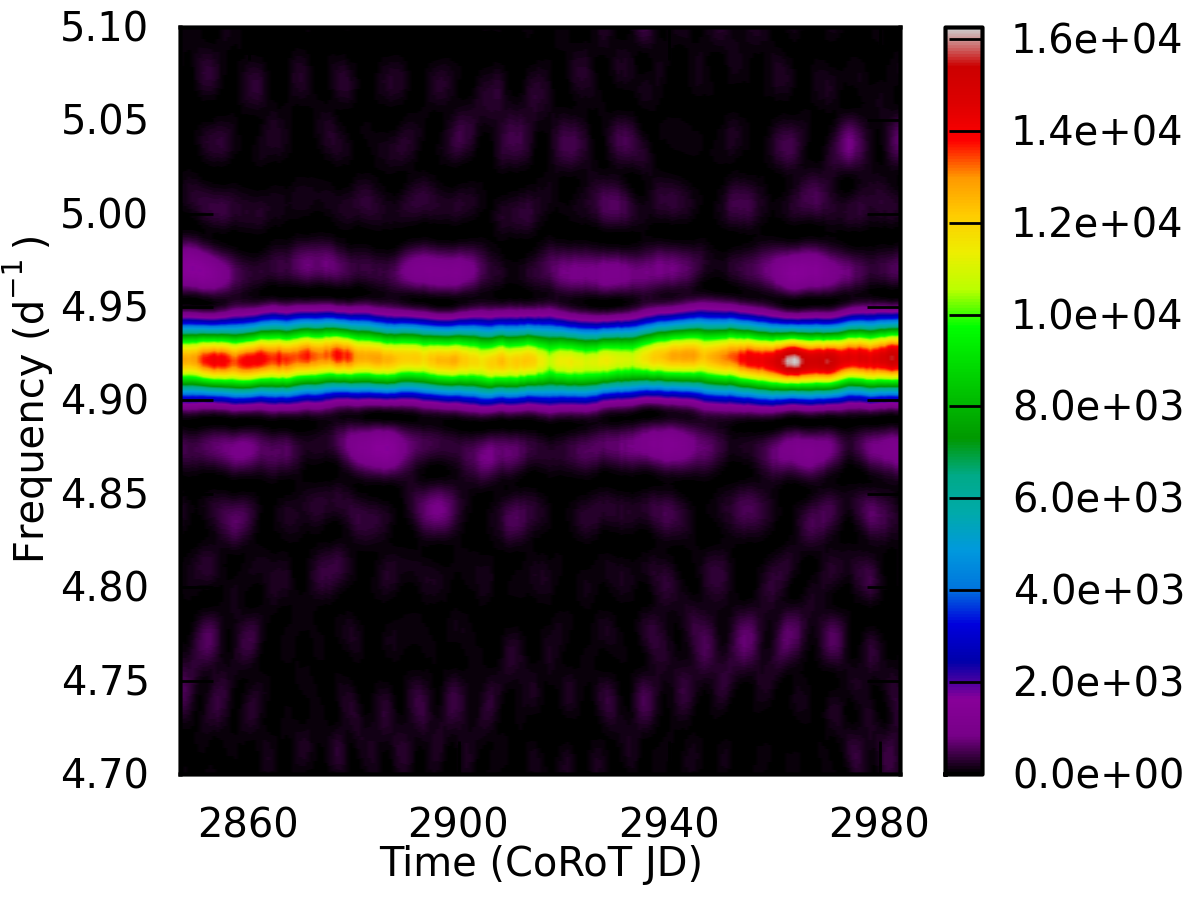}

 \includegraphics[width=0.495\columnwidth]{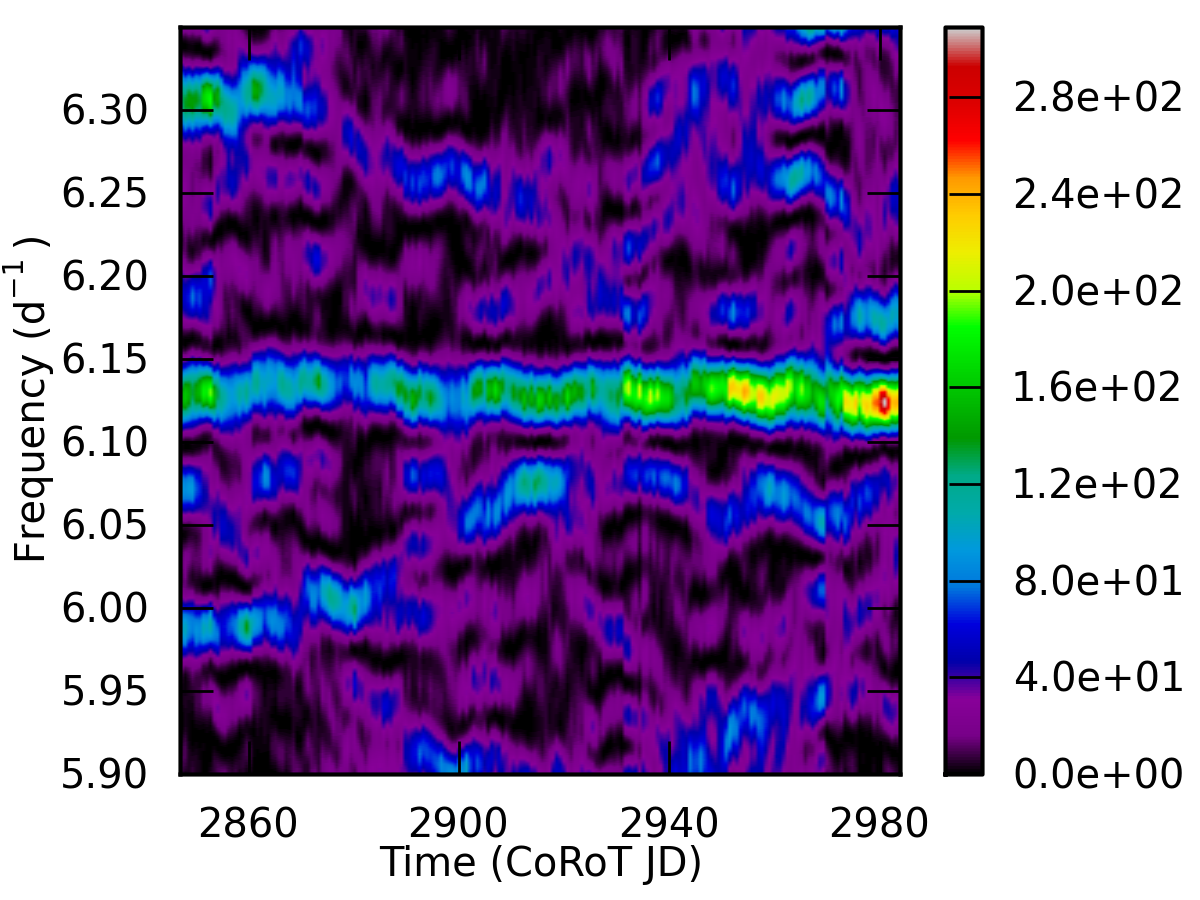}
 \includegraphics[width=0.495\columnwidth]{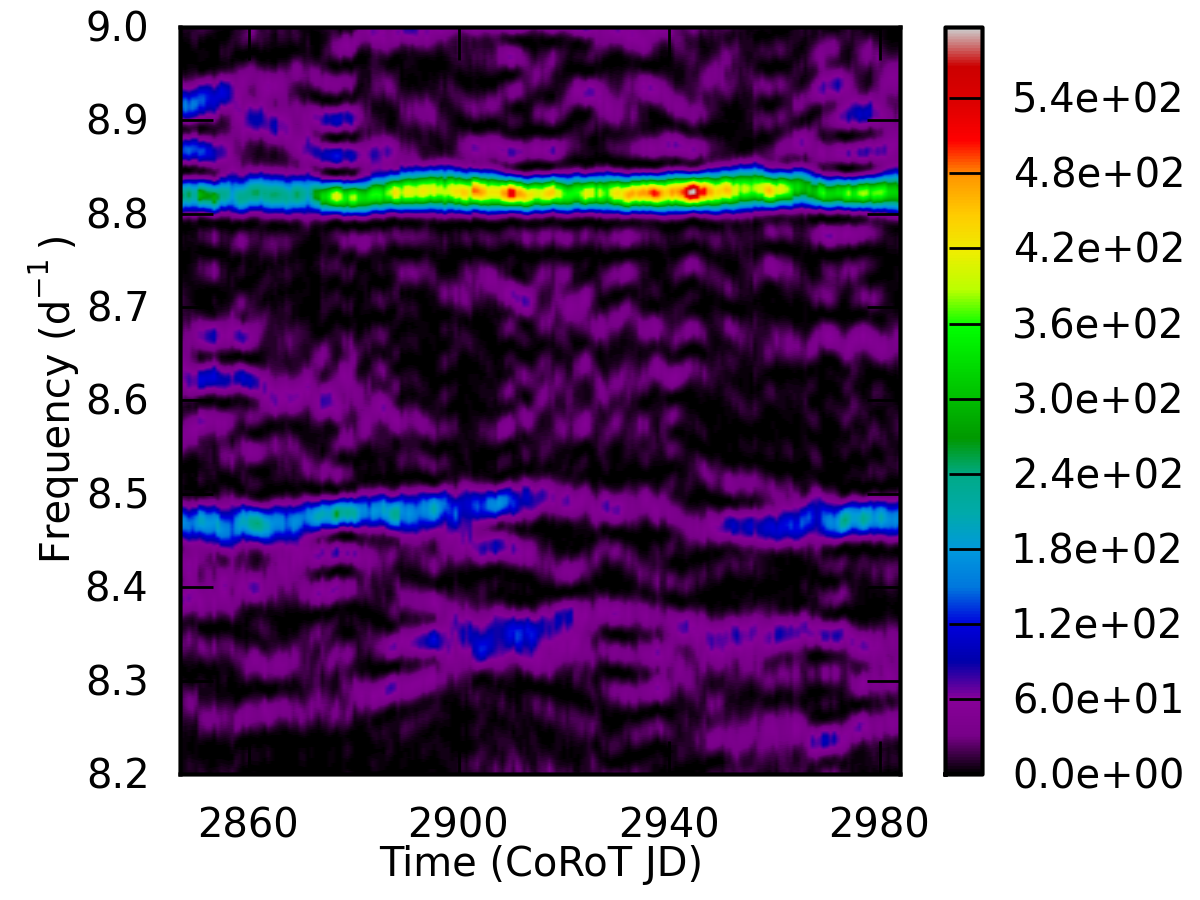}
\caption{Short-time Fourier transformations of four isolated frequencies in the high-frequency region (rectangular window of width 30\,d$^{-1}$, colours denote squared amplitude (ppm$^2$)). In contrast to Fig.\,\ref{fig:stft_lowfreqs}, the frequencies appear throughout the whole light curve.}\label{fig:stft_midfreqs}
\end{figure}

\begin{figure}
 \includegraphics[width=\columnwidth]{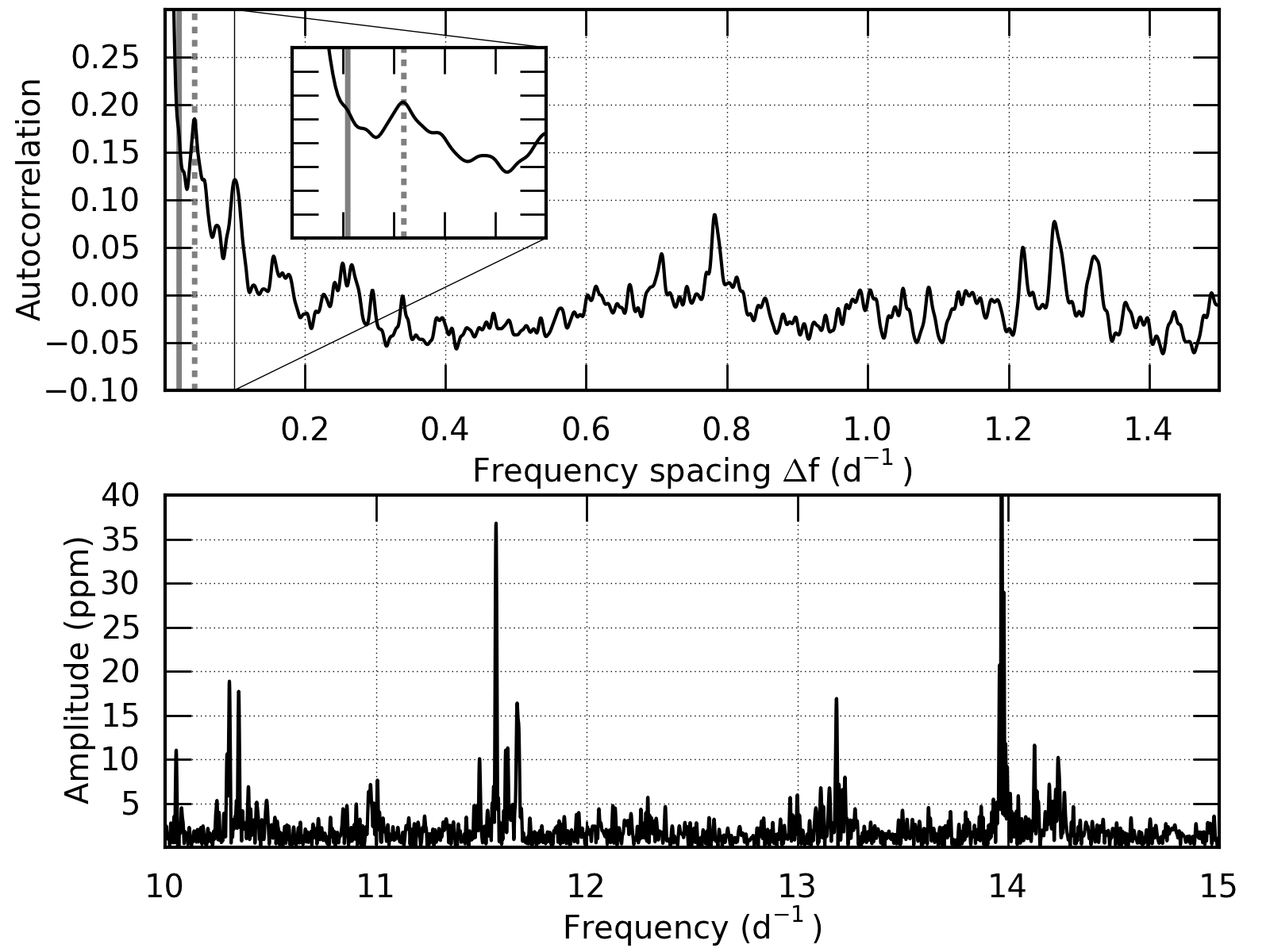}
\caption{Autocorrelation (upper panel) of the frequency spectrum (lower panel) between 10 and 15\,d$^{-1}$ (with the frequencies outside this region removed) uncovers a multiplet spacing of $\Delta_f=0.044$\,d$^{-1}$ that was measured between the peaks near $\sim~10.4$\,d$^{-1}$ and $\sim~11.6$\,d$^{-1}$. The peak around $\sim~0.8$\,d$^{-1}$ originates from the spacing between the structures around 13.2\,d$^{-1}$ and 14\,d$^{-1}$ (bottom panel), and those at 10.3\,d$^{-1}$ and 11\,d$^{-1}$. The thick grey line is plotted at a separation of 3/$T$ for comparison with the frequency resolution.}\label{fig:ac}
\end{figure}

\begin{figure}
 \includegraphics[width=\columnwidth]{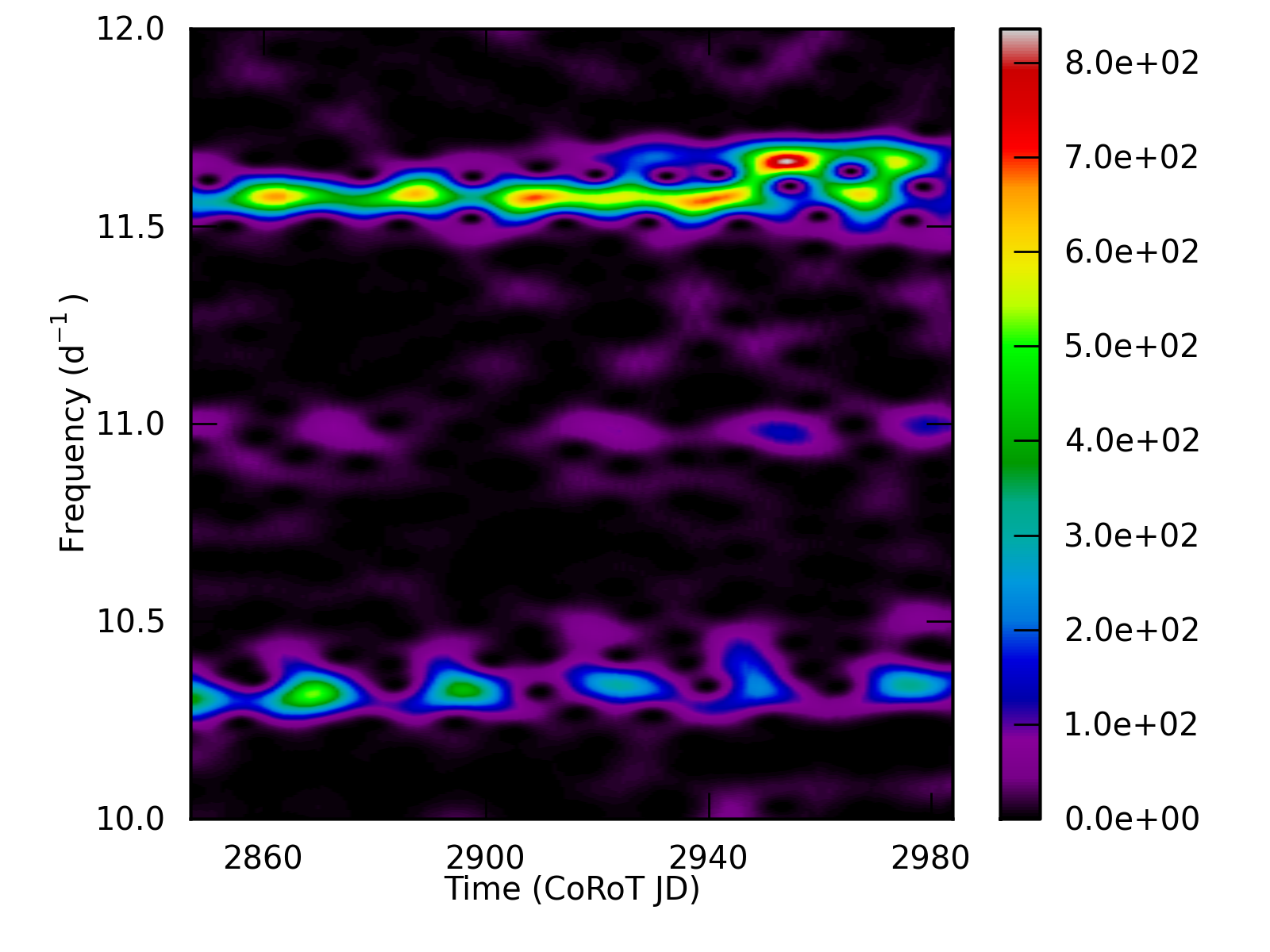}
\caption{The short-time Fourier transformation of the frequency region between 10 and 12\,d$^{-1}$ (with a Hamming window to improve the dynamical range, window width of 20\,d$^{-1}$). The typical succession of constructive and destructive interference are typical signs of a barely resolved multiplet structure. The separation between the frequencies is of the order of $6/T$.}\label{fig:stft_highfreq}
\end{figure}

In addition, some time-independent frequencies clearly stand out, and are therefore good candidate pulsation modes (Fig.\,\ref{fig:bardiagram}). This concerns 17 frequencies, marked in boldface in Table\,\ref{tbl:freqanal}.  Some of these candidate pressure modes are isolated (Fig.\,\ref{fig:stft_midfreqs}), while others have a clear multiplet structure (Fig.\,\ref{fig:ac}, lower panel). In the STFT (Fig.\,\ref{fig:stft_highfreq}), the latter show the typical interference patterns connected to beatings: when the spacing between two frequencies is $x/T$, we expect to see $x$ times of constructive or destructive interference.

The autocorrelation of the periodogram between 10 and 15\,d$^{-1}$ (Fig.\,\ref{fig:ac}) reveals a well-resolved splitting $\Delta f_\mathrm{obs}=0.044\pm0.007$\,d$^{-1}$, possibly related to rotational splitting. For slow rotators, the rotational splitting of pressure modes can be approximated with \citep{ledoux1951}
\[\Delta f_\mathrm{rot} = m\beta_{n\ell}\Omega,\]
where $\beta_{n\ell}$ is dependent on the stellar structure, number of nodes $n$ and degree $\ell$, but typically close to unity. $\Omega$ is the rotation frequency of the star, and $m$ the azimuthal order of the mode. The value of $\Delta f_\mathrm{obs}$ is too low to be a large frequency separation between pressure modes of equal degree and consecutive radial order in the asymptotic approximation. Also, it is unlikely that this value is the low-frequency separation between modes of different degree, since the exact same pattern emerges around 10.4\,d$^{-1}$ and 11.6\,d$^{-1}$, which would imply that the large spacing is exactly the same for all degrees and that all modes of a given radial order but different degree cluster around the same value. The first-order interpretation of $\Delta f_\mathrm{obs}$ as the rotation frequency $\Omega$ is compatible (but delivers no additional constraints) with stellar models between 5 and 10\,$M_\odot$ and stellar radii between 2 and 9\,$R_\odot$ and a projected rotational equatorial velocity $v_\mathrm{eq}\approx 6.9$\,km\,s$^{-1}$, with a restraint of the inclination angle $i$ to be $i>20^\circ$.

A hint for a larger spacing around 0.8\,d$^{-1}$  is also discernible in the two panels of Fig.\,\ref{fig:ac}. The peak in the autocorrelation at 0.8\,d$^{-1}$ in the top panel of Fig.\,\ref{fig:ac} is linked to the multiplets around 13.2\,d$^{-1}$ and 14\,d$^{-1}$ and the multiplets around 10.3\,d$^{-1}$ and 11\,d$^{-1}$. The peak at 1.25\,d$^{-1}$ is tied to the multiplets around 10.3\,d$^{-1}$ and 11.56\,$d^{-1}$. Detailed seismic modelling is needed to interpret this as a large frequency separation between equal-degree modes, though the radial order of these modes is probably too low to be in the asymptotic regime.

\section{Fundamental parameters}

\subsection{Multicolour photometry: an early B-type main sequence star}
An estimate of the fundamental parameters of HD\,50230 is obtained by fitting model atmospheres to observed multicolour photometry. We chose this approach over the application of existing calibration schemes \citep[e.g., ][]{cramer1984}, because fitting model atmospheres directly allows for an easy insertion of up-to-date atmosphere models, and allows for the combined use of different photometric systems. For early-type stars, the shape of the spectral energy distribution (SED) in the blue part, before (and including) the Balmer jump, is most sensitive to changes in the fundamental parameters. In general, SEDs are not very sensitive to binarity, because the luminosity scales with the square of the radius of the star and fourth power of effective temperature. For main sequence binaries, the brightest star is expected to be the hottest and the largest. To cover a wide spectral range and to emphasise the UV region, we collected TD1 \citep{cat_td1}, ANS \citep{cat_ans}, Geneva \citep{cat_gcpd}, and 2MASS \citep{cat_2mass} measurements. Although there are also Johnson \citep{johnson} measurements available, we decided not to include them here. The Johnson filter system is often used to denote a wide range of slightly different response curves, and including them does not increase the wavelength coverage.

To convert the Vega-system magnitudes to fluxes, we used the calibration of \citet{calib_geneva}, \citet{maizapellaniz2006}, and \citet{calib_ans} in combination with the Vega model of \citet{bohlin2004}. We chose to fit the TLUSTY model atmospheres of \citet{lanz2007}, because they include non-LTE effects, which can influence the shape of the SED of early-type stars.

The comparison between model atmospheres and observations for the determination of fundamental parameters was made via a grid-based approach as described in \citet{degroote2011}. The goodness-of-fit and confidence intervals (CI) of the parameters were determined using a $\chi^2$ statistic with five degrees of freedom (the effective temperature $T_\mathrm{eff}$, surface gravity $\log g$, metallicity $Z$, interstellar reddening $E(B-V)$, and angular diameter $\theta$). Because of the grid-based approach, correlation effects on the parameter uncertainties are naturally taken into account.

\begin{table*}
\centering \caption{Fundamental parameters of HD\,50230 and their $95\%$ CI.}
\label{tbl:photfundpars}
\begin{tabular}{lcccccc}
\hline\hline
                                               & \multicolumn{3}{c}{Photometric}                      & Spectroscopic      & Adopted$^c$\\
Parameter                                      & CI$_\mathrm{low}$ & CI$_{50\%}$ & CI$_\mathrm{high}$ &                             \\\hline
$T_{1,\mathrm{eff}}$ (K)\tablefootmark{a}      & 15800             & 18000       & 21500              & $18500\pm1000$     & $18000\pm1500$\\
$T_{2,\mathrm{eff}}$ (K)                       & -                 & -           & -                  & $\leq16000$        &  $\leq16000$\\
$\log [g_1$ (cm/s$^2$] (dex)                   & 3.00$^d$          & -           & 4.50               & $3.8\pm0.3$        & $3.8\pm0.3$\\
$\log [g_2$ (cm/s$^2$] (dex)                   & -                 & -           & -                  & -                  & $\sim4$\\
$\log Z/Z_\odot$\tablefootmark{b} (dex)        & -1.0              & 0.3         & 0.3                & -                  & -\\
$E(B-V)$ (mag)                                 & 0.12              & 0.19        & 0.27               & -                  & $0.19\pm0.08$\\
$\theta$ (mas)                                 & 0.037             & 0.040       & 0.045              & -                  & $0.040\pm0.005$\\
\hline\hline
\end{tabular}
\tablefoot{\tablefoottext{a} Subscript `1' and `2' denote primary and secondary, respectively. The secondary component is not detectable from multicolour photometry.\tablefoottext{b} $Z_\odot=0.02$ dex.
\tablefoottext{c} The adopted values are compromises between photometric and spectroscopic determinations where both are available.
\tablefoottext{d} The $\log g$ is photometrically not constrained.}
\end{table*}

\begin{figure}
 \includegraphics[width=\columnwidth]{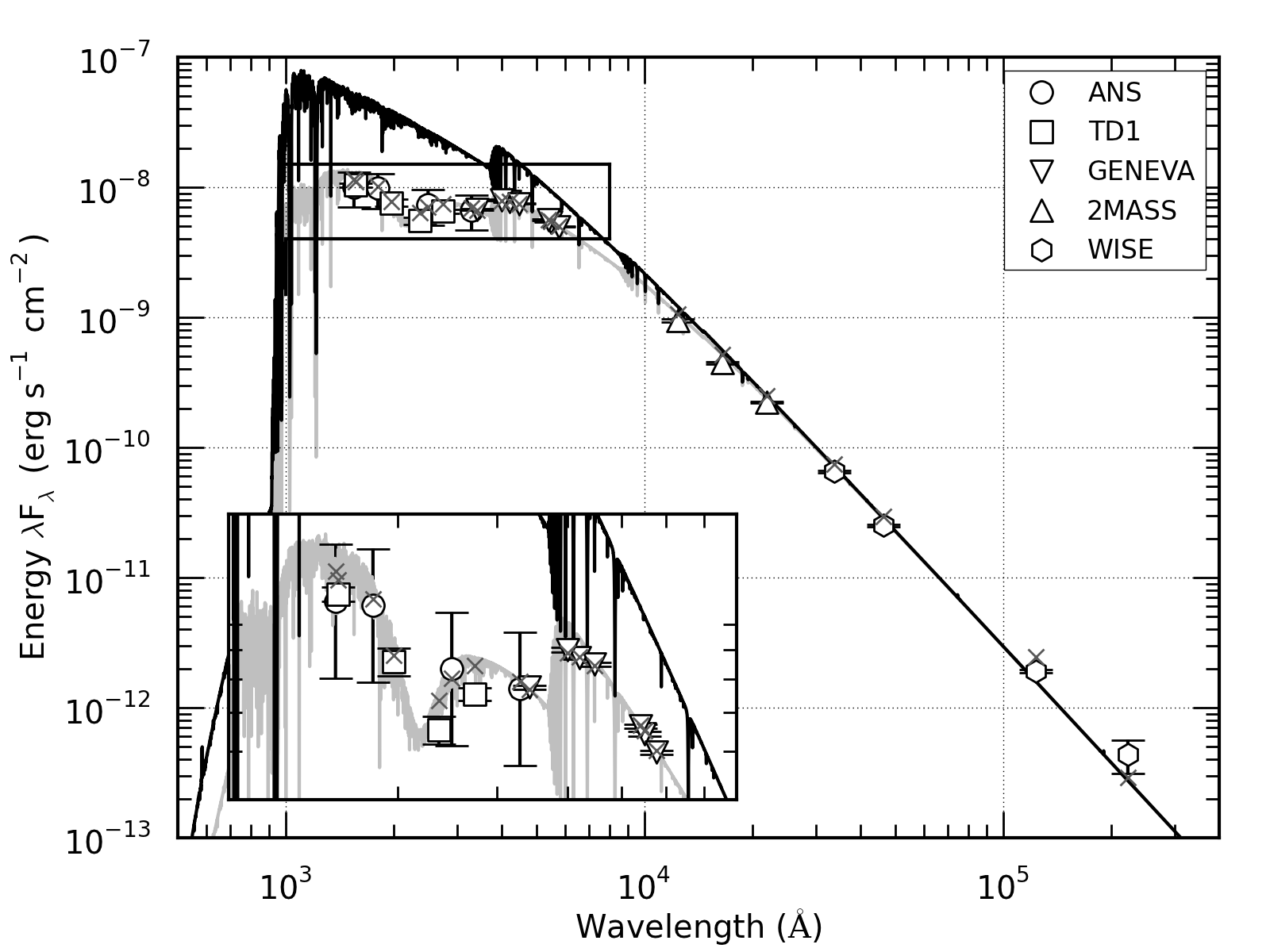}
\caption{Spectral energy distribution of HD\,50230. Observations and their associated errors are shown with different open markers (see legend). Synthetic photometry from the best-fitting model (grey solid line, the unreddened model is shown as the black solid line) are marked with crosses. The inset is a zoom on the ultraviolet region.}\label{fig:sed}
\end{figure}

After fixing the fundamental parameters of HD\,50230, we searched for near-infrared photometry obtained by the WISE satellite \citep{cat_wise}, which we calibrated following \citet{calib_wise}. We observed no significant infrared excess. The results are listed in Table\,\ref{tbl:photfundpars} and the final SED is shown in Fig\,\ref{fig:sed}.

\subsection{Spectroscopy: a suspected binary}
Between 11 October 2008 and 10 January 2012, we have gathered 60 high-resolution spectra from three different instruments: 17 spectra from the \textsc{CORALIE} spectrograph
\citep[$R\approx 50\,000$, 3870-6900\AA, ][]{coralie} mounted on the Euler telescope at La Silla (Chile), 40 spectra from the \textsc{HERMES} spectrograph
\citep[$R\approx 85\,000$, 3770-9000\AA, ][]{hermes} on the Mercator telescope on La Palma (Spain), and four spectra from the \textsc{HARPS} spectrograph in the high-efficiency mode
\citep[EGGS, $R\approx 80\,000$, 3800-6800\AA, ][]{harps} on the ESO 3.6m telescope at La Silla (Chile). A full log of the observations can be found in Table\,\ref{tbl:logbook}. The first column shows the UT of mid-exposure, the second column contains the name of the instrument, and the last two columns give the exposure time and S/N of the spectra around 4500\,$\AA$.

A first look at the spectra shows many narrow and shallow lines, suggesting a low projected rotational velocity. The high resolution of the \textsc{HARPS} spectra allows the Fourier decomposition method \citep[see, e.g., ][]{simondiaz2007} to measure $v_\mathrm{eq}\sin i$ values as low as $\sim$2 km\,s$^{-1}$ irrespective of any additional pulsational broadening effects. To obtain the best possible value, we averaged the two highest signal-to-noise ratio spectra and applied the Fourier decomposition on a set of 78 lines (from ions AlII, AlIII, ArII, CII, FeII, FeIII, NII, NeI, OI, OII, SII, SIII, SiII, and SiIII), with depth and isolation as selection criteria. We derive $v_\mathrm{eq}\sin i=6.9\pm1.5$\,km\,s$^{-1}$, assuming a linear limb-darkening law with $\epsilon=0.35$. This low value is consistent with the observation that the two peaks of the narrow MgII\,4481 doublet are visibly separated. HD\,50230 is therefore a slow rotator.

\begin{figure}
 \includegraphics[width=\columnwidth]{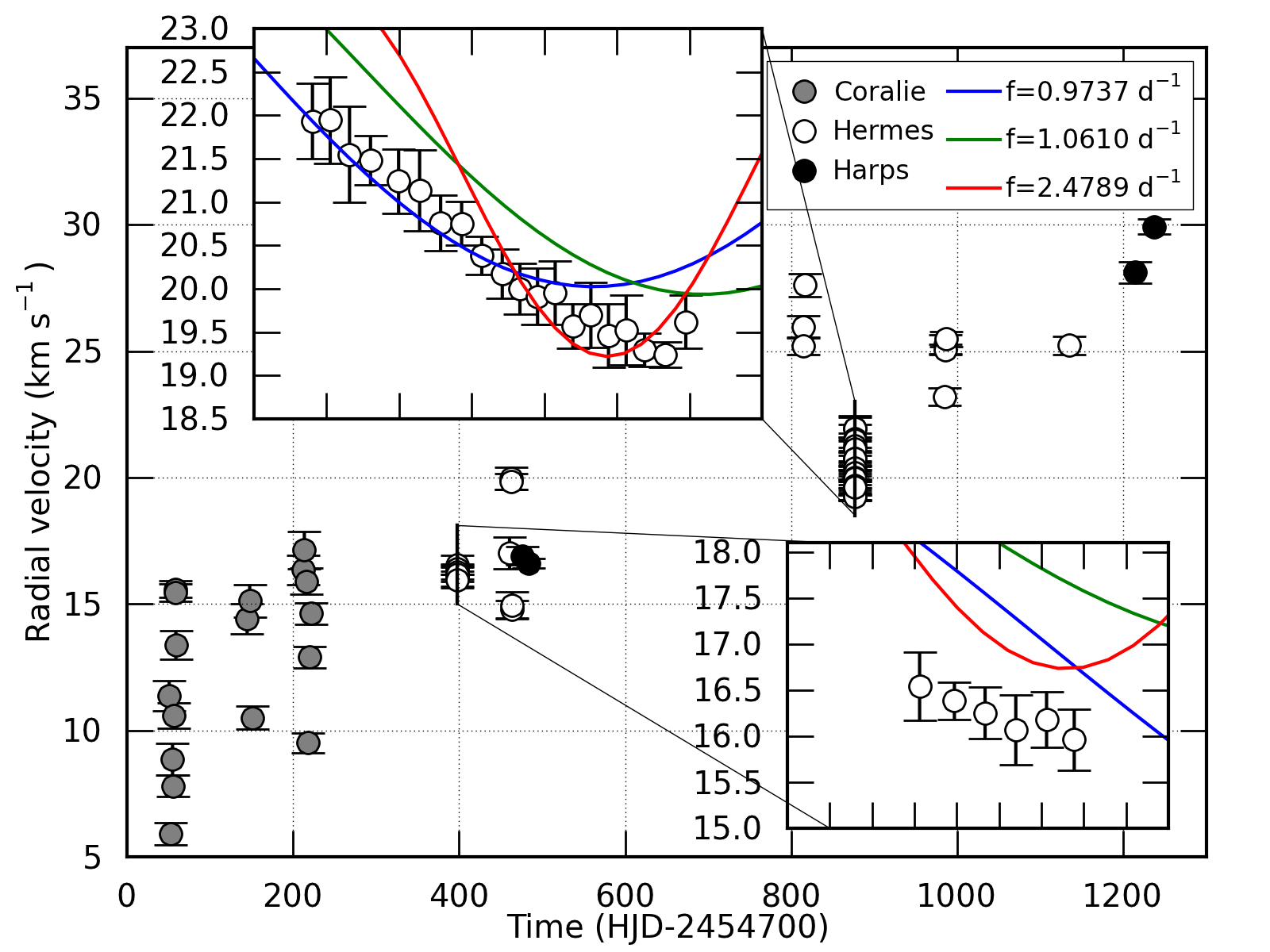}
\caption{Radial velocity derived from each individual spectrum via the LSD method. The inset in the upper left corner shows a zoom on an eight hour continuous monitoring of HD\,50230,
the inset in the lower right corner shows a zoom on a two-hour continuous monitoring. In colours, three fits with different frequencies are shown, vertically displaced to remove the day-to-day radial velocity shift.}\label{fig:rv}
\end{figure}

\begin{figure}
 \includegraphics[width=\columnwidth]{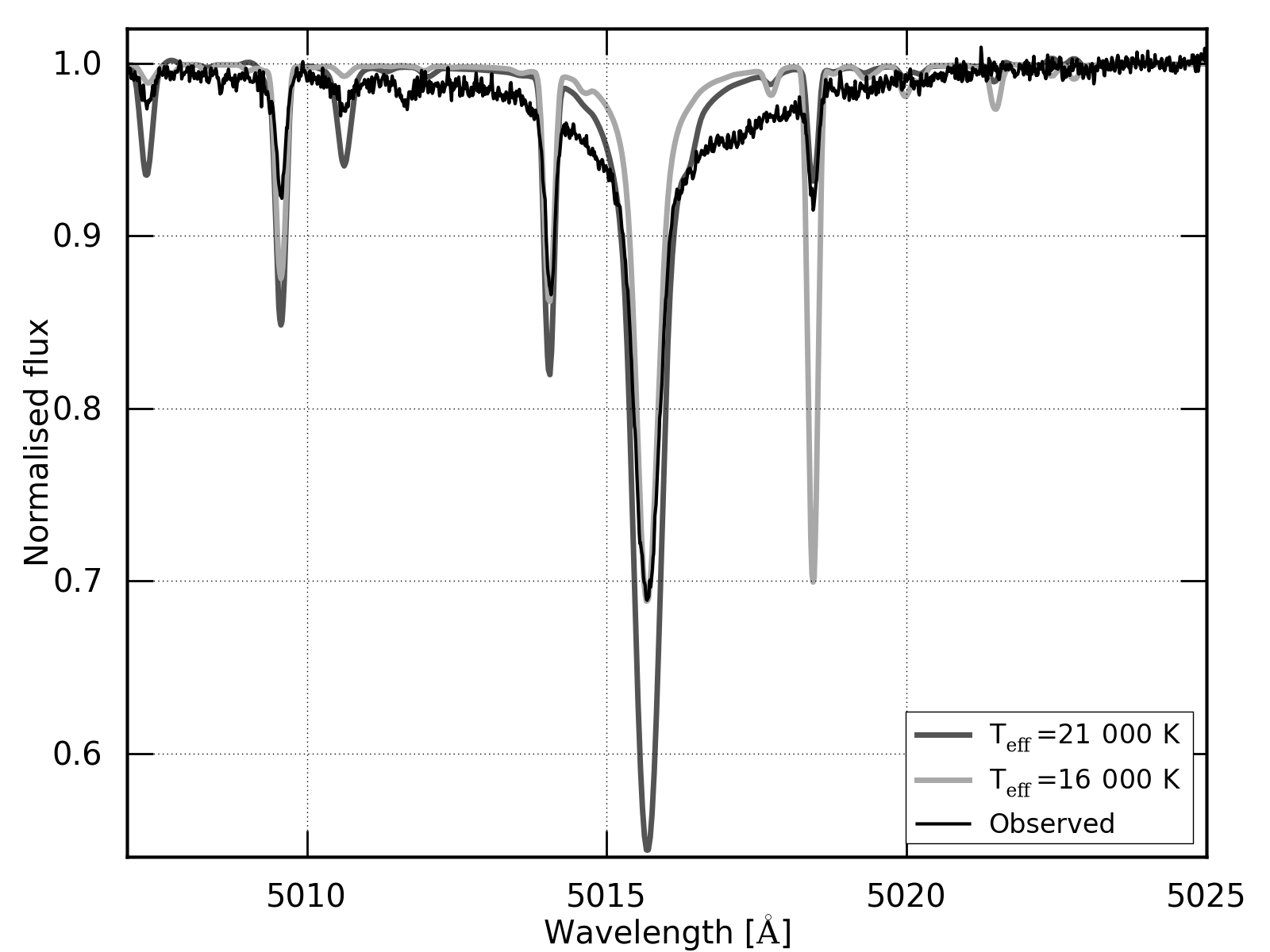}
\caption{HeI profile (black) of the averaged spectra. In grey, two single-star model HeI profiles are shown with different effective temperatures. No single adopted temperature can explain the wings of the profile, suggesting the presence of a broad-lined companion.}\label{fig:companion}
\end{figure}

To improve the estimation of the fundamental parameters based on the spectral features, we constructed an average spectrum. The individual spectra were normalised via spline fits to the continuum, and were corrected for radial velocity shifts. The radial velocity of each spectrum was computed via the least-squares deconvolution algorithm (LSD) of \citet{donati1997}, with varying assumptions on the template spectra and selected lines. The measurements show a clear continuously rising trend during the three years of monitoring (Fig.\,\ref{fig:rv}). To check whether the variability is caused by pulsations or binarity, we used the generalised least-square method of \citet{zechmeister2009} to fit sines and Keplerian orbits to the data. This method has the advantage over other period-finding algorithms such as the Scargle periodogram \citep{scargle1982} that it also fits a constant term to the data, avoiding the problem of determining the average radial velocity in case of poor sampling. At the same time, the method still has the benefits of fitting functions directly to the data, in contrast to phase-binning methods such as the phase dispersion minimisation technique \citep{stellingwerf1978}. No significant periods are found. We conclude that the rising trend is most likely caused by a companion in a wide binary system. Instrumental zeropoint offsets, which can be of the order of 1.5\,km\,s$^{-1}$ \citep{uytterhoeven2008}, are ruled out because the peak-to-peak difference is much larger and the trend is also visible in the spectra from \textsc{Harps} or \textsc{Hermes} only. This is confirmed in the time-average, radial-velocity-corrected spectrum, where broad absorption features are discernible. These are most apparent around the He and Si lines (Fig.\,\ref{fig:companion}), but also in the red wing of the H$\alpha$ line, where a small asymmetry is observed (Fig.\,\ref{fig:hd50230:linefits4}, bottom right panel). It is highly unlikely that the broad features are due to pulsationally induced line profile variations, since they are not symmetric in the averaged spectra, and the deviation is only apparent close to the continuum.

\subsection{Spectroscopic fundamental parameters}

The fundamental parameters of HD\,50230 as determined from photometry are listed in Table\,\ref{tbl:photfundpars}. The surface gravity is not constrained except for a lower limit of $\log g\approx\,3$\,dex. Alternatively, the fundamental parameters can be determined with the help of spectra in two ways: by fitting the line profiles directly with the help of synthesised spectra \citep[e.g., ][]{lehmann2011}, or by comparing observed and predicted equivalent widths (EWs) \citep[e.g.,][]{morel2006}. The former method has the advantage of a more concise way of handling blends, while the latter benefits from a reduction in the number of free parameters (i.e., $v_\mathrm{eq}\sin i$ and the radial velocity). We combined the two approaches by using the EW method \citep[for a description of the grid, see ][]{morel2006} to estimate the effective temperature of the primary component based on carefully selected lines, and turned to the spectral fitting method to incorporate the secondary component.

\begin{figure}
 \includegraphics[width=\columnwidth]{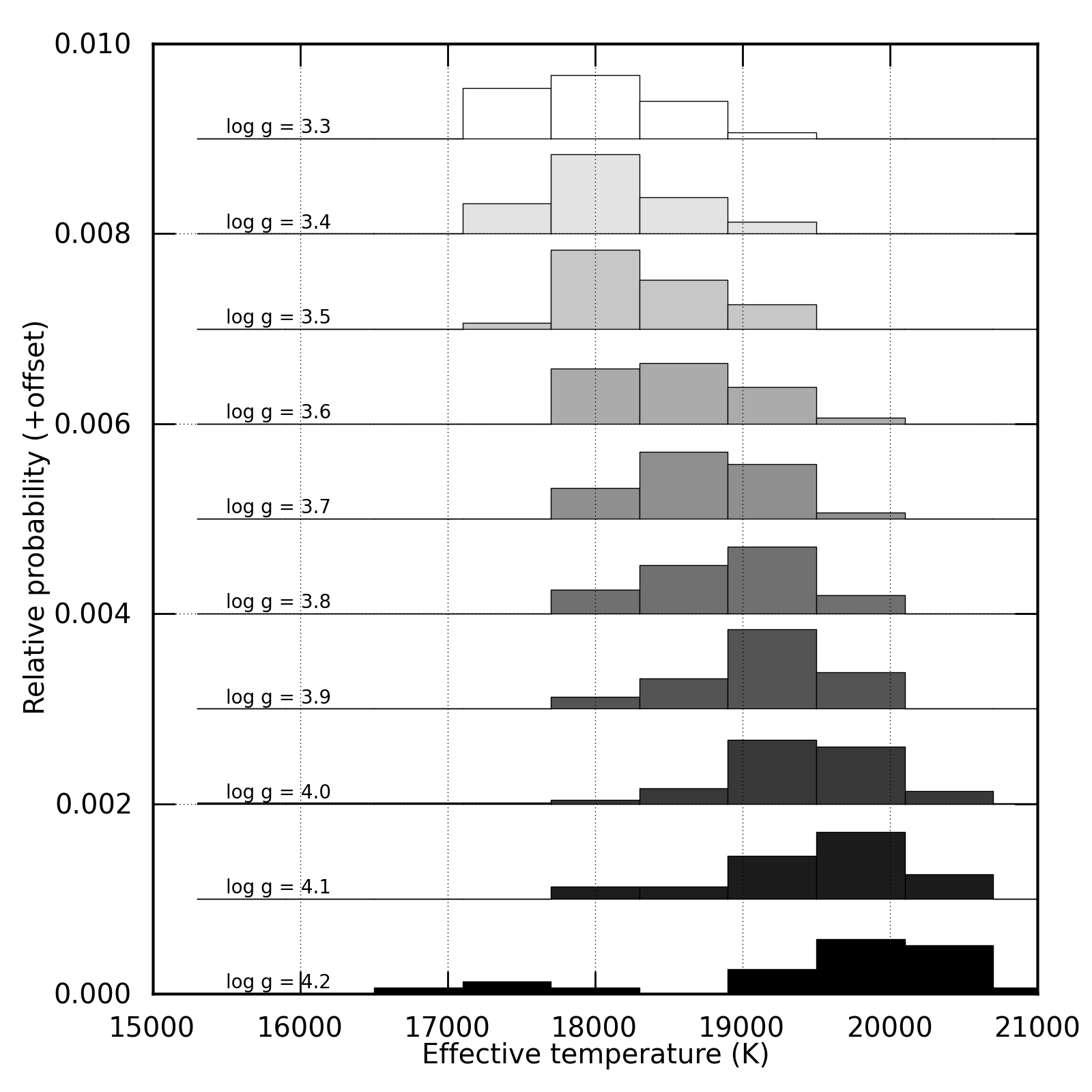}
\caption{Histograms of effective temperatures of the primary component derived from line ratios of SII/SIII and SiII/SiIII, assuming different values for the surface gravity $\log g$.}\label{fig:ew_calib}
\end{figure}

The EW-ratio of ions belonging to different ionisation stages of the same element are highly sensitive to changes in $T_\mathrm{eff}$, but are fairly insensitive to binarity, as long as the continuum flux ratio of both components is similar in the wavelength range under study, or trivially when the secondary component is not visible in the spectrum. In the temperature range between 16\,000 and 22\,000\,K, useful ions are SiII and SiIII, and SII and SIII. For the Si lines, this includes the ratios of SiII6371 to SiIII4567, SiIII4574, SiIII5739 and SiIII4552. The SiII5041, SiII5056, SiII4128, and SiII4130 lines were discarded because of the uncertainties in the atomic modelling of these high-energy transitions \citep{simondiaz2010}. For the sulfur atom, we selected SII5032, SII4815, SII4230, SII4162, SII4716, SII4991, SII4294, SII4463, SII4824, SII4656, SII4792, SII4278, and SII4217 from the first ionisation stage, and SIII4284 and SIII4253 from the second ionisation stage. The observed EWs of the selected unblended features were calculated by numerical integration of the time-averaged normalised spectra (weighted with the S/N), and compared to the synthetic EWs. We find $17500<T_\mathrm{eff}<19500$, and notice that lower surface gravity values imply lower effective temperatures (Fig.\,\ref{fig:ew_calib}). We attempted to derive a value for the surface gravity and individual abundances by calculating the abundance of each element predicted by each individual line, adopting each $\log g$ parameter in the grid, and then minimising the scatter. These efforts failed, largely because of the unknown scaling factor that has to be included to compensate for the presence of the companion.

Subsequently, we use the $\chi^2$ fitting method to match the observed HARPS spectra with a grid of TLUSTY synthetic spectra \citep{lanz2007} when $T_\mathrm{eff}\geq15000$, and ATLAS synthetic spectra \citep{palacios2010} when $T_\mathrm{eff}<15000$. Because the secondary component is not visible in all selected lines, we first performed a grid search using a template spectrum of the HeI6678 line and varying the radial velocity of the component $v_r'$ and projected rotational velocity of the component $v_\mathrm{eq}'\sin i'$. From this, we fixed the values to $v_r'=40$\,km\,s$^{-1}$ and $v_\mathrm{eq}'\sin i'=117$\,km\,s$^{-1}$, though the uncertainties are several tens of km\,s$^{-1}$ on both quantities. We assumed equal metallicities for both components, and demanded that the companion is less evolved than the primary ($\log g'>\log g$). The $\chi^2$ minimisation method was performed with effective temperatures of the  primary between 16\,000\,K and 22\,000\,K and between 15\,000\,K and 22\,000\,K for the secondary, surface gravities between 3.5 and 4.5 for the primary and secondary. The metallicity was chosen to be 0.1, 0.2, 0.5, or 1.0 times the solar value. For all 23 lines, the synthetic spectrum was calculated by taking the average of the synthetic profiles from the primary and secondary, weighted with the synthetic Eddington flux. Each $i$th profile was fitted to the data together with a linear continuum via
\[\chi_i'^2 = \frac{1}{N}\sum_{\lambda=\lambda_0}^{\lambda_n} \left(\frac{F_{i,\mathrm{obs}}(\lambda)}{a_i\lambda +b_i} - F_{i,\mathrm{syn}}(\lambda)\right)^2,\]
and the total $\chi^2$ value was calculated as the squared sum of the residuals of all line profiles,
\[\chi_1^2 = \frac{1}{N}\sum_{\lambda\in P} \left(\frac{F_{i,\mathrm{obs}}(\lambda)}{a_i\lambda +b_i} - F_{i,\mathrm{syn}}(\lambda)\right)^2,\]
where the index $\lambda$ runs over all wavelengths in all profiles P. As an alternative test statistic, the $\chi^2$ was also calculated as the average of the $\chi_i'^2$ values of all $N_p$ profiles,
\[\chi_2^2 = \frac{1}{N_p}\sum_i \chi_i'^2.\]
In the former statistic, the broad Balmer and helium profiles dominate because they cover a broader wavelength range. The latter statistic is designed to give equal weight to all profiles, regardless of the intrinsic width.

The fit procedure does not result in an accurate determination of the two components of the binary system, because the fits to several lines are not satisfactory (e.g., Figs\,\ref{fig:hd50230:linefits1}-\ref{fig:hd50230:linefits6}). This could possibly be explained by shortcomings in atomic physics (e.g., in Si), or by metal mixtures significantly different from the solar mixture that is adopted in the models. A histogram weighted with the $\chi^2$-values sets the effective temperature and surface gravity of the primary at $T_\mathrm{eff}=17\,000\,K$ and $\log g=3.5$\,(cgs), while the minimum $\chi^2$ is attained at $T_\mathrm{eff}=19\,000\,K$ and $\log g=4.0$\,(cgs). This range is compatible with the temperature determination from the SED fit and the EW method, which are less influenced by the fainter companion. The effective temperature of the companion is $T_\mathrm{eff}\leq 15\,000$\,K according to the $\chi^2$ fitting (the minimum is obtained at the edges of the grid), while the weighted histogram sets it to $T_\mathrm{eff}=16\,000$\,K. The surface gravity is found to be between $\log g=4.0-4.5$\,(cgs). The most likely value for the metallicity is half the solar value, but cannot be accurately determined from the used fitting procedure. A fit to all lines with $T_\mathrm{eff}=19\,000$\,K and $\log g=4.0$\,(cgs) for the primary, and $T_\mathrm{eff}=15\,000$\,K and $\log g=4.0$\,(cgs), is shown in Figs\,\ref{fig:hd50230:linefits1} to \ref{fig:hd50230:linefits6}. While the binary fit procedure fails at characterising the secondary, the binary line profile fits result in a better explanation for the asymmetry or broad line components in some of the lines. We conclude that $T_\mathrm{eff}=18\,500\pm1000$\,K and $\log g=3.8\pm0.3$\,(cgs) is an acceptable range for the fundamental parameters of the primary, and $T_\mathrm{eff}\leq16\,000$\,K and $\log g\sim 4$\,(cgs) for the secondary.

\section{Line profile variability}\label{sect:lpv}

Pulsational variability can cause radial velocity changes or variability in the line profile itself, depending on the degree and azimuthal order of the mode. They can be detected and identified via a pixel-to-pixel frequency analysis of the profiles \citep[e.g., ][]{zima2006}, or via characterisation of the line profile's moments \citep{aerts1992}. Given the narrow lines, we opted for the latter method.

The combination of the scarce time sampling of the spectroscopic measurements and the continuously rising trend in radial velocities (Fig.\,\ref{fig:rv}) impedes the detection of sinusoidal variability due to pulsations in the line profiles. For that reason, we applied a linear fit to both features simultaneously, using a function of the form
\begin{equation}M(t_i) = a_0 + a_1 t_i + a_2\sin[2\pi(ft_i + a_3)].\label{eq:lin_plus_sine}\end{equation}
Here, $a_0$ and $a_1$ represent the parameters of the straight-line fit, and $a_2$, $f$, and $a_3$ are the amplitude, frequency and phase of the sinusoidal component. We fitted the linear parameters $a_j$ to the time series of moments calculated at all observing times $t_i$, for a grid of test frequencies $f$ that includes all frequencies detected in the CoRoT light curve. To improve the SNR of the individual profiles, we used the LSD method \citep{donati1997} to compute the average profile of the deepest metallic spectral lines. From a fit evaluation via the $F$-statistic we conclude that there are no significant frequencies. A similar procedure excluding a linear trend also yields no significant frequencies.

\section{Discussion and conclusions}

We classified HD\,50230 as a wide double-lined spectroscopic binary consisting of a component with a low projected rotational velocity, and a component with a moderate projected rotation velocity. From the EW ratios, we established an effective temperature around 18500\,K for one component, which agrees with multicolour photometry. From this, we conclude that the narrow-lined component is the primary, and the broad-line component the (cooler) secondary. This is confirmed by a spectral line fitting procedure, where an upper limit for the effective temperature of 16\,000\,K is found. We estimate both stars to be main-sequence stars, and rule out the possibility of a subdwarf companion based on luminosity arguments, although their surface gravities could not be determined accurately from the wings of the Balmer lines, because of the blend of the two components.

The identification of HD\,50230 as a spectroscopic binary complicates the interpretation of the light curve compared to the single-star hypothesis. However, the period spacing discovered in \citet{degroote2010a} is unlikely to be affected by the binary light curve, partially because of the high amplitudes of the modes considered, but mainly because the chance of observing such a regular pattern as the combination of two independent frequency patterns, are slim. Less obvious is the connection between the gravity modes and the pressure modes as originating from the same star. The companion is less luminous (the flux ratio is below 0.7) and cooler than the primary, placing it far below the $\beta$\,Cep instability strip. It is therefore unlikely to exhibit this family of pressure modes. Moreover, there are strong hints for rotationally split multiplets in the pressure mode regime, of which the narrow spacing is incompatible with fast rotation as occurs in the secondary component. Possibly, the low-frequency region is contaminated by the signal of the companion at low amplitudes because it resides in the SPB strip and may have a flux as high as half that of the primary. Even so, we do not expect that the gravity-mode spectrum, and in particular the gravity mode period spacing, originates from the companion, since cool B-type stars do not typically exhibit a variability spectrum this dense \citep[see, e.g., ][]{balona2011,degroote2011}. We finally remark that, although we give a list of 556 frequencies detected in the CoRoT light curve, the second half of that list should be taken with caution, because they do not satisfy the BIC criterion.

The spectroscopic line profiles show obvious signs of variability on short time scales. In particular, an eight-hour continuous monitoring of the star with HERMES shows a continuously \textit{declining} trend in the radial velocities, in contrast to the previously described orbital \textit{rising} trend. The data set at hand did not suffice, however, to find clear periodicities. The origin of this problem could lie in the dense gravity mode region, making the resulting complex beating pattern impossible to separate from the noise. It is hard to imagine, however, that we could detect frequencies in the line profile variations which are not present in the CoRoT light curve at photometric micromagnitude precision. This makes $f_{034}\approx2.48$\,d$^{-1}$ and $f_{050}=4.92$\,d$^{-1}$ the most likely candidates for causing LPV. The latter frequency is excluded as the main contributor to the radial velocity dispersion, based on the eight-hour continuous observation run, which should cover at least one cycle. If we focus on the measurements taken in close succession phased to $f_{034}$ (Fig.\,\ref{fig:rv}, red curve), we can see that this frequency is indeed likely to be present in the data, but other frequencies are equally successfull at fitting the observations (e.g., Fig.\,\ref{fig:rv}, blue and green curve).

In conclusion, we have gathered observational information to begin an in-depth seismic study of the hybrid B-type pulsator HD\,50230. Despite its slow surface rotation of the order of $v_\mathrm{eq}\approx10$\,km\,s$^{-1}$, the seismic modelling will have to take into account rotational effects when interpreting small deviations from uniform period spacings \citep{aerts2011}. The ultimate future goal is to reconcile the observational information from the period spacings with that of the lower-order mode spectrum, which will allow a description of the whole interior of the star, from the convective core to the radiative envelope.

\begin{acknowledgements} 
PD and CA are grateful for the hospitality during their stay at the Kavli Institute for Theoretical Physics in the framework of the Research Programme Asteroseismology in the Space Age. The research leading to these results has received funding from the European Research Council under the European Community's Seventh Framework Programme (FP7/2007--2013)/ERC grant agreement n$^\circ$227224 (PROSPERITY), from the Belgian PRODEX Office under contract C90309: CoRoT Data Exploitation, and by the National Science Foundation of the United States under Grant No. NSF PHY05-51164.
\end{acknowledgements}

\bibliographystyle{aa} 


\Online

\begin{appendix}
\section{Appendix}
 
\begin{table}

 \caption{Logbook of spectroscopic observations.}

\begin{tabular}{ccccc}\hline\hline
Calendar day (UT)$^a$ & instrument & $T_\mathrm{exp}$ (s) & SNR$^b$\\\hline
2008-10-11 08:14:16  & \textsc{Coralie} & 1801 & 56\\
2008-10-13 07:57:30  & \textsc{Coralie} & 1801 & 50\\
2008-10-15 08:07:28  & \textsc{Coralie} & 1801 & 52\\
2008-10-16 08:16:20  & \textsc{Coralie} & 1801 & 58\\
2008-10-17 08:48:56  & \textsc{Coralie} & 1801 & 64\\
2008-10-18 08:11:23  & \textsc{Coralie} & 1801 & 61\\
2008-10-18 08:42:42  & \textsc{Coralie} & 1801 & 58\\
2008-10-19 08:33:39  & \textsc{Coralie} & 1500 & 58\\
2009-01-13 04:26:21  & \textsc{Coralie} & 1801 & 54\\
2009-01-16 06:33:07  & \textsc{Coralie} & 1801 & 55\\
2009-01-19 04:45:44  & \textsc{Coralie} & 1801 & 54\\
2009-03-22 02:07:09  & \textsc{Coralie} & 1801 & 58\\
2009-03-23 01:57:27  & \textsc{Coralie} & 1801 & 56\\
2009-03-25 02:35:33  & \textsc{Coralie} & 1801 & 44\\
2009-03-27 03:16:36  & \textsc{Coralie} & 1801 & 48\\
2009-03-29 02:16:51  & \textsc{Coralie} & 1801 & 52\\
2009-03-31 02:14:55  & \textsc{Coralie} & 1801 & 55\\
2009-09-23 04:22:44  & \textsc{Hermes} & 800 & 63\\
2009-09-23 04:46:24  & \textsc{Hermes} & 1200 & 71\\
2009-09-23 05:07:17  & \textsc{Hermes} & 1200 & 79\\
2009-09-23 05:28:11  & \textsc{Hermes} & 1200 & 74\\
2009-09-23 05:49:07  & \textsc{Hermes} & 1200 & 76\\
2009-09-23 06:07:56  & \textsc{Hermes} & 950 & 70\\
2009-11-25 04:49:04  & \textsc{Hermes} & 1500 & 74\\
2009-11-27 01:22:10  & \textsc{Hermes} & 900 & 67\\
2009-11-27 01:38:09  & \textsc{Hermes} & 900 & 75\\
2009-11-28 06:11:23  & \textsc{Hermes} & 1200 & 58\\
2009-11-28 06:32:13  & \textsc{Hermes} & 1200 & 58\\
2009-12-10 06:43:32  & \textsc{Harps} & 1200 & 147\\
2009-12-18 06:36:40  & \textsc{Harps} & 1400 & 148\\
2010-11-14 04:05:54  & \textsc{Hermes} & 1800 & 101\\
2010-11-14 05:17:00  & \textsc{Hermes} & 1057 & 74\\
2010-11-16 04:51:22  & \textsc{Hermes} & 2200 & 103\\
2011-01-15 21:22:00  & \textsc{Hermes} & 1000 & 55\\
2011-01-15 21:39:35  & \textsc{Hermes} & 1000 & 42\\
2011-01-15 21:58:45  & \textsc{Hermes} & 1200 & 56\\
2011-01-15 22:19:36  & \textsc{Hermes} & 1200 & 59\\
2011-01-15 22:47:28  & \textsc{Hermes} & 1200 & 59\\
2011-01-15 23:08:19  & \textsc{Hermes} & 1200 & 59\\
2011-01-15 23:29:09  & \textsc{Hermes} & 1200 & 62\\
2011-01-15 23:49:59  & \textsc{Hermes} & 1200 & 66\\
2011-01-15 00:09:59  & \textsc{Hermes} & 1100 & 65\\
2011-01-15 00:30:18  & \textsc{Hermes} & 1000 & 65\\
2011-01-15 00:47:48  & \textsc{Hermes} & 1000 & 71\\
2011-01-15 01:05:19  & \textsc{Hermes} & 1000 & 64\\
2011-01-15 01:22:49  & \textsc{Hermes} & 1000 & 61\\
2011-01-15 01:40:19  & \textsc{Hermes} & 1000 & 62\\
2011-01-15 01:58:12  & \textsc{Hermes} & 1000 & 69\\
2011-01-15 02:15:43  & \textsc{Hermes} & 1000 & 65\\
2011-01-15 02:33:13  & \textsc{Hermes} & 1000 & 62\\
2011-01-15 02:51:33  & \textsc{Hermes} & 1100 & 65\\
2011-01-15 03:11:33  & \textsc{Hermes} & 1200 & 73\\
2011-01-15 03:32:24  & \textsc{Hermes} & 1200 & 71\\
2011-05-03 21:19:35  & \textsc{Hermes} & 2200 & 66\\
2011-05-04 20:52:21  & \textsc{Hermes} & 1800 & 95\\
2011-05-04 21:23:11  & \textsc{Hermes} & 1800 & 93\\
2011-05-05 21:14:42  & \textsc{Hermes} & 1800 & 79\\
2011-09-30 05:39:25  & \textsc{Hermes} & 2400 &116\\
2011-12-18 05:16:13  & \textsc{Harps}  & 1200 &132\\
2012-01-10 05:00:07  & \textsc{Harps}  & 1800 &148\\
\hline\hline\end{tabular}\label{tbl:logbook}
\begin{list}{}{}       
\item[a] \footnotesize{Time of mid-exposure}
\item[b] \footnotesize{Measured around 4550\,$\AA$}
\end{list}             
\end{table}

\begin{figure*}
\centering  \includegraphics[width=0.42\textwidth]{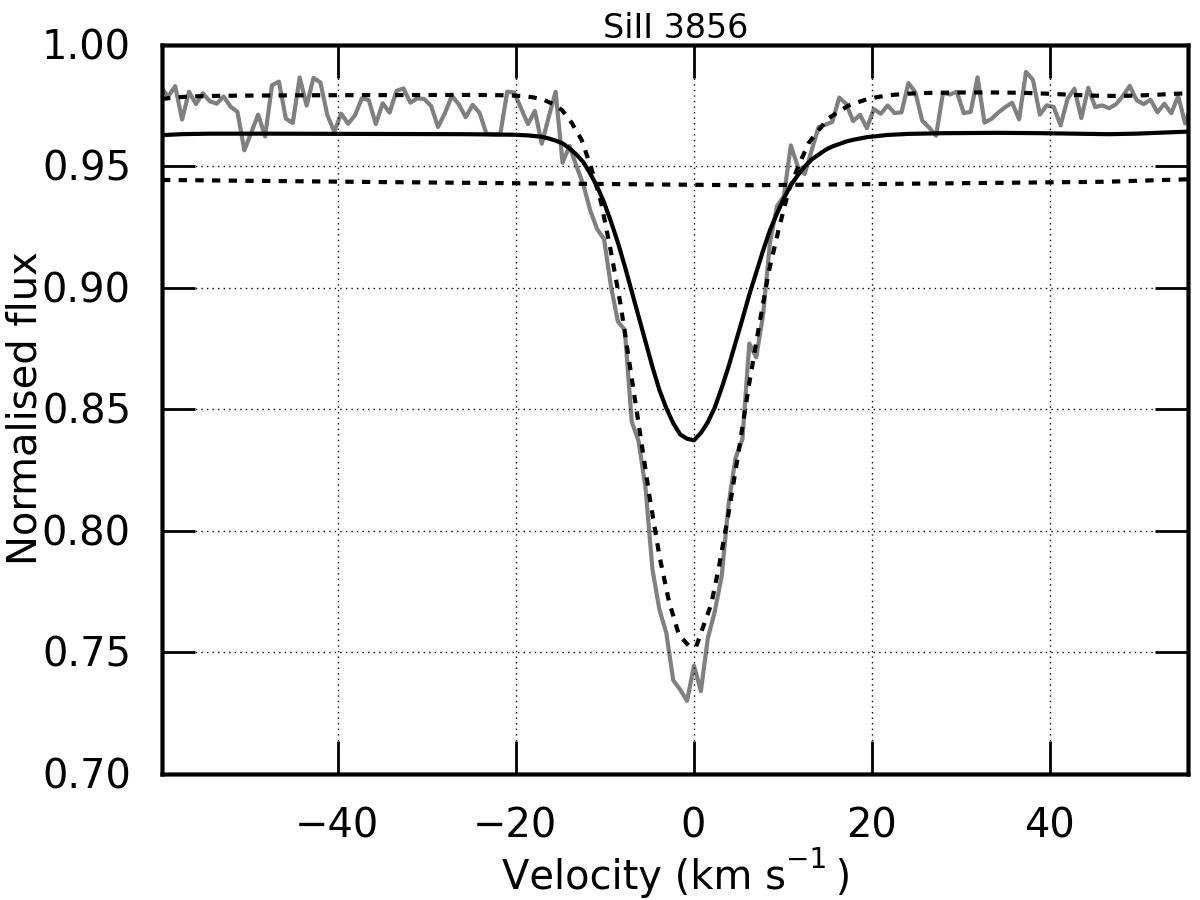}
            \includegraphics[width=0.42\textwidth]{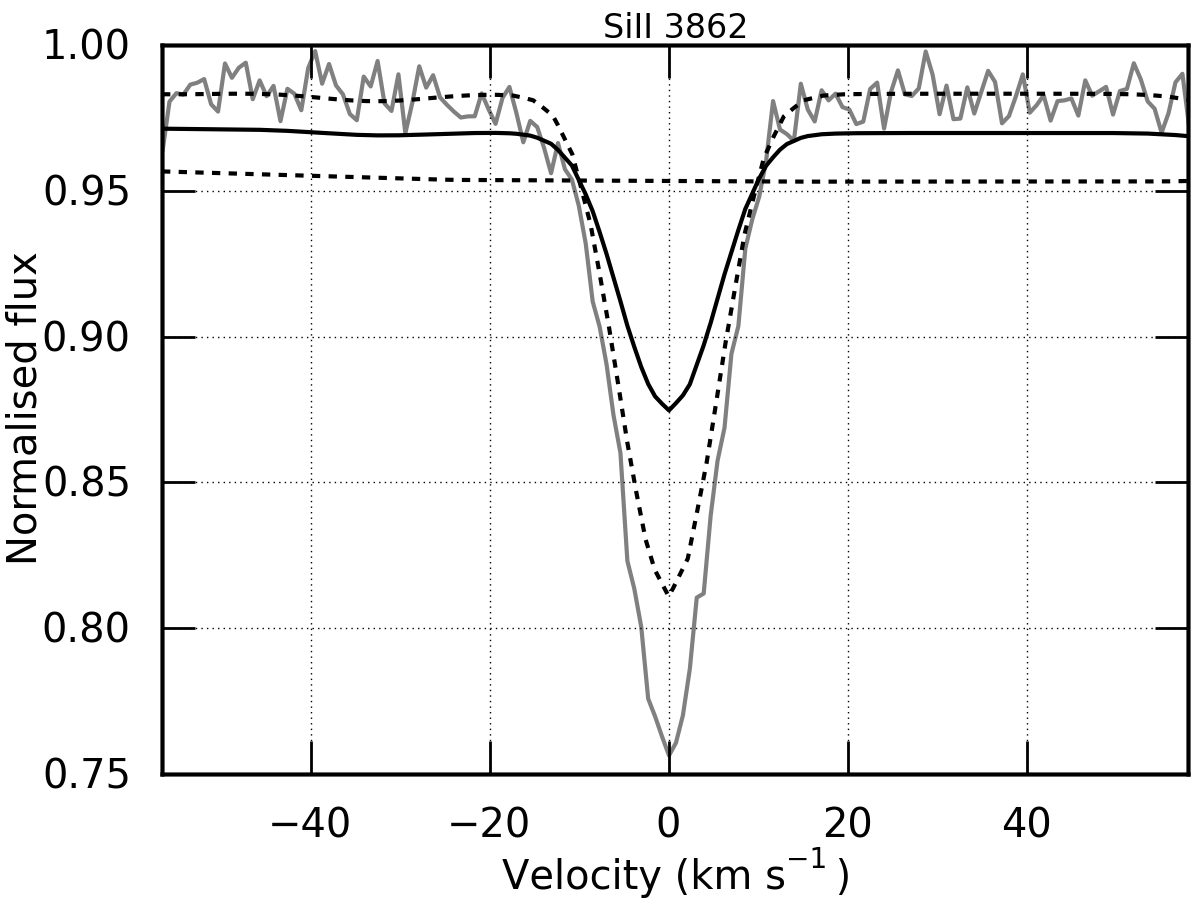}

            \includegraphics[width=0.42\textwidth]{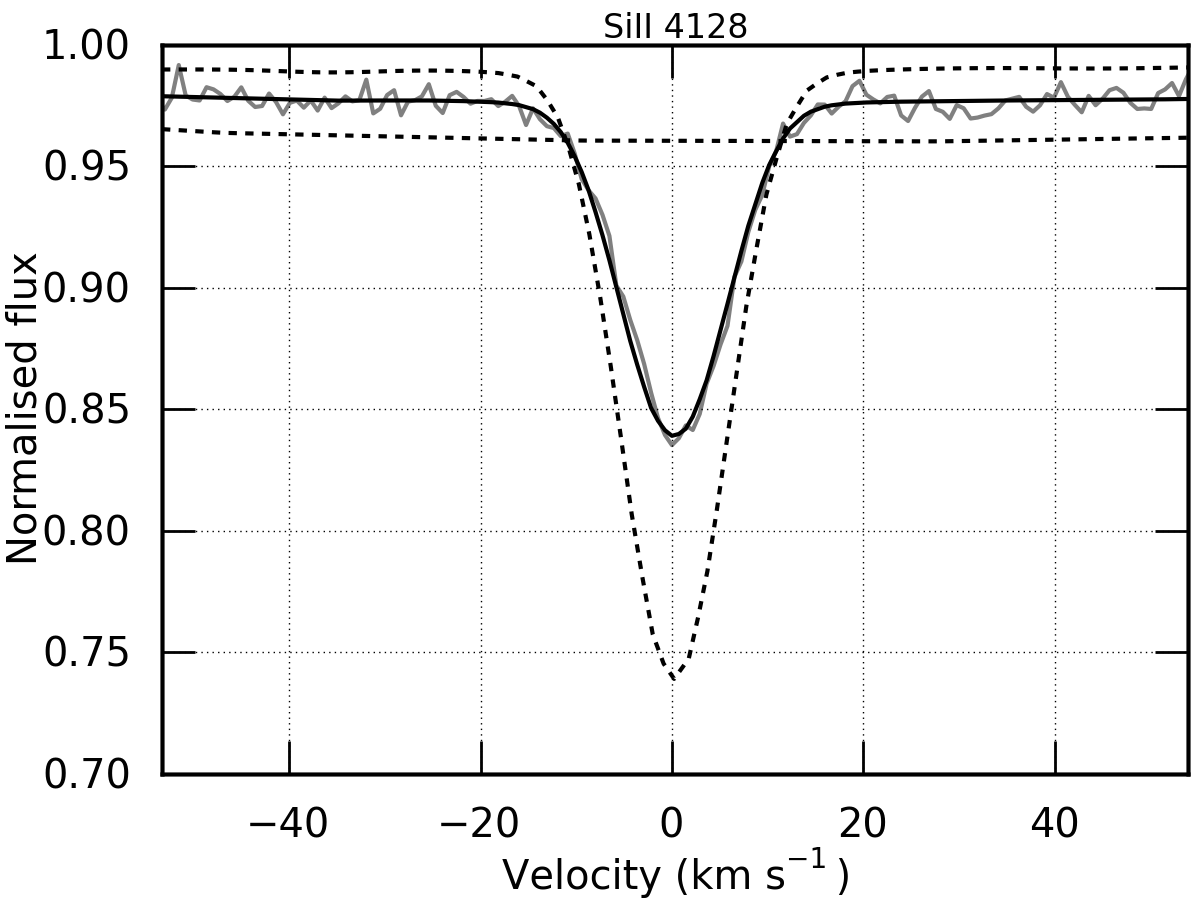}
            \includegraphics[width=0.42\textwidth]{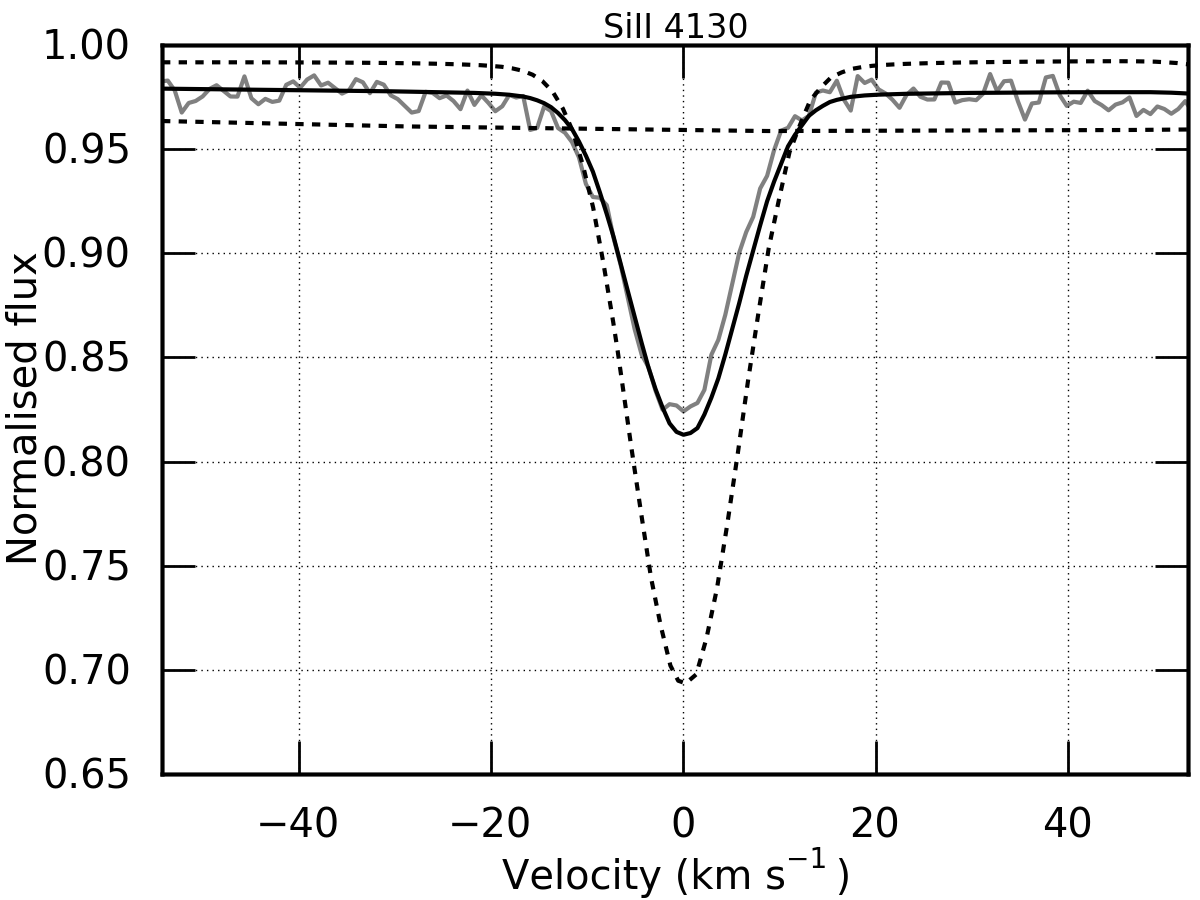}

            \includegraphics[width=0.42\textwidth]{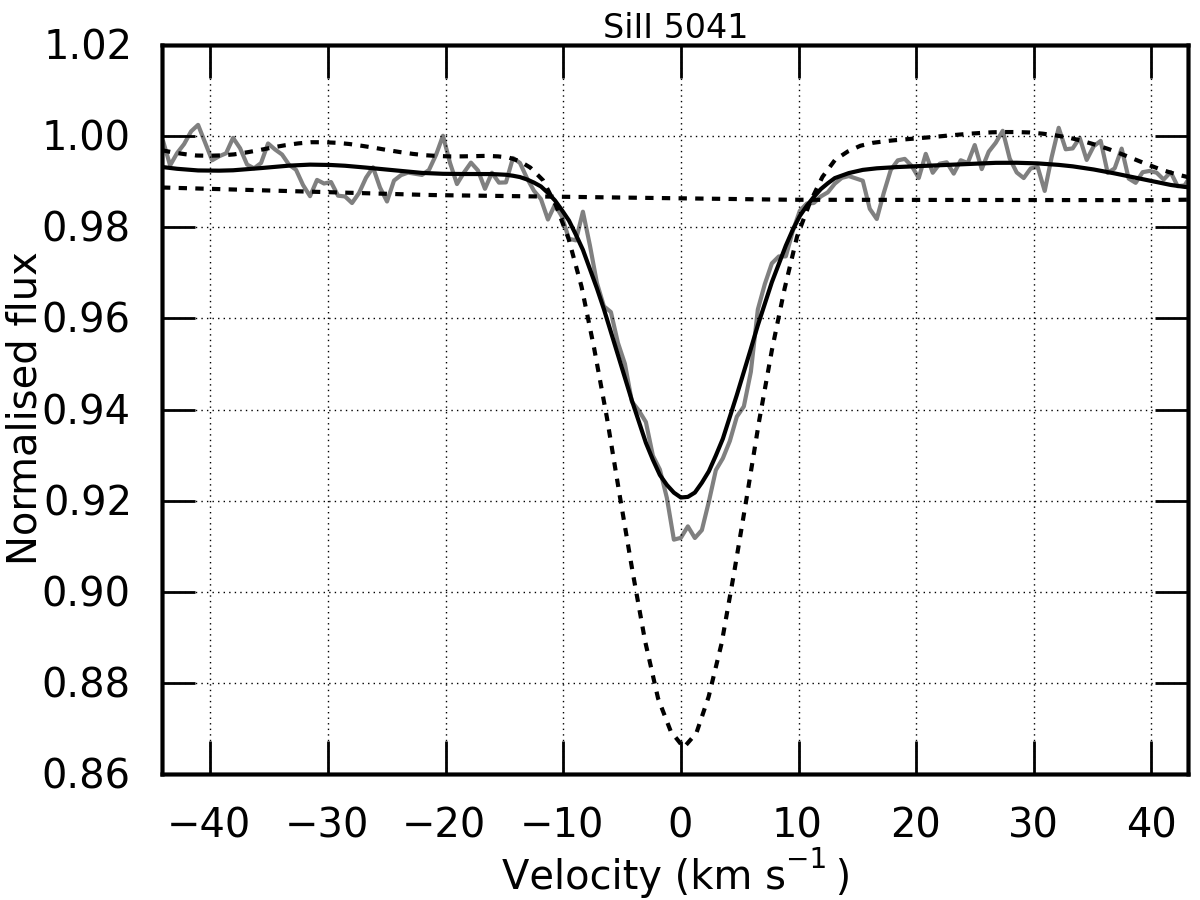}
            \includegraphics[width=0.42\textwidth]{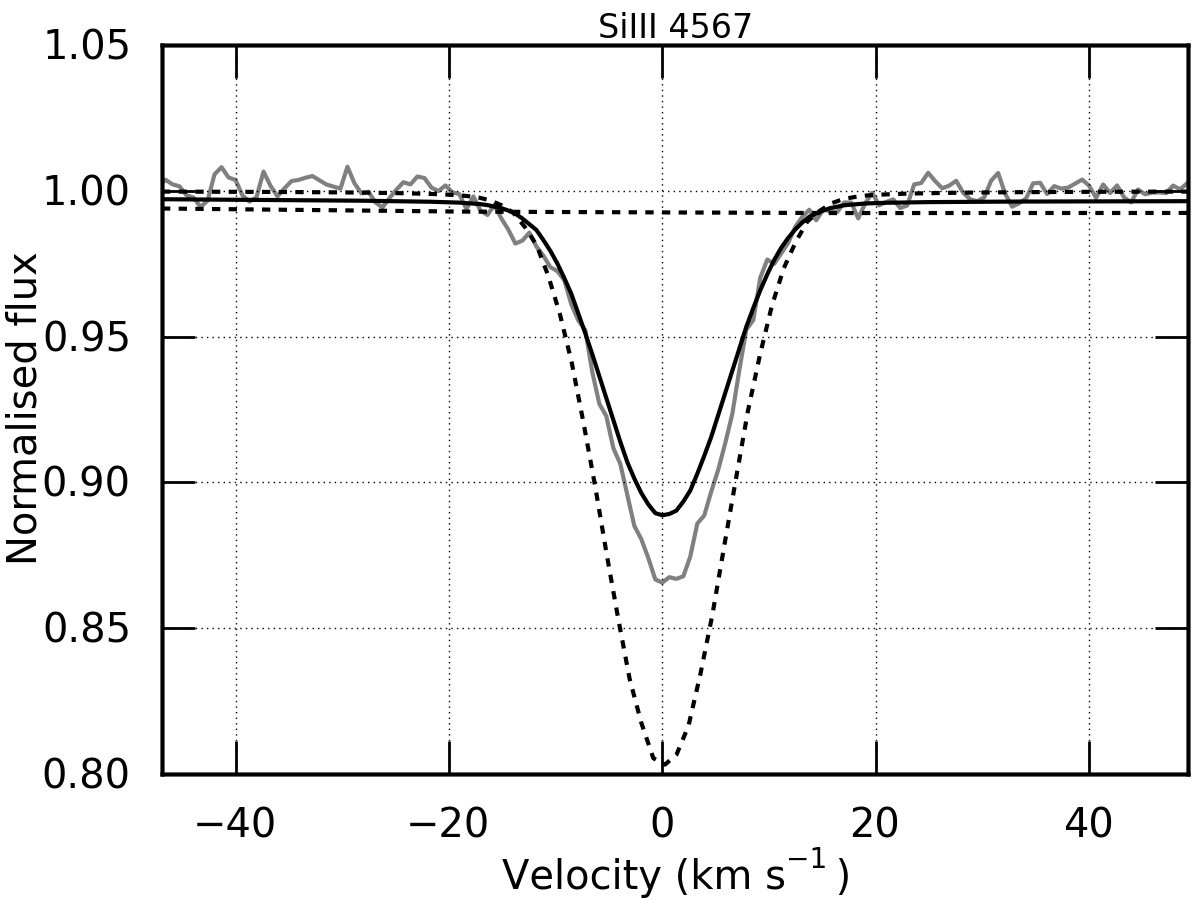}

            \includegraphics[width=0.42\textwidth]{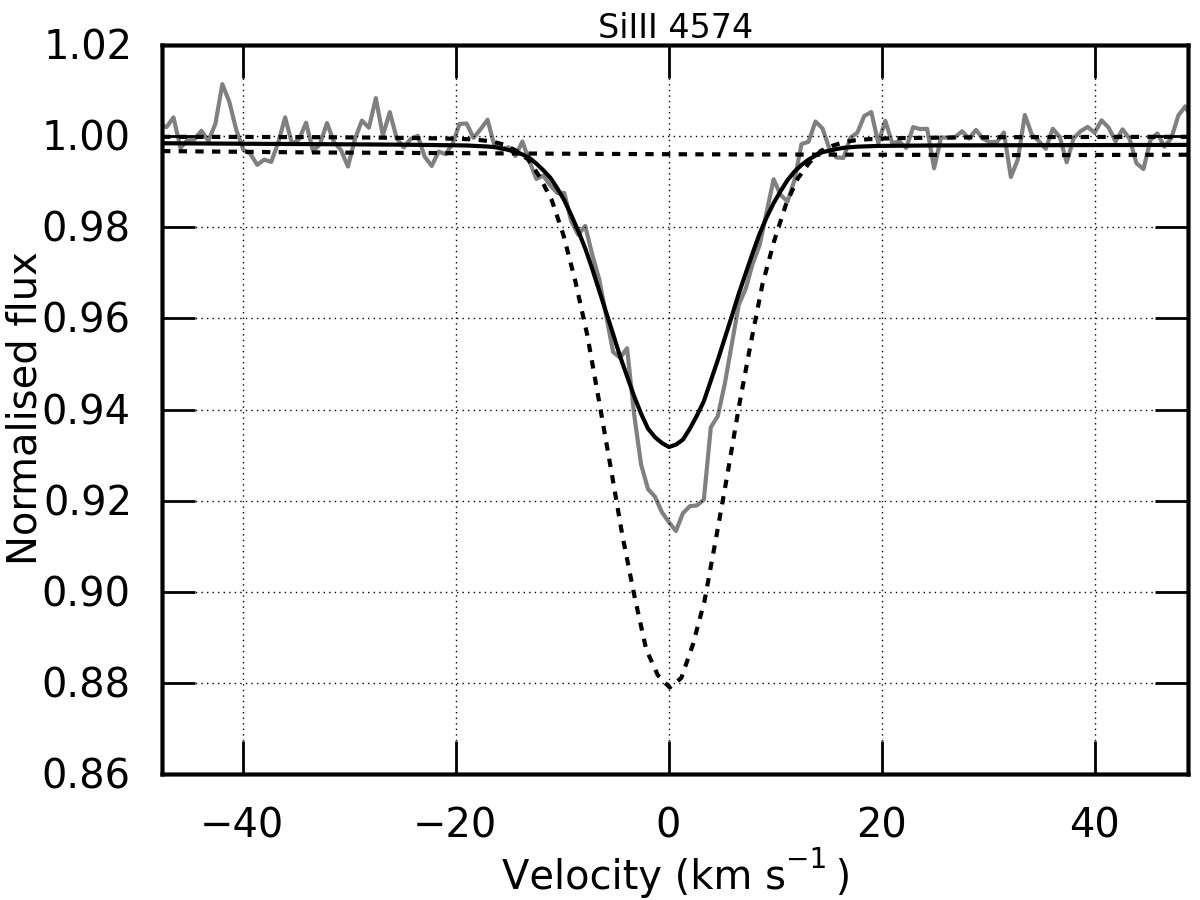}
            \includegraphics[width=0.42\textwidth]{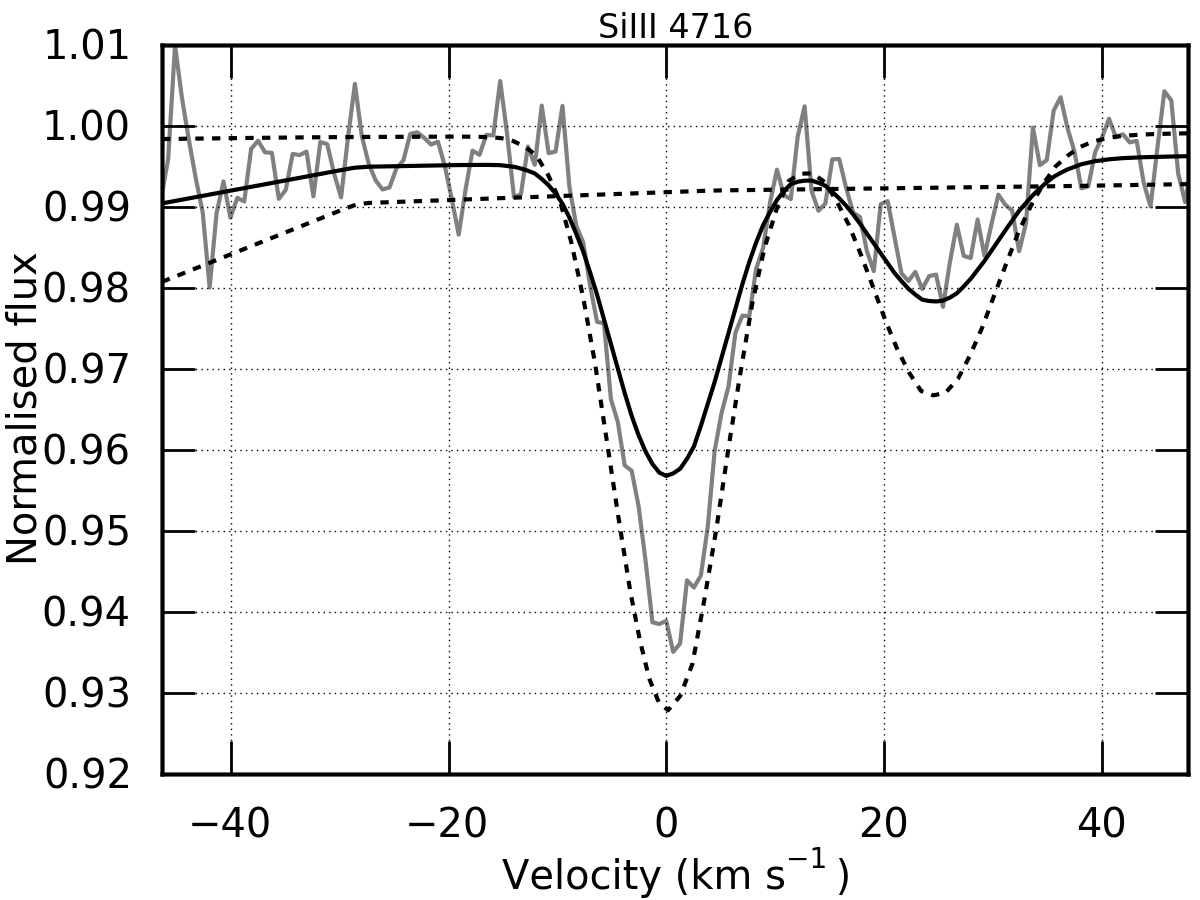}
\caption{Binary fit to a selection of spectral lines. In grey, the average of all observed HARPS spectra of HD\,50230 is plotted. Dashed black lines denote the contributions of the separate companions, the solid black line shows the obtained binary fit. Note that due to the simultaneous fitting of the continuum with the lines, the line continuum level can be off if the line does not fit well (i.e, the average quadratic distance between the observed and synthetic line profile is minimised). The fit consists of a primary component with $T_\mathrm{eff}=18\,500$\,K and $\log g=4.0$\,(cgs), and $T_\mathrm{eff}=15\,000$\,K and $\log g=4.0$\,(cgs) for the secondary.}\label{fig:hd50230:linefits1}
\end{figure*}

\begin{figure*}
\centering  \includegraphics[width=0.45\textwidth]{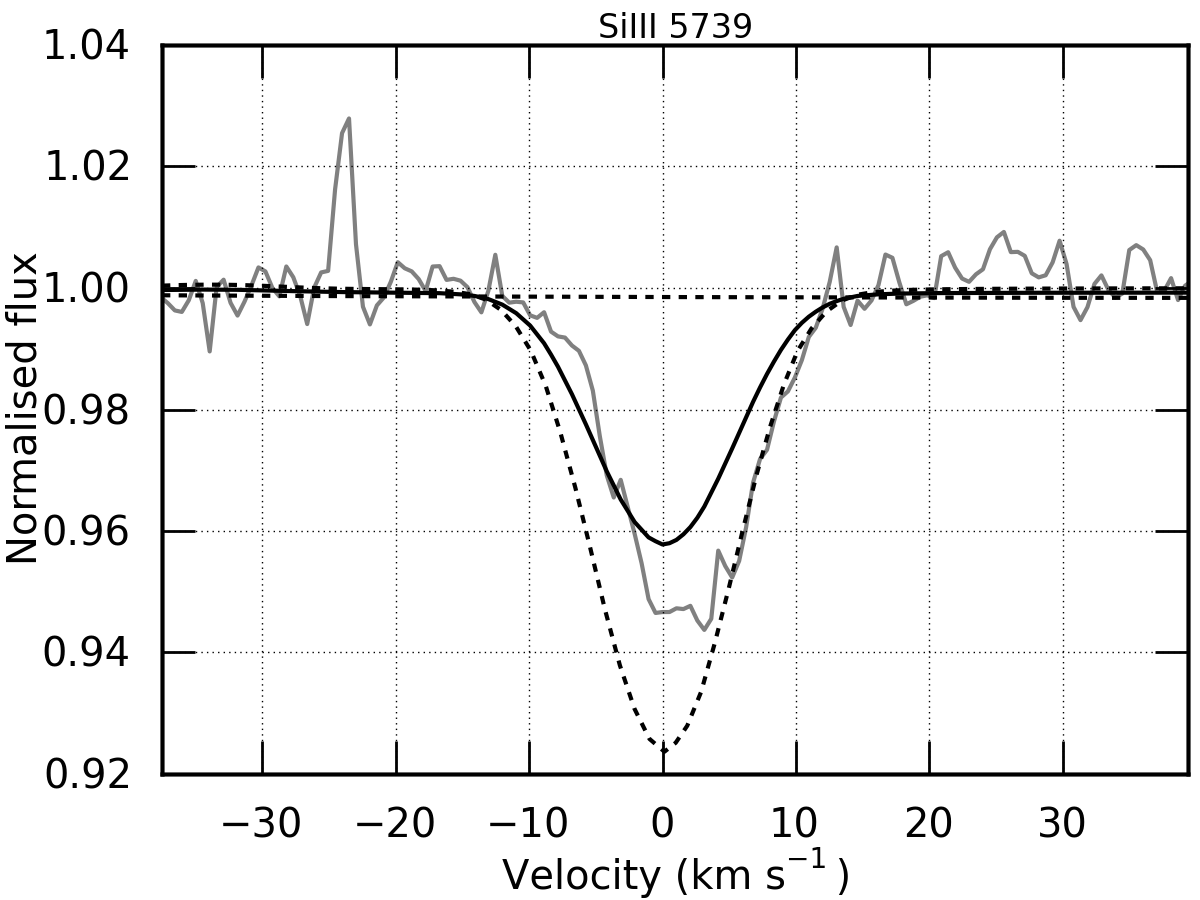}
            \includegraphics[width=0.45\textwidth]{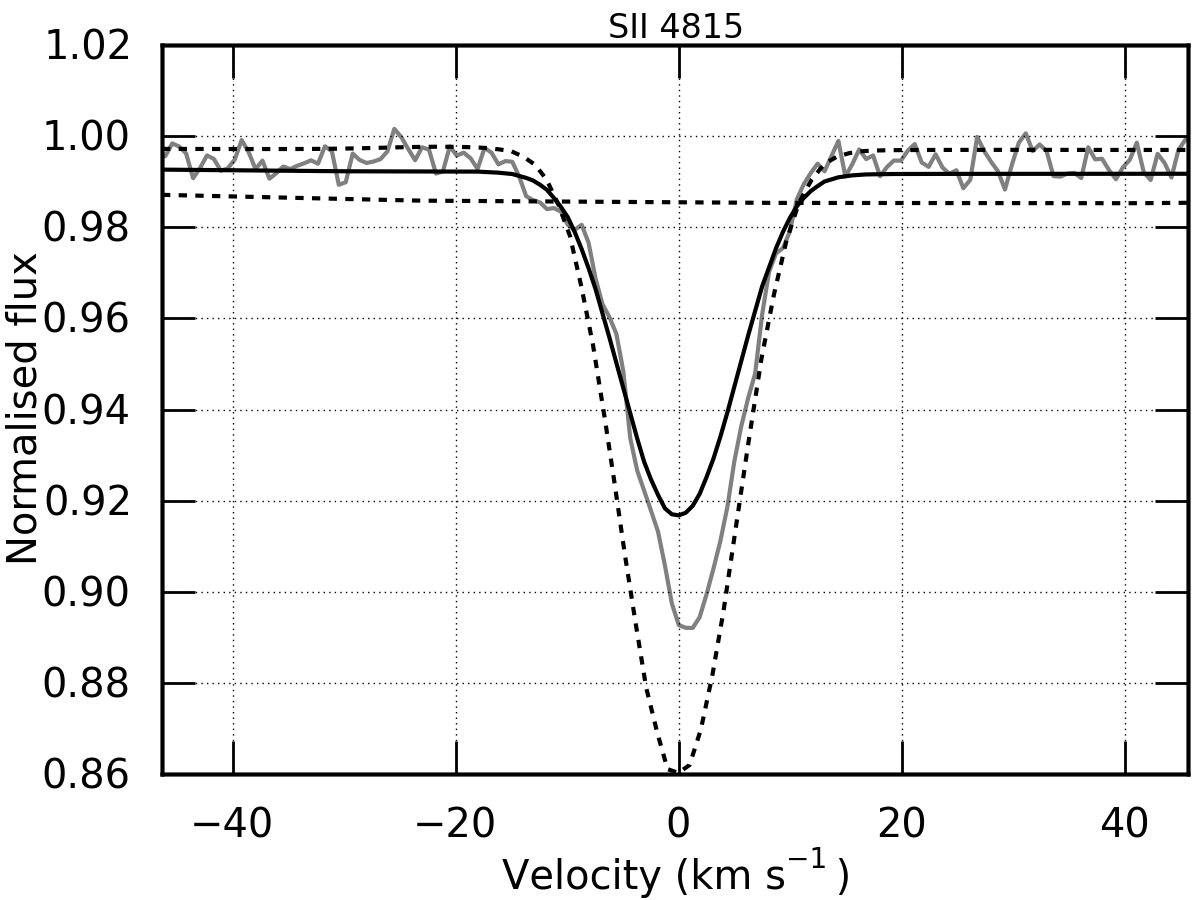}

            \includegraphics[width=0.45\textwidth]{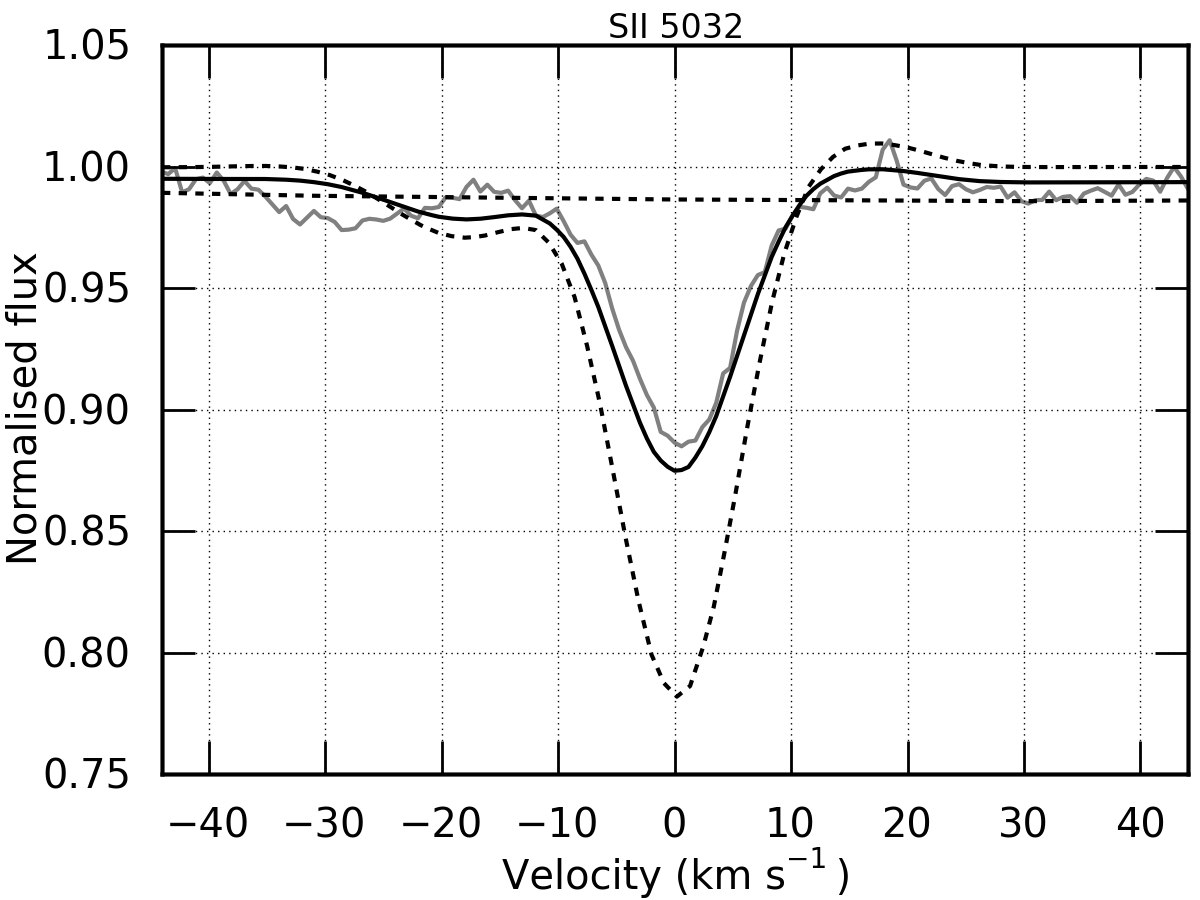}
            \includegraphics[width=0.45\textwidth]{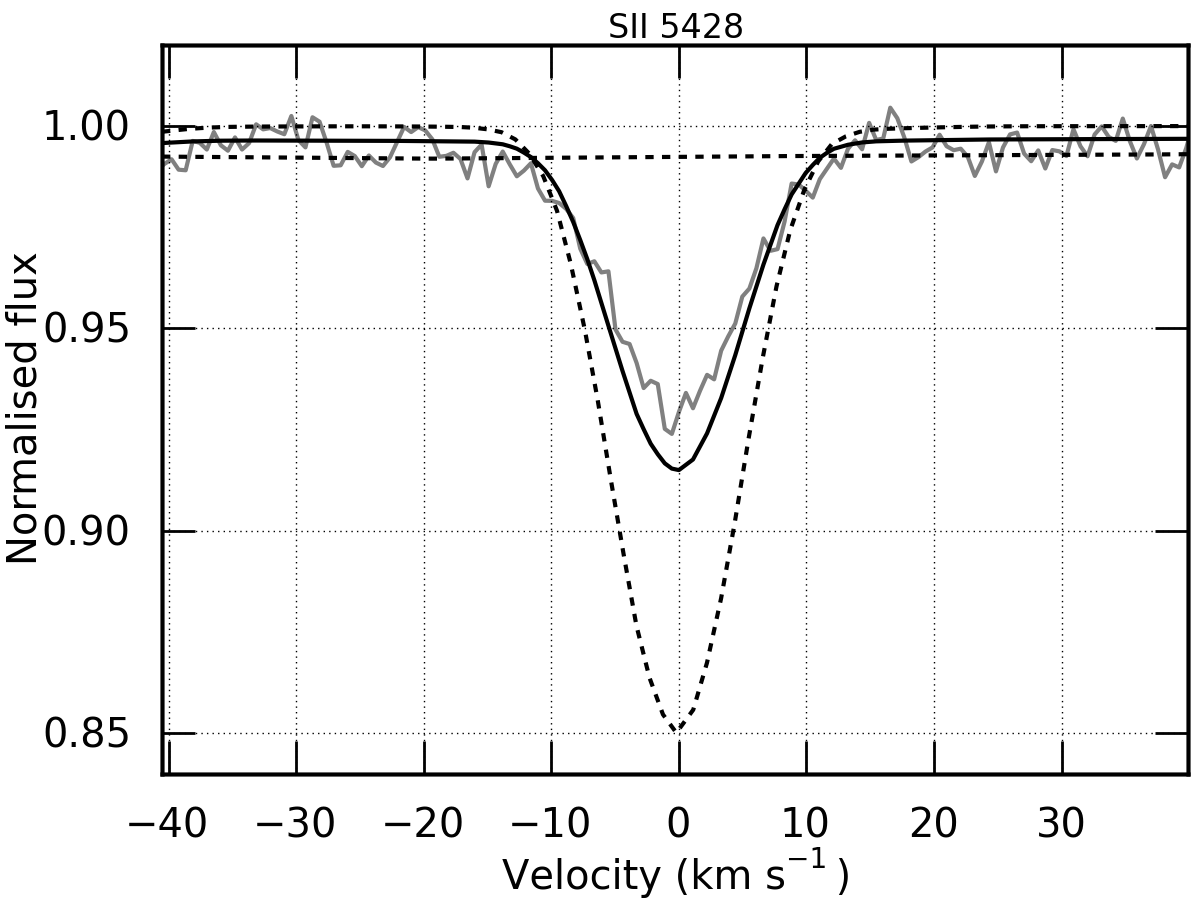}

            \includegraphics[width=0.45\textwidth]{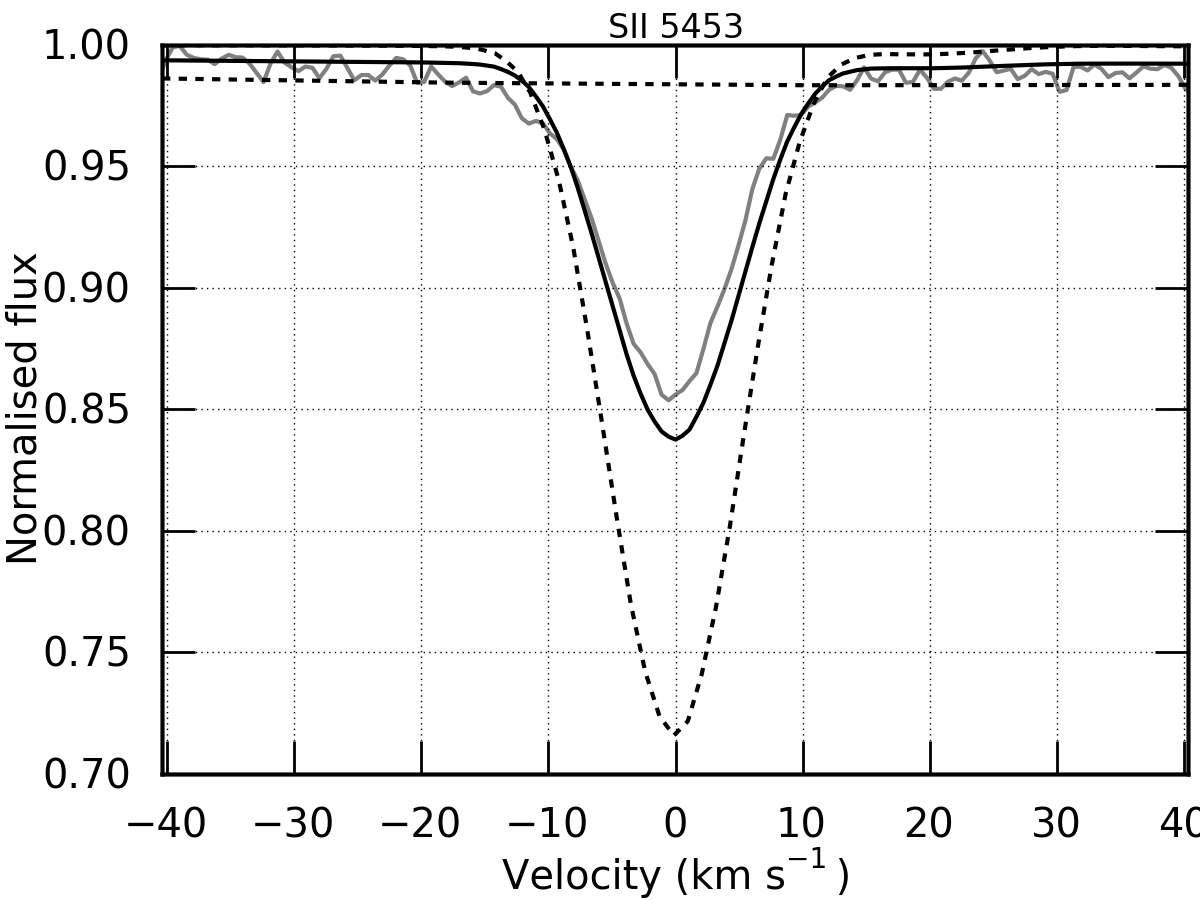}
            \includegraphics[width=0.45\textwidth]{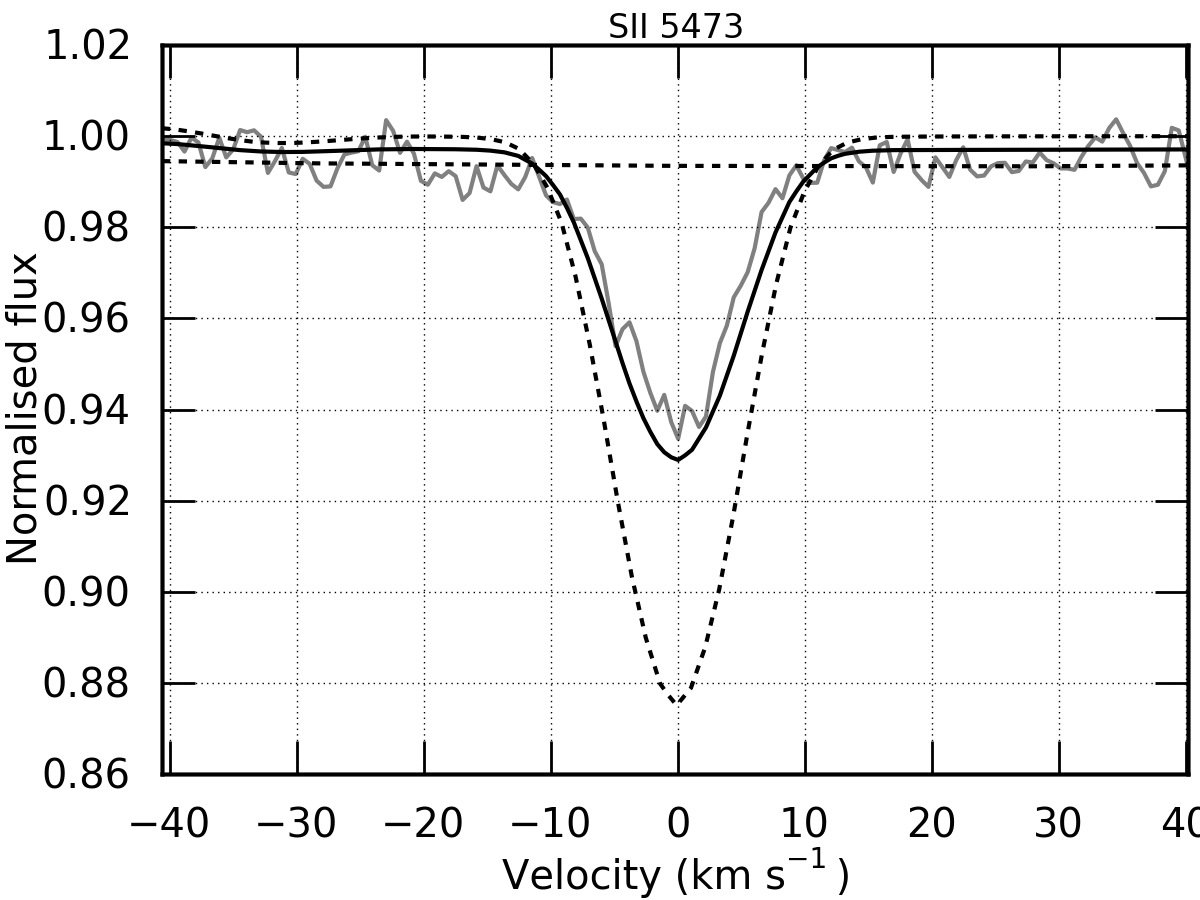}

            \includegraphics[width=0.45\textwidth]{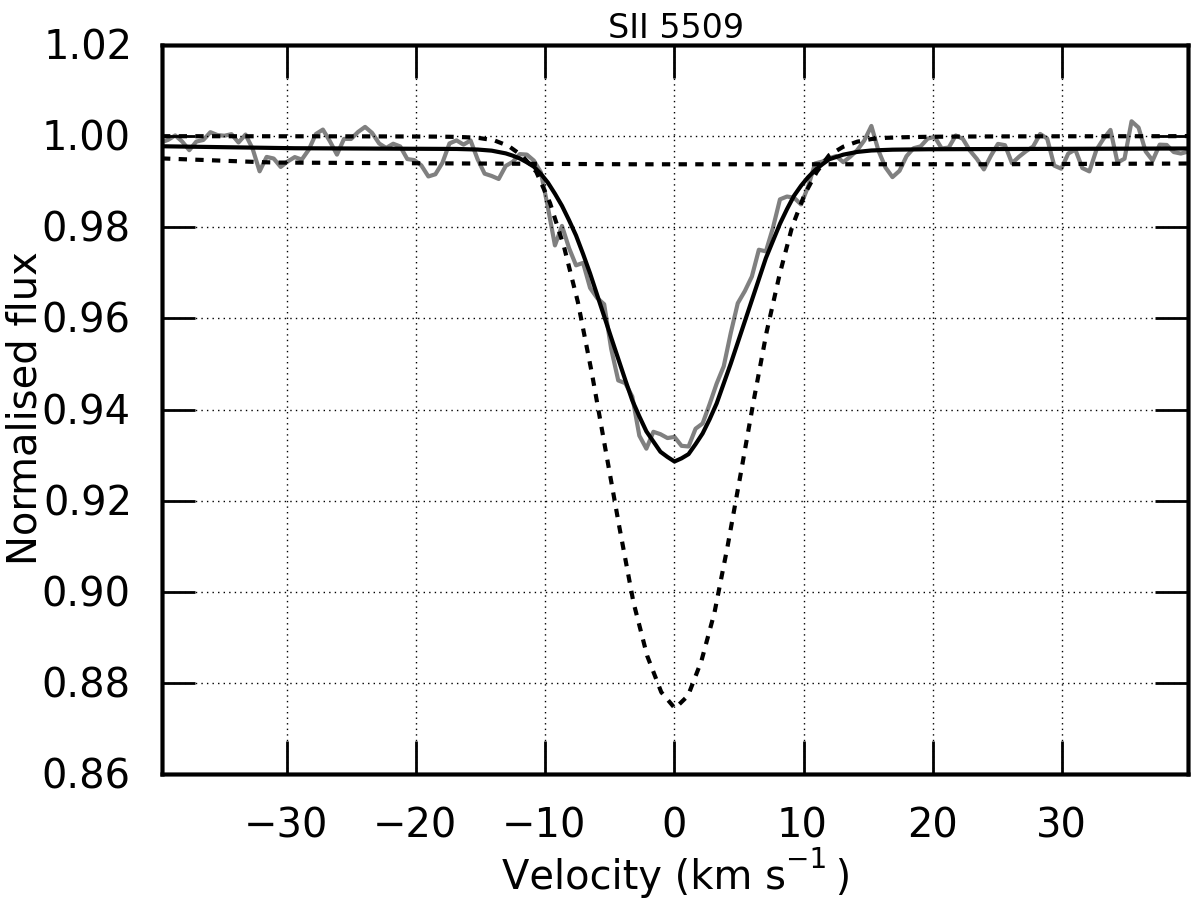}
            \includegraphics[width=0.45\textwidth]{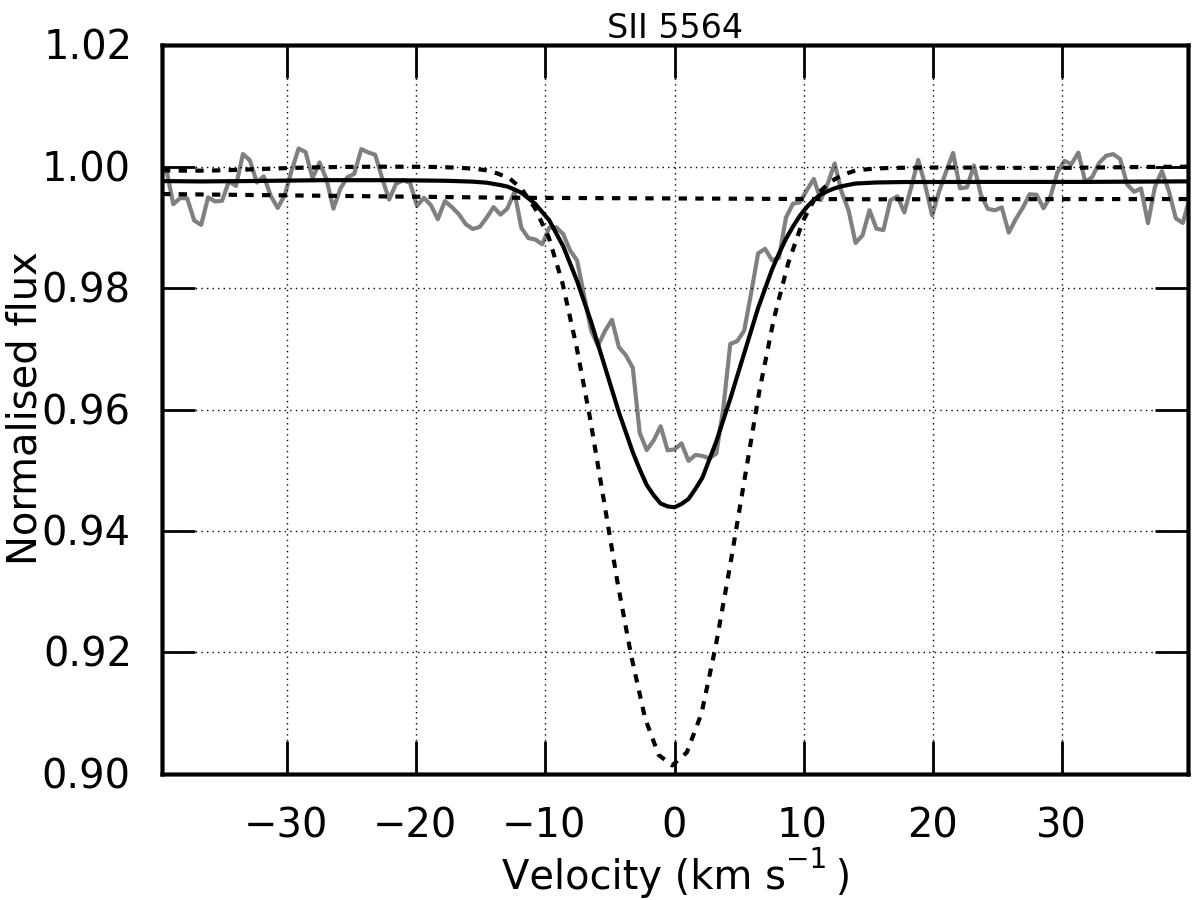}
\caption{Same as Fig.\,\ref{fig:hd50230:linefits1}, but for additional spectral lines.}\label{fig:hd50230:linefits2}
\end{figure*}

\begin{figure*}
\centering  \includegraphics[width=0.45\textwidth]{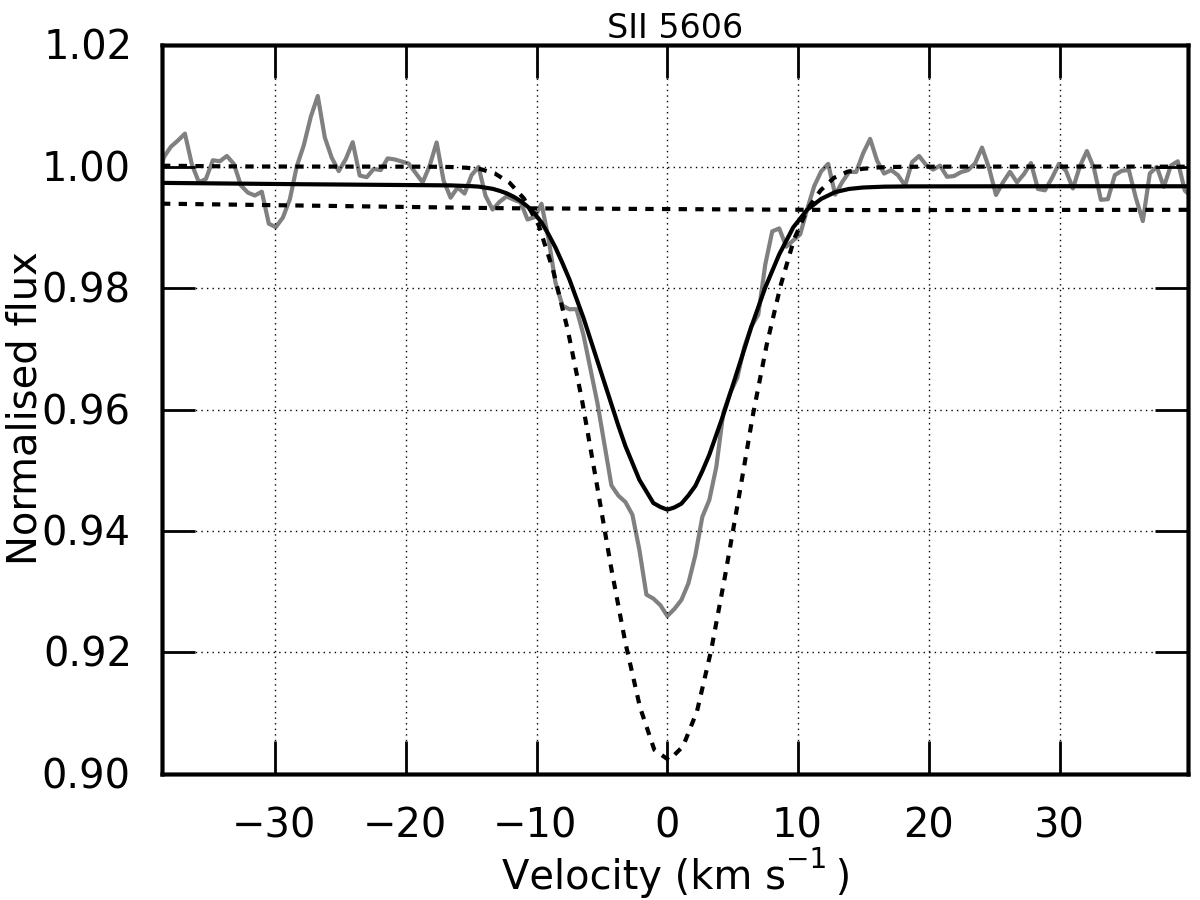}
            \includegraphics[width=0.45\textwidth]{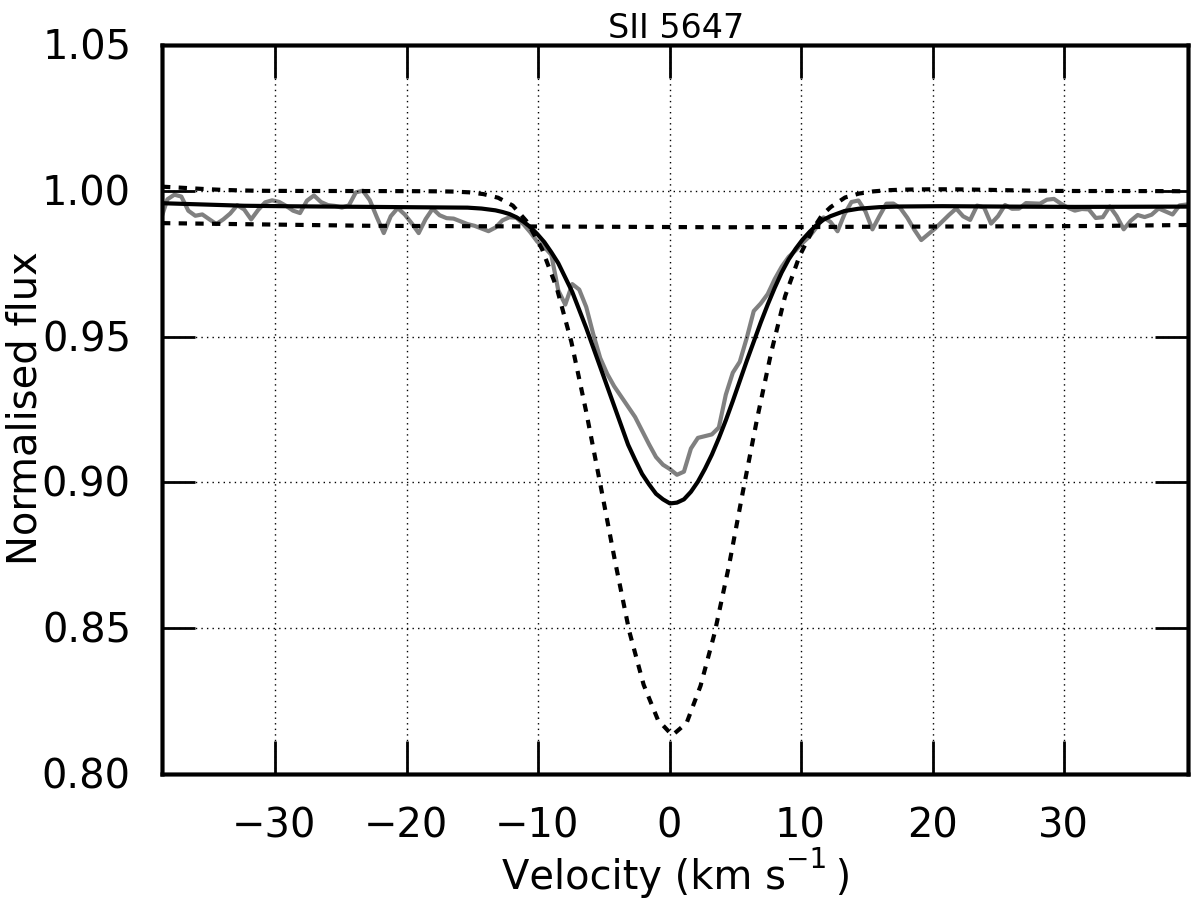}

            \includegraphics[width=0.45\textwidth]{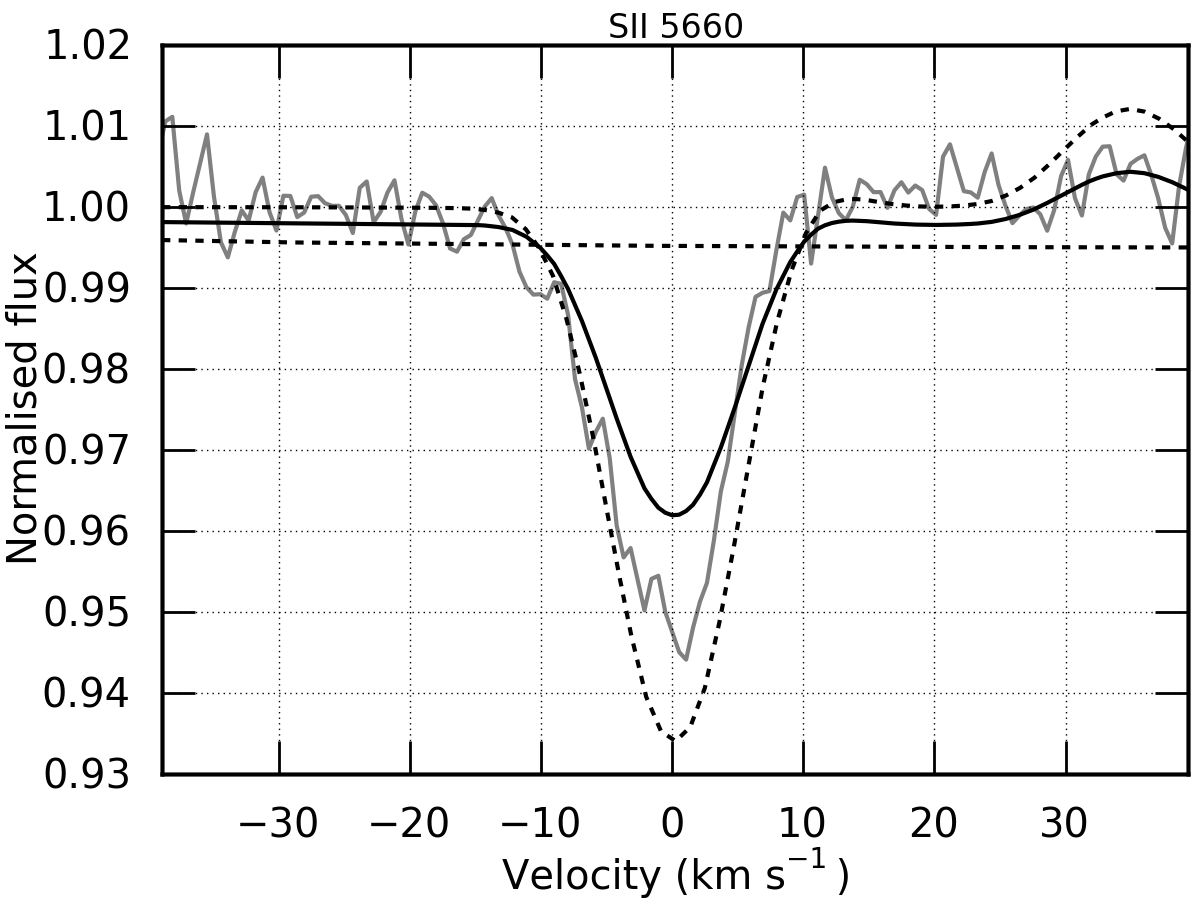}
            \includegraphics[width=0.45\textwidth]{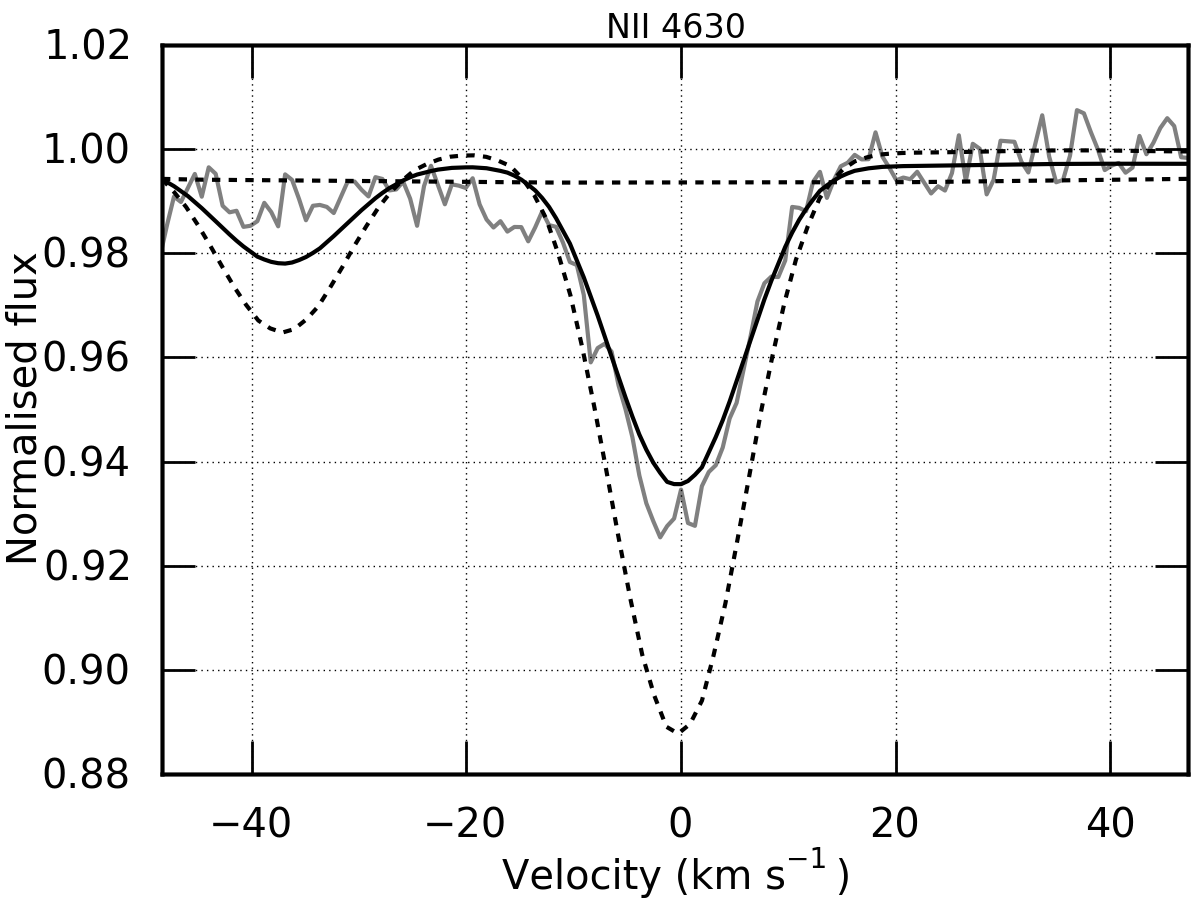}

            \includegraphics[width=0.45\textwidth]{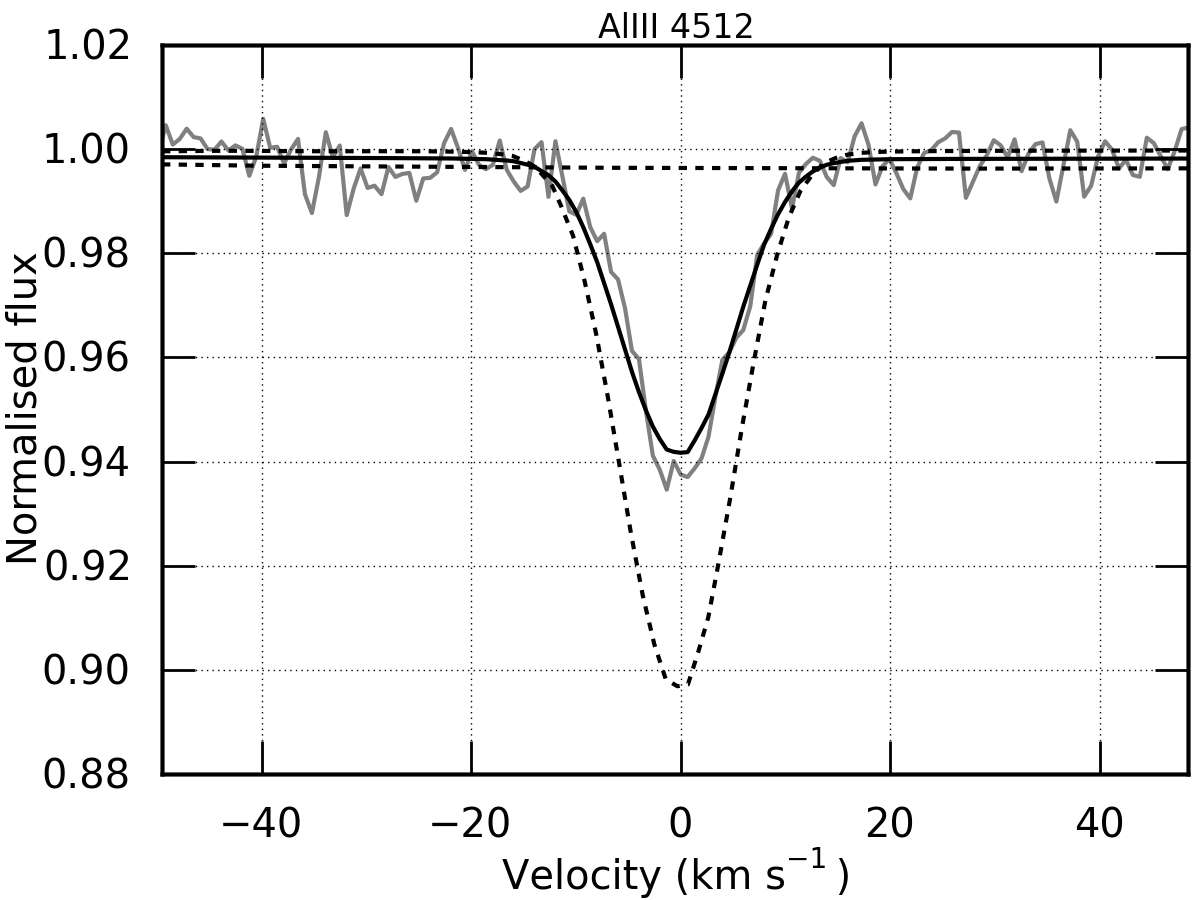}
            \includegraphics[width=0.45\textwidth]{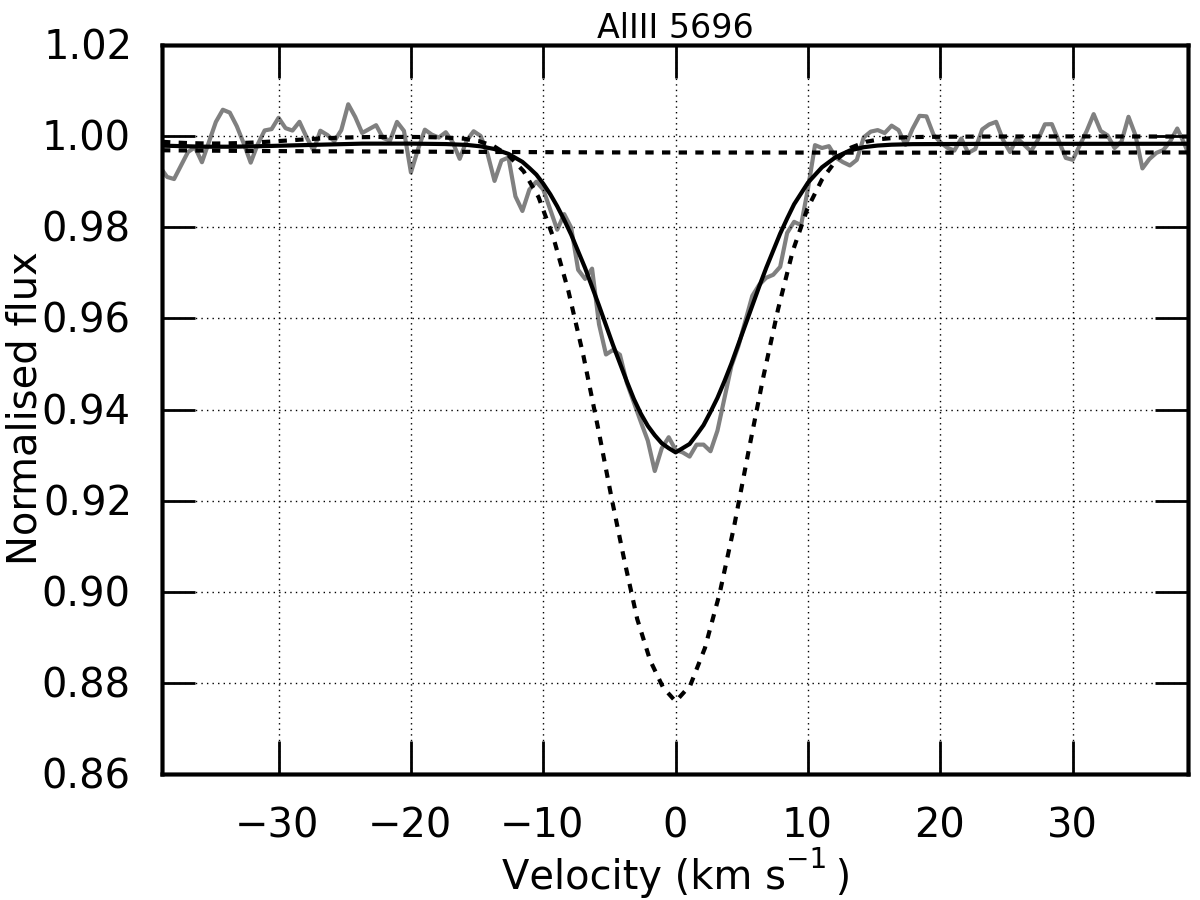}

            \includegraphics[width=0.45\textwidth]{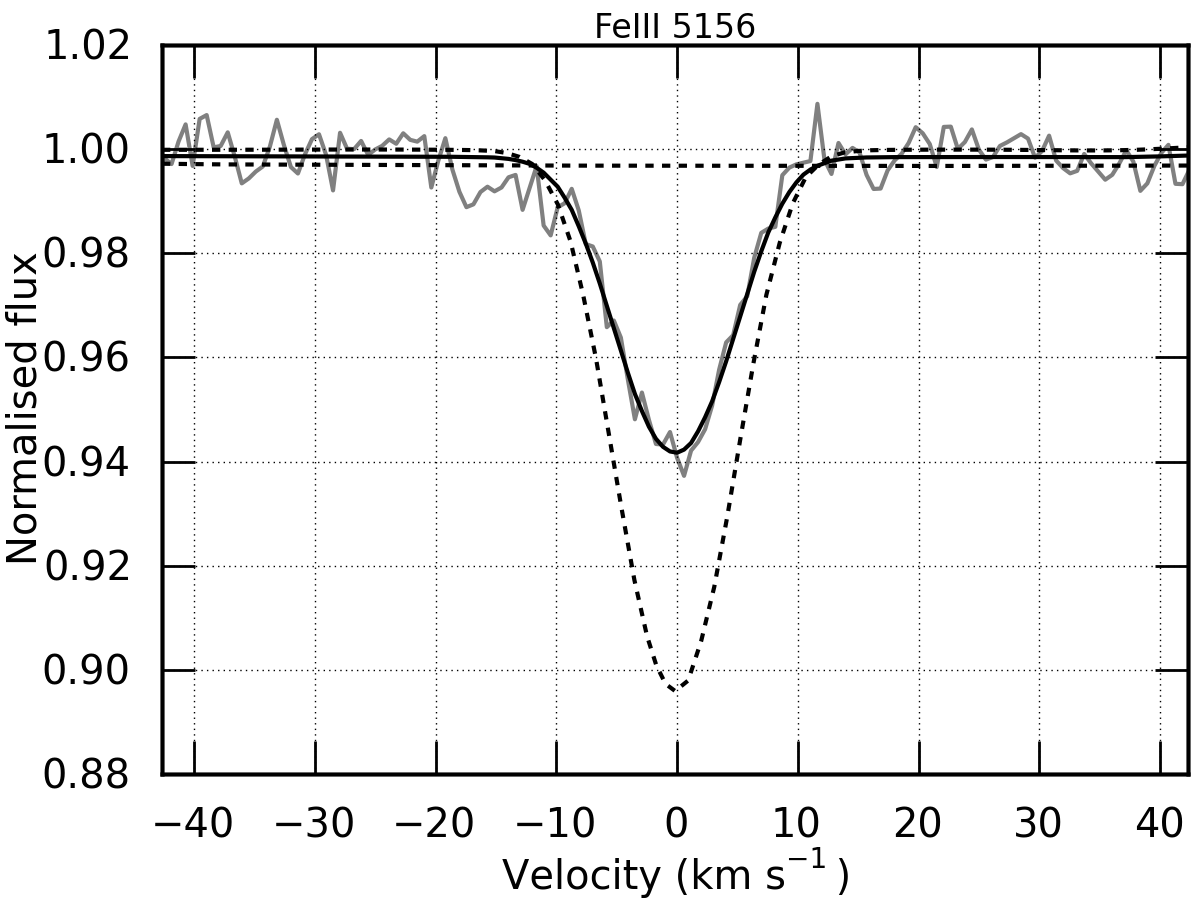}
            \includegraphics[width=0.45\textwidth]{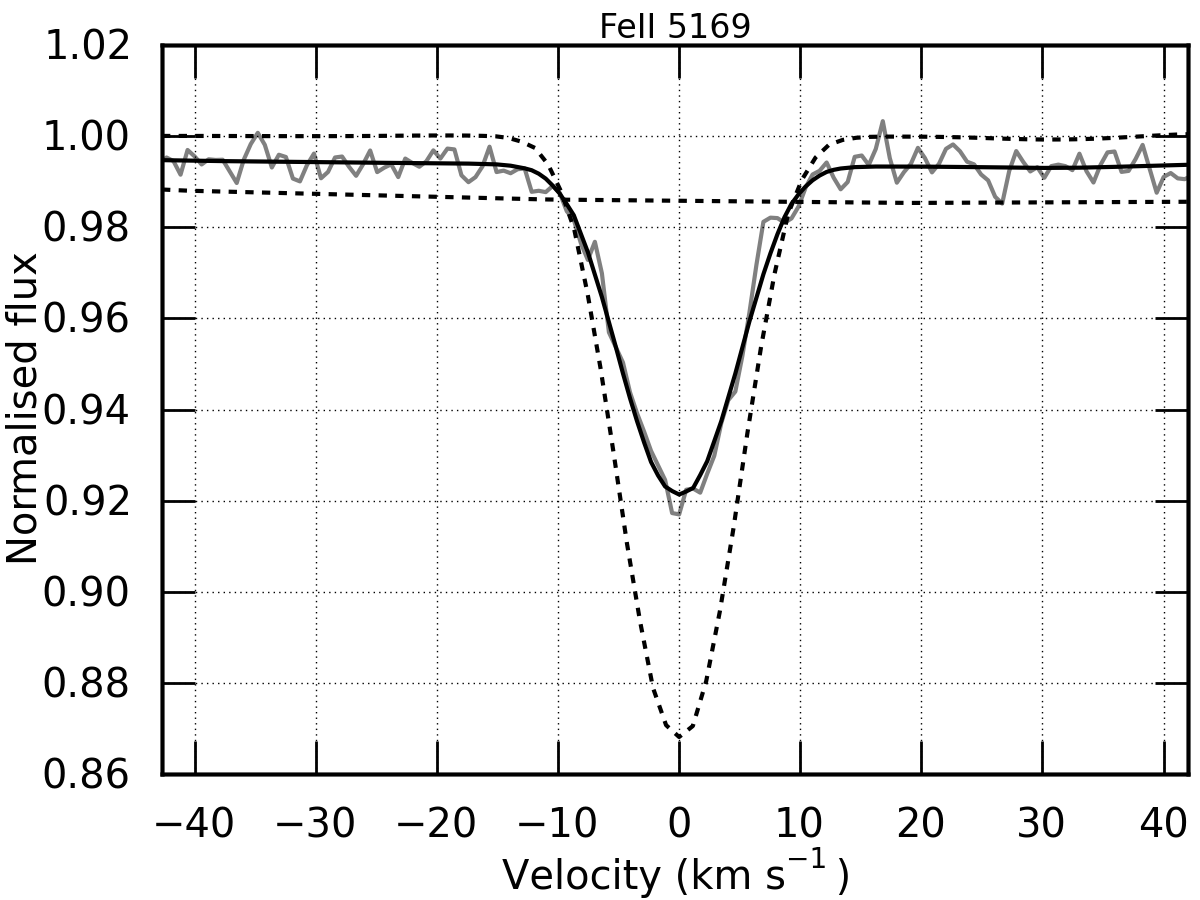}
\caption{Same as Fig.\,\ref{fig:hd50230:linefits1}, but for additional spectral lines.}\label{fig:hd50230:linefits3}
\end{figure*}

\begin{figure*}
\centering  \includegraphics[width=0.45\textwidth]{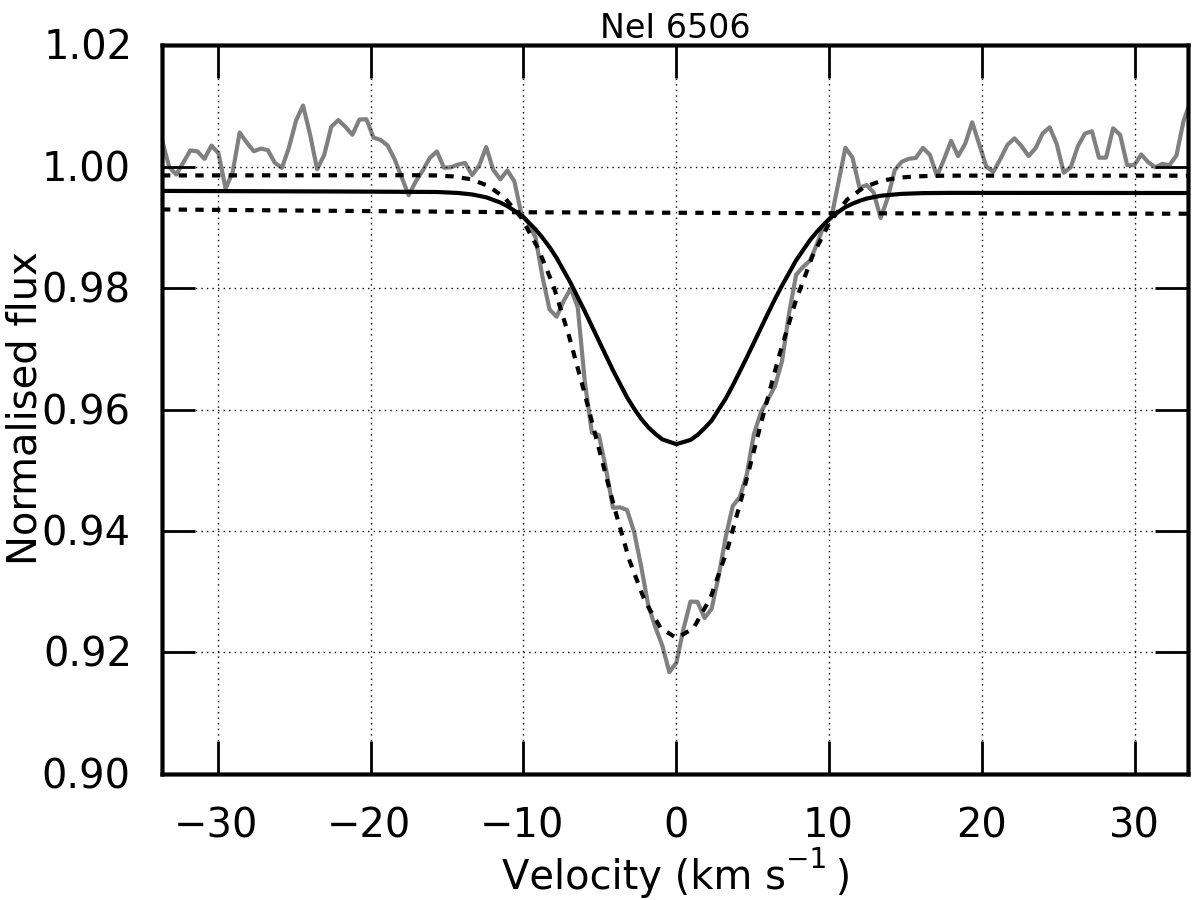}
            \includegraphics[width=0.45\textwidth]{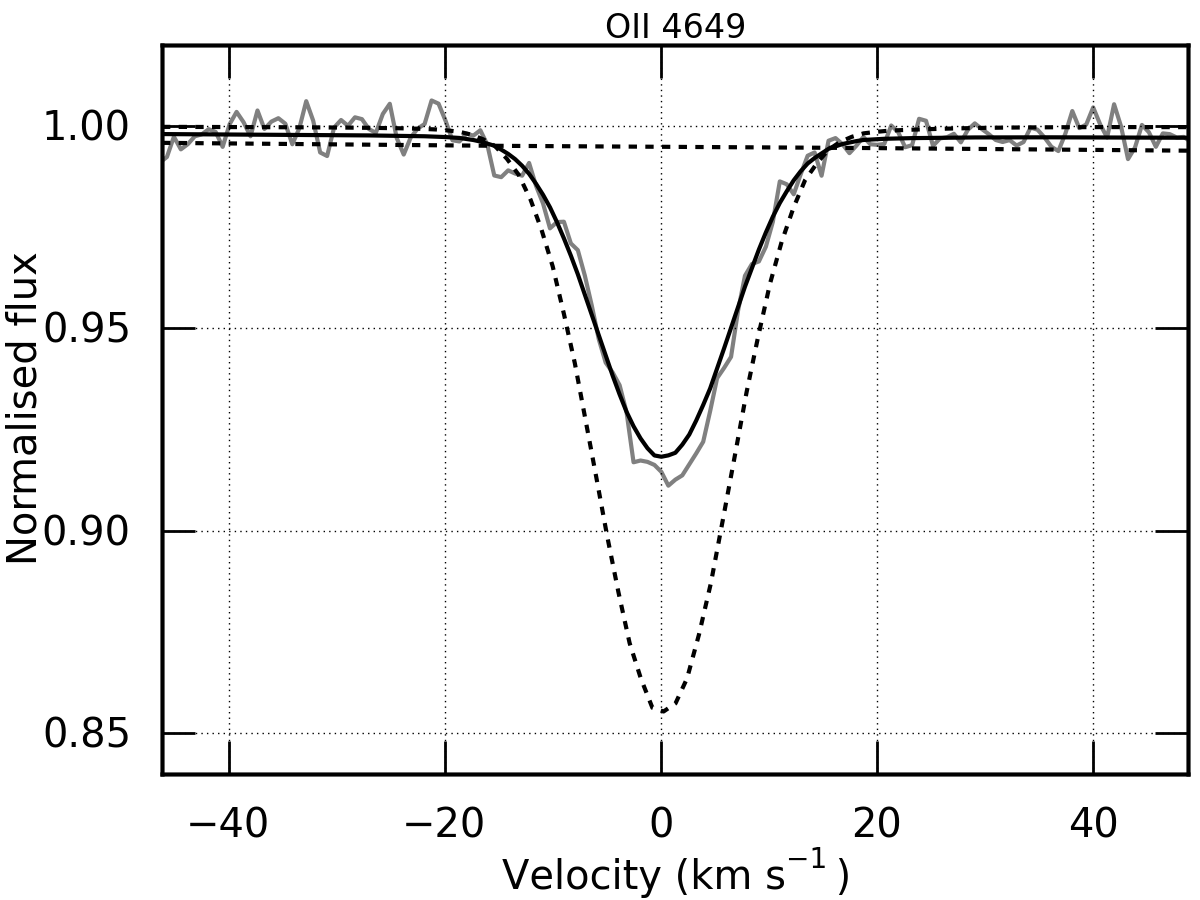}

            \includegraphics[width=0.45\textwidth]{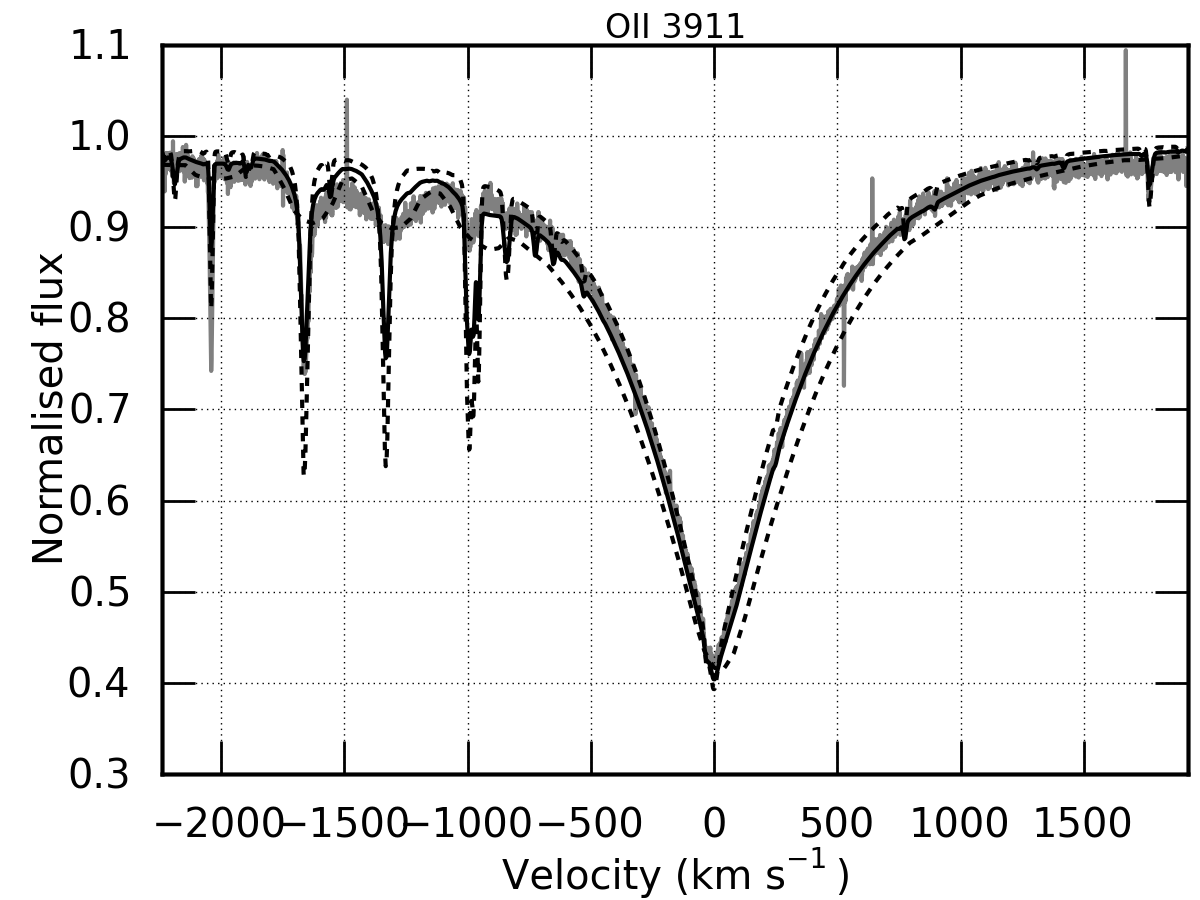}
            \includegraphics[width=0.45\textwidth]{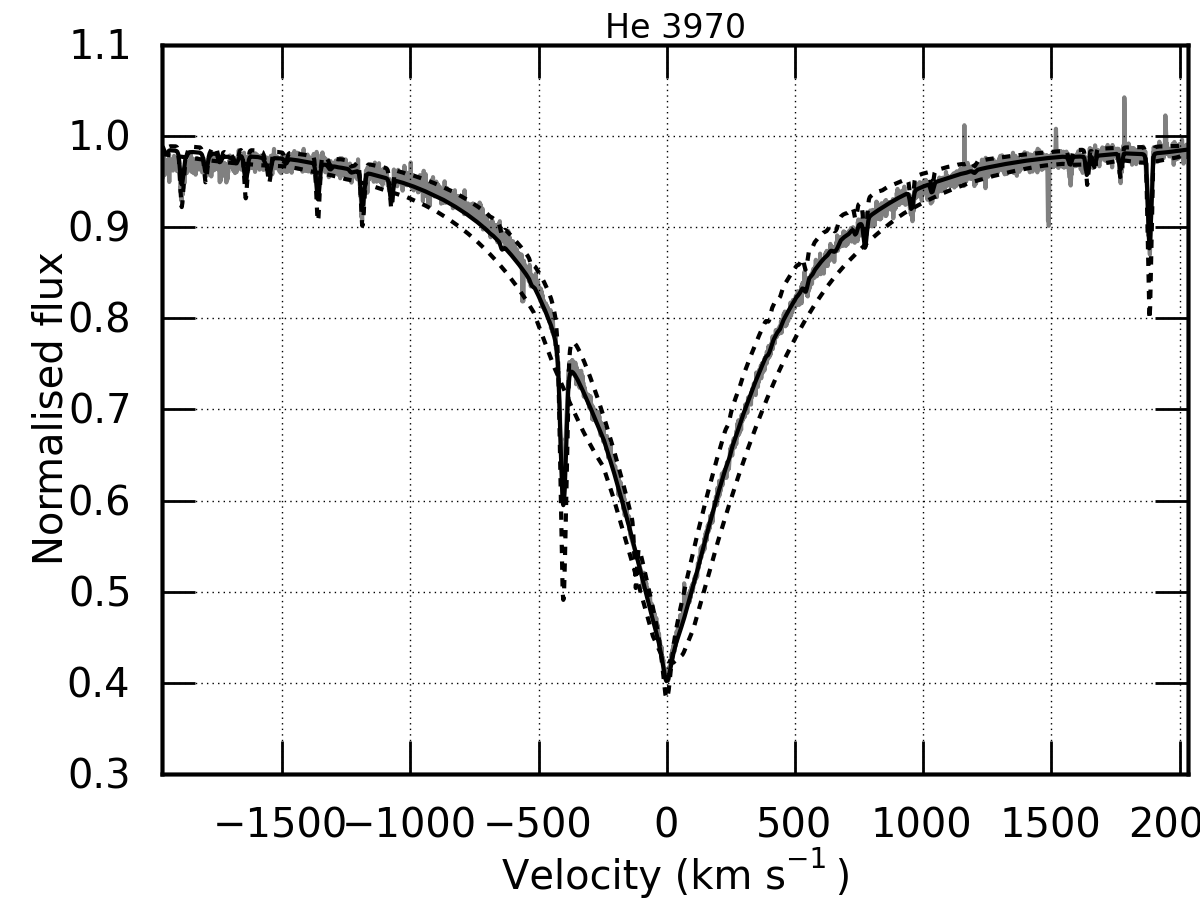}

            \includegraphics[width=0.45\textwidth]{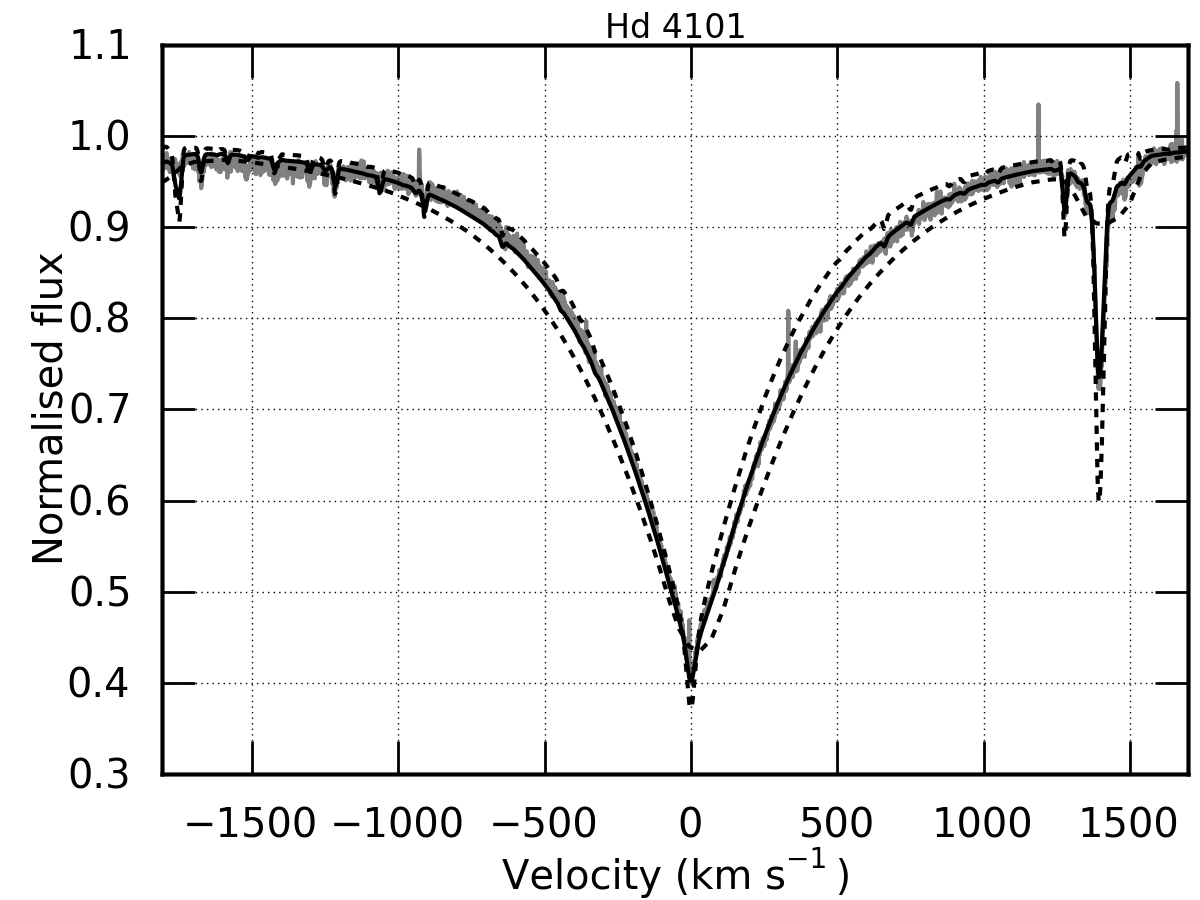}
            \includegraphics[width=0.45\textwidth]{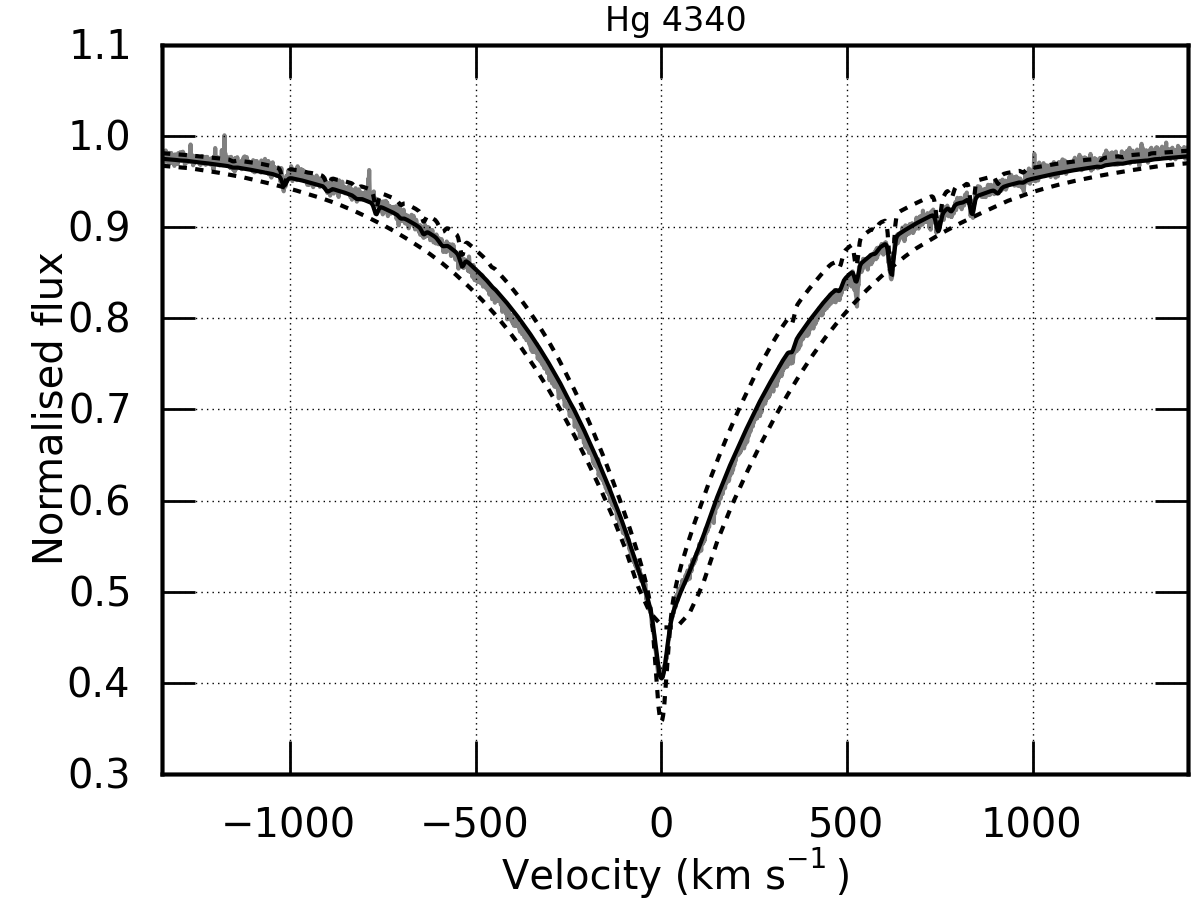}

            \includegraphics[width=0.45\textwidth]{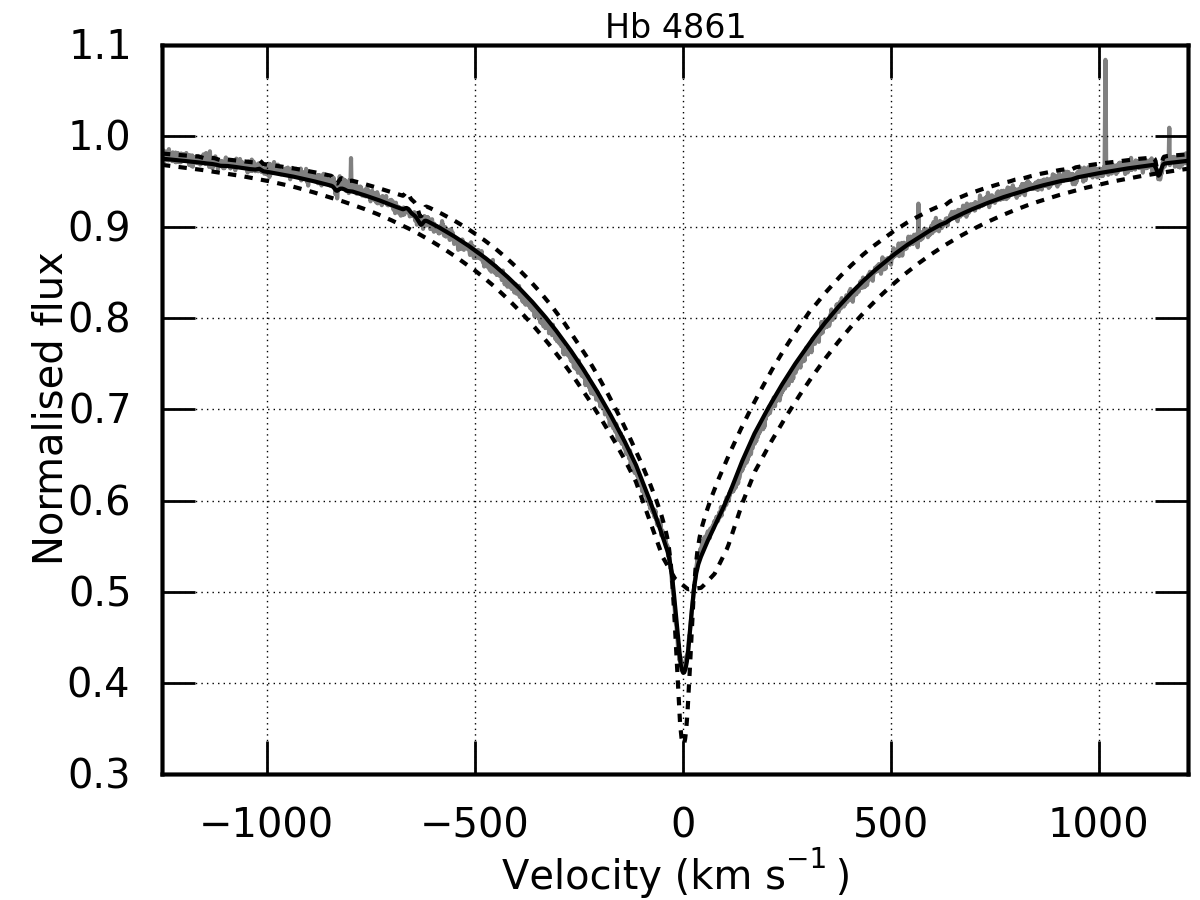}
            \includegraphics[width=0.45\textwidth]{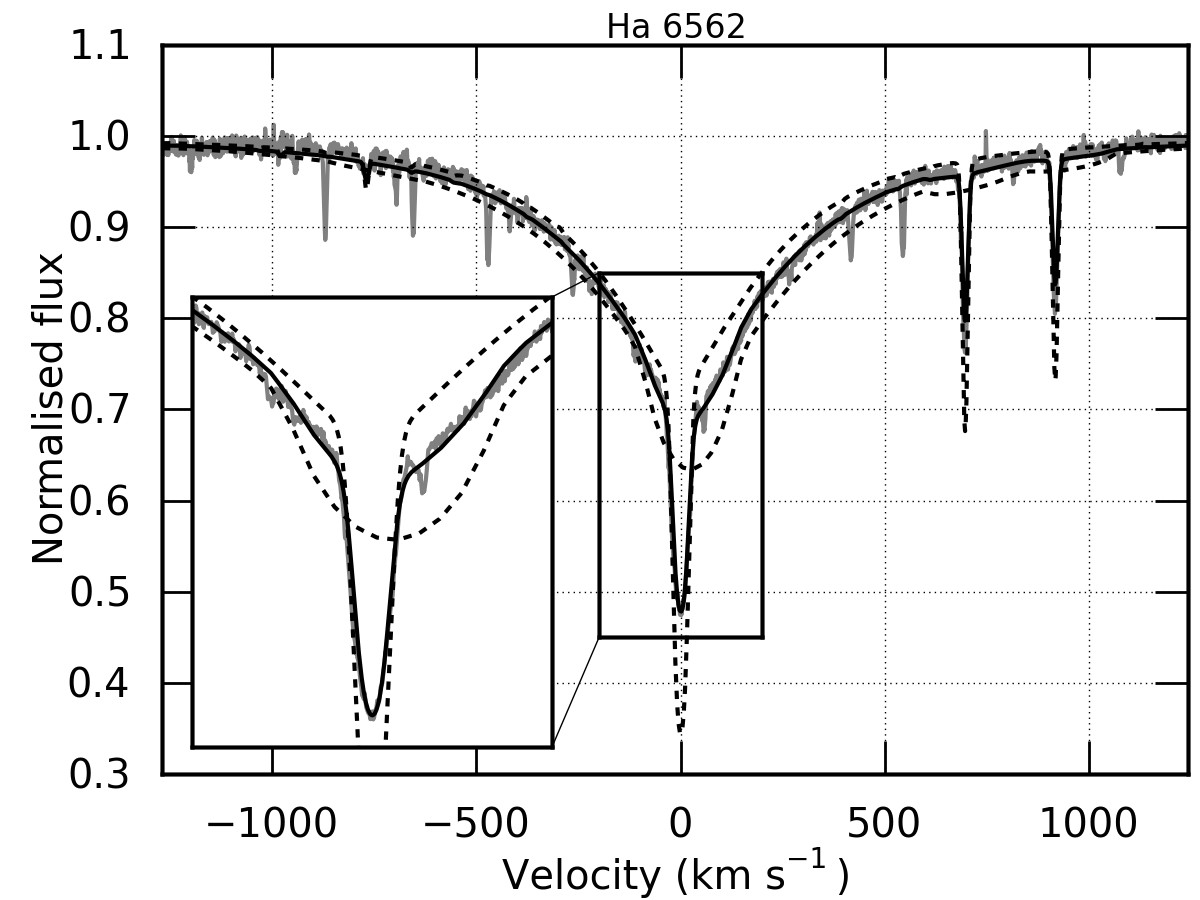}
\caption{Same as Fig.\,\ref{fig:hd50230:linefits1}, but for additional spectral lines.}\label{fig:hd50230:linefits4}
\end{figure*}

\begin{figure*}
\centering  \includegraphics[width=0.45\textwidth]{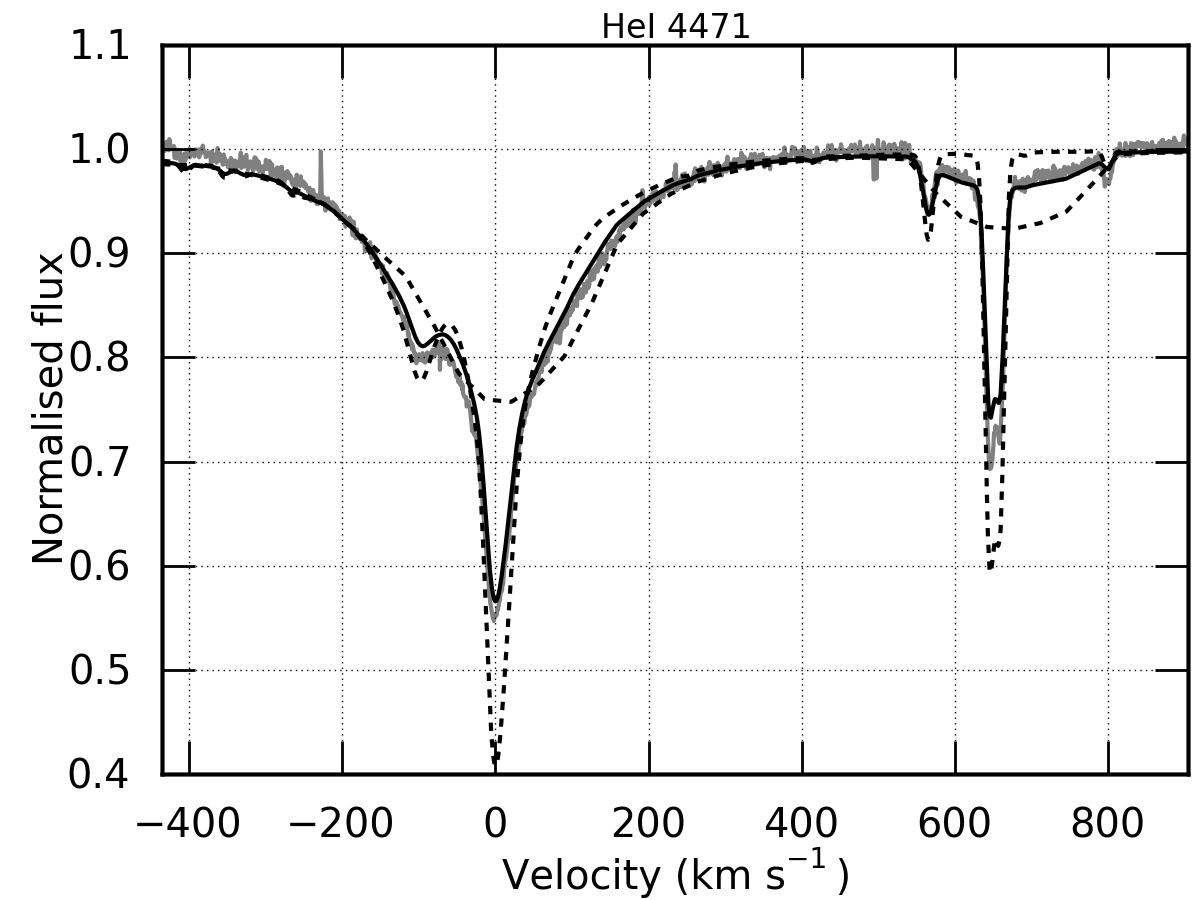}
            \includegraphics[width=0.45\textwidth]{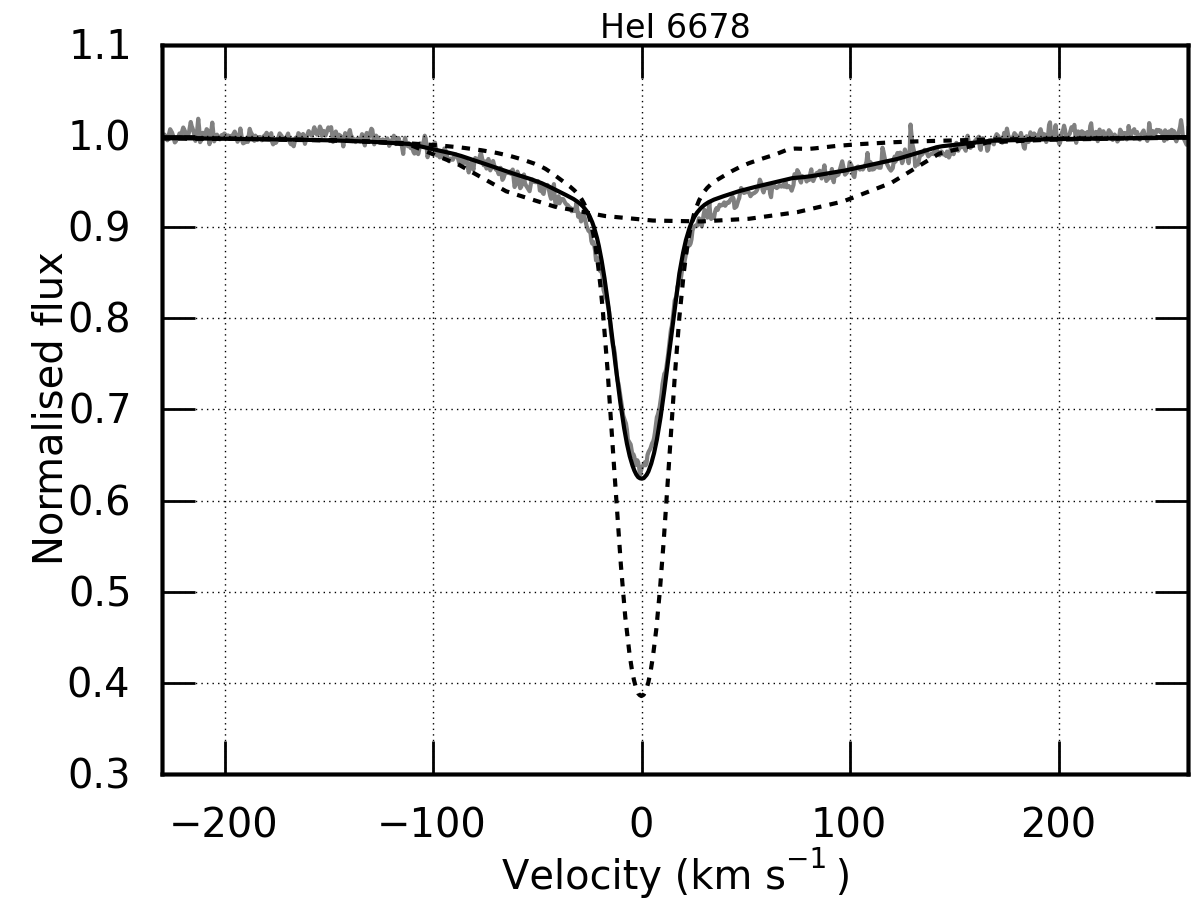}

            \includegraphics[width=0.45\textwidth]{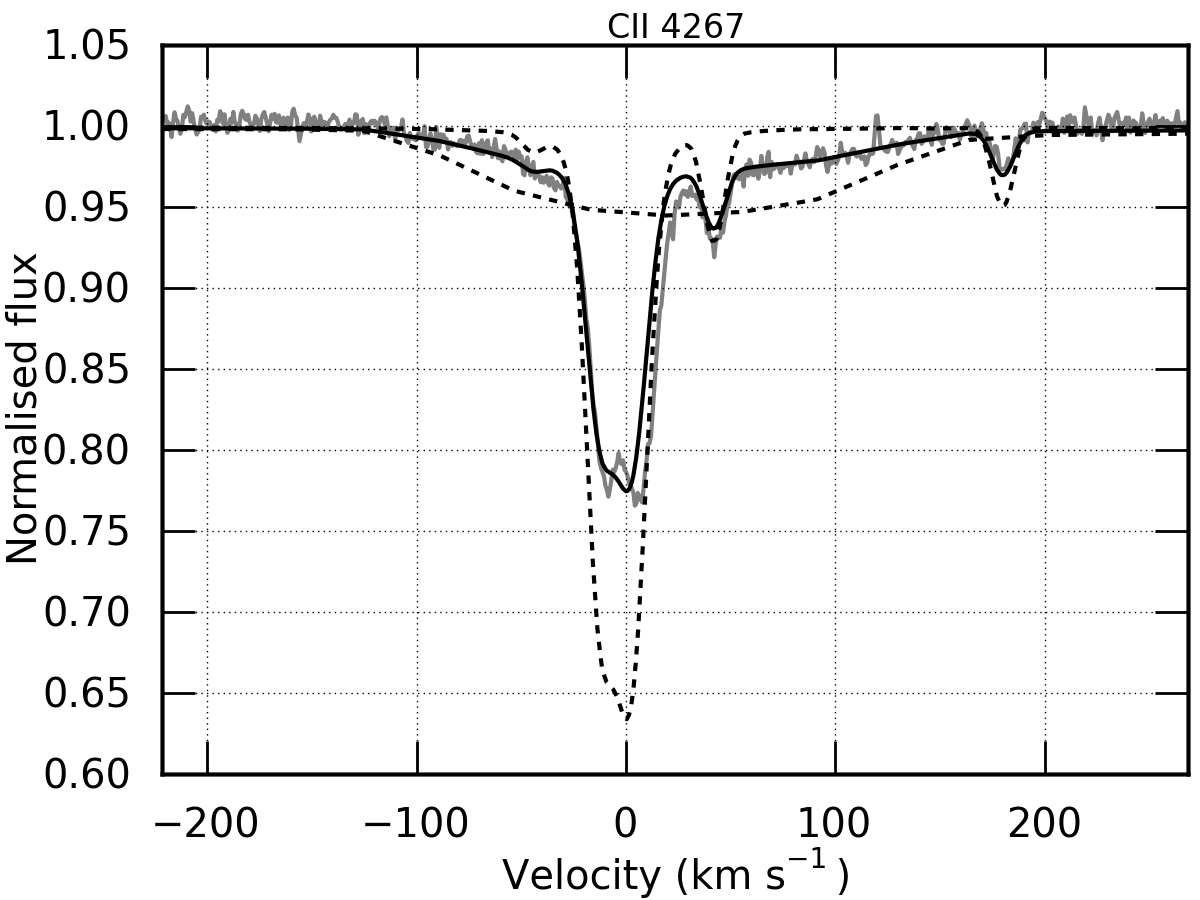}
            \includegraphics[width=0.45\textwidth]{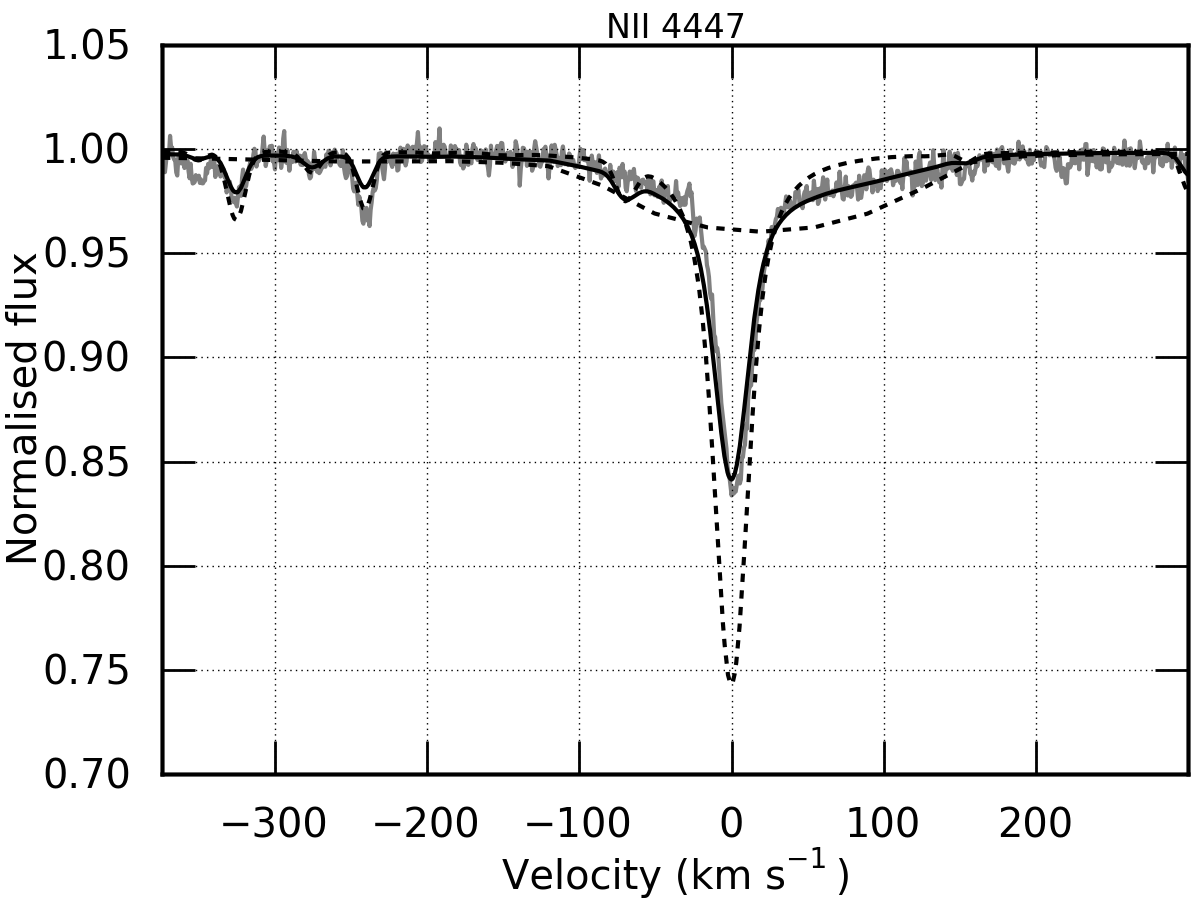}

            \includegraphics[width=0.45\textwidth]{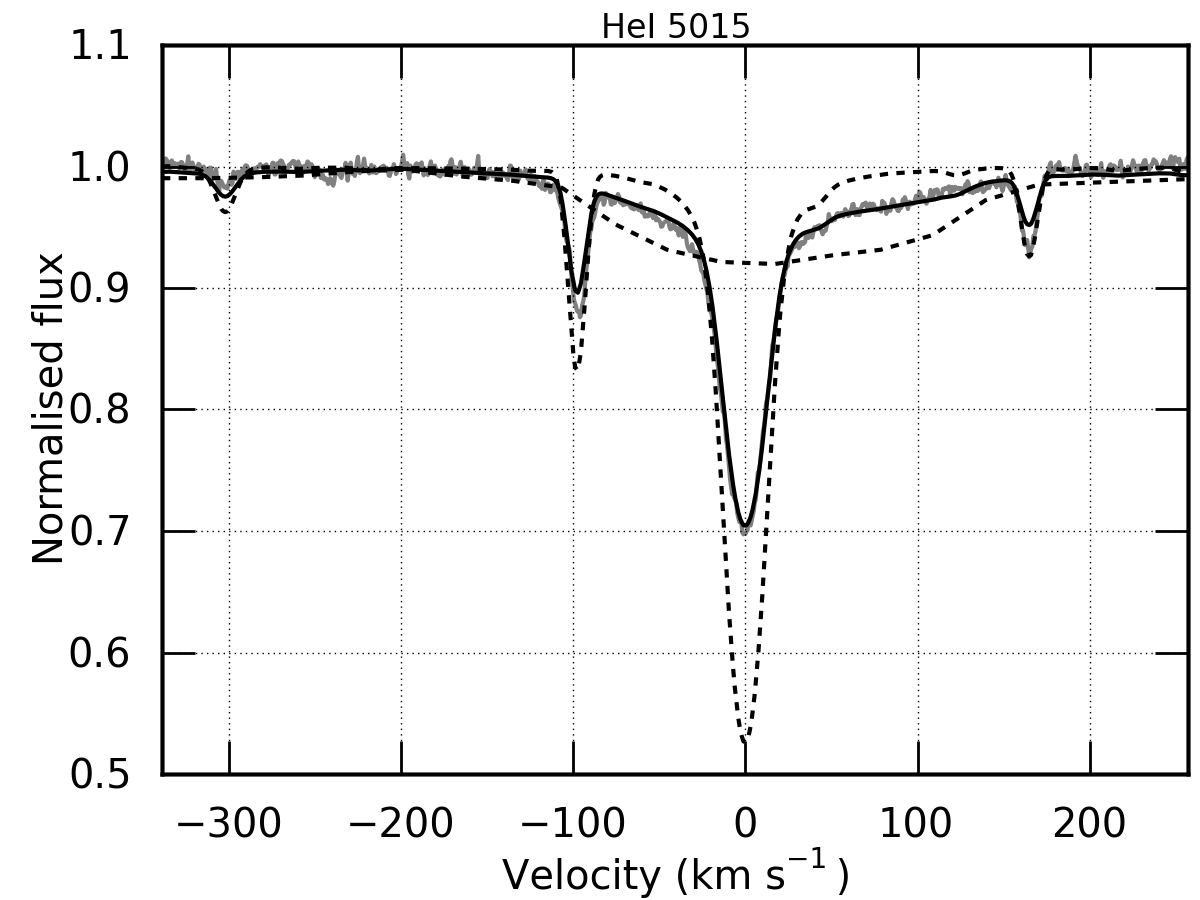}
            \includegraphics[width=0.45\textwidth]{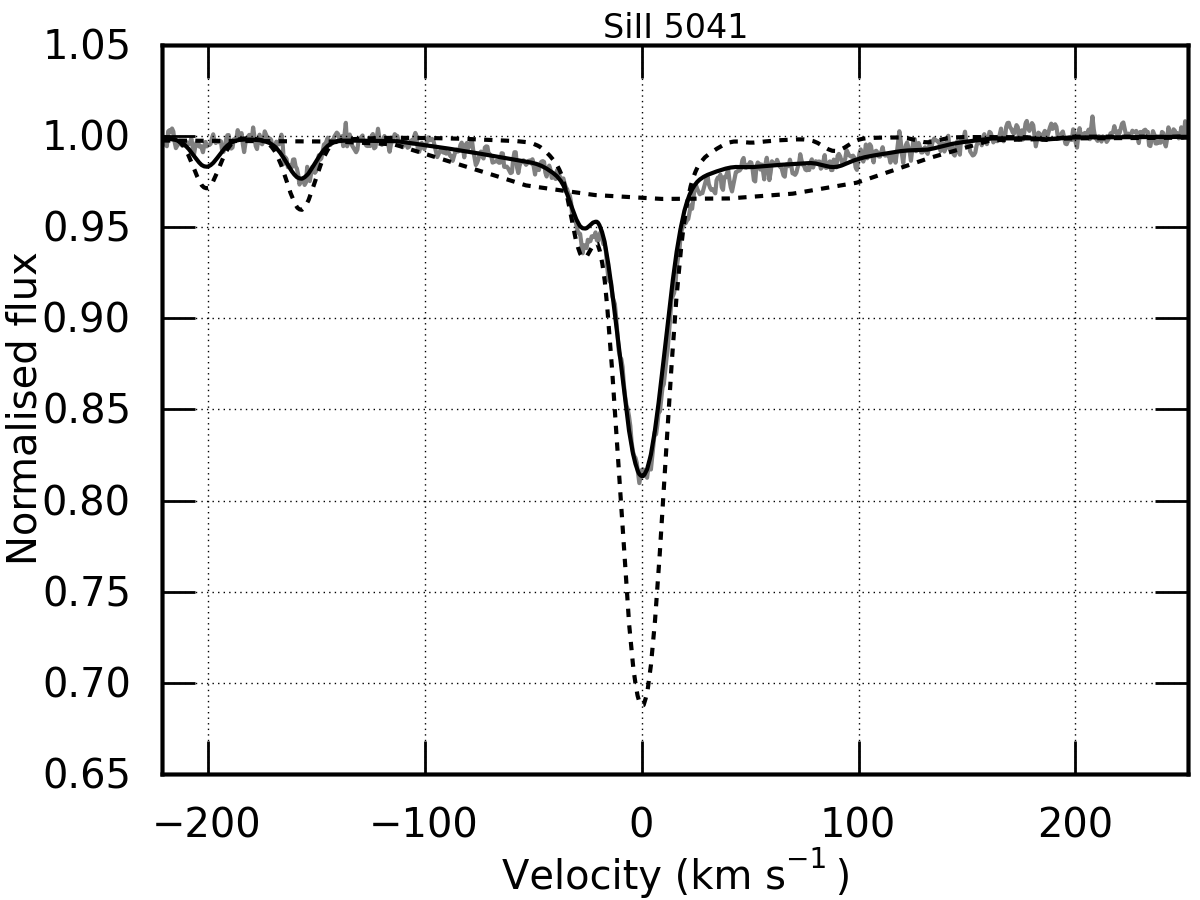}

            \includegraphics[width=0.45\textwidth]{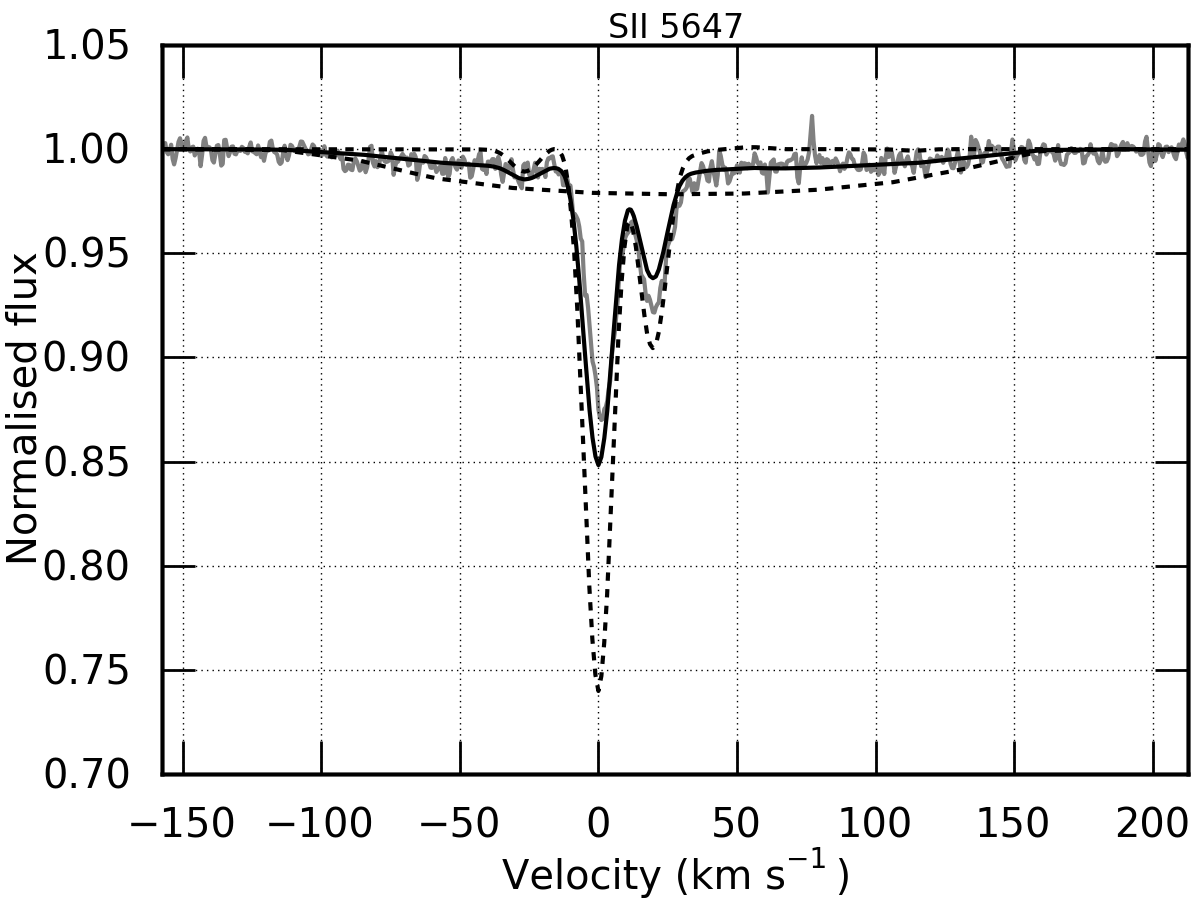}
            \includegraphics[width=0.45\textwidth]{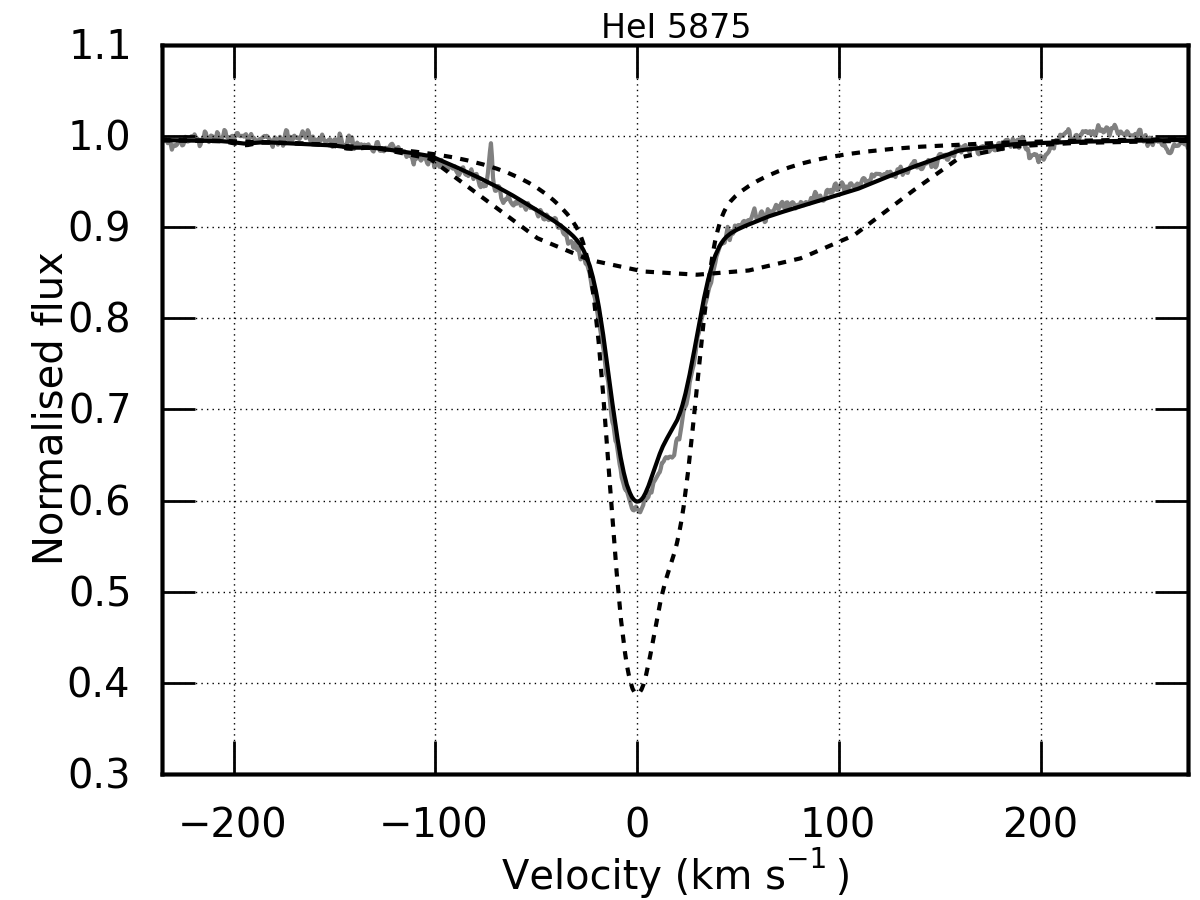}
\caption{Same as Fig.\,\ref{fig:hd50230:linefits1}, but for additional spectral lines.}\label{fig:hd50230:linefits5}
\end{figure*}

\begin{figure*}
\centering  \includegraphics[width=0.45\textwidth]{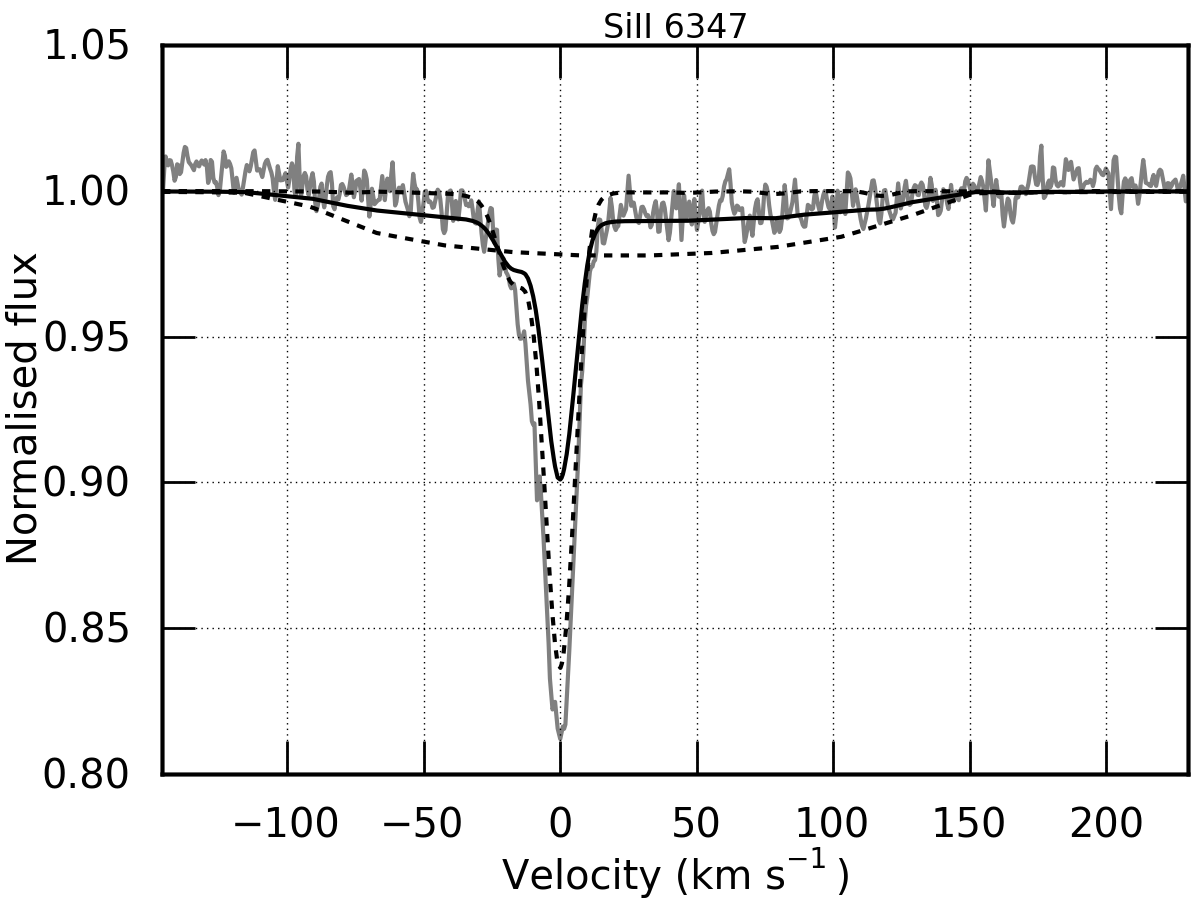}

\caption{Same as Fig.\,\ref{fig:hd50230:linefits1}, but for additional spectral lines.}\label{fig:hd50230:linefits6}
\end{figure*}

\longtab{2}{
\begin{longtable}{rrrrrrrrrl}
\caption{\label{tbl:freqanal} Frequency list and statistics from a Fourier
analysis with iterative nonlinear updates of the parameters.}\\
\hline\hline
nr & $A$ & $\sigma(A)^1$ & $f^2$ & $\sigma(f)^1$ &
$\phi$ & $\sigma(\phi)^1$ & SNR$^3$ & BIC$^4$ & 
Remark$^5$s\\
   & [ppm] & [ppm] & [d$^{-1}$] & [$10^{-3}$d$^{-1}$] & [2$\pi$] & [2$\pi$] \\\hline

\endfirsthead
\caption{continued.}\\
\hline\hline
nr & $A$ & $\sigma(A)^1$ & $f^2$ & $\sigma(f)^1$ &
$\phi^4$ & $\sigma(\phi)^1$ & SNR$^3$ & BIC$^4$ & 
Remarks$^5$\\
& [ppm] & [ppm] & [d$^{-1}$] & [$10^{-3}$d$^{-1}$] & [2$\pi$] & [2$\pi$] \\\hline
\endhead
\hline
\endfoot
$f_{001}$ & 1746.3 & 8.05 & \textit{0.68440} & 0.02 & -0.417 &  0.005 & 26.9 & 1.00000 &   \\
$f_{002}$ & 1607.6 & 7.57 &  0.59342 & 0.02 &  0.228 &  0.005 & 26.5 & 0.99938   & \\
$f_{003}$ & 1518.2 & 7.13 &  0.84085 & 0.02 &  0.330 &  0.005 & 26.0 & 0.98990   & \\
$f_{004}$ & 1457.3 & 6.66 &  0.56340 & 0.02 &  0.228 &  0.005 & 27.5 & 0.98873   & \\
$f_{005}$ & 1495.0 & 6.14 & \textit{0.87931} & 0.02 &  0.361 &  0.004 & 29.2 & 0.98372  & \\
$f_{006}$ & 1384.3 & 5.64 & \textit{0.97367} & 0.02 & -0.469 &  0.004 & 30.2 & 0.97772  & \\
$f_{007}$ & 1417.3 & 5.09 &  0.65088 & 0.01 & -0.124 &  0.004 & 32.5 & 0.98052   & \\
$f_{008}$ & 1307.5 & 4.55 &  0.62213 & 0.01 &  0.485 &  0.004 & 32.6 & 0.96622   & \\
\hline\hline           
\end{longtable}        
\begin{list}{}{}       
\item[1] \footnotesize{$\sigma$ denotes the standard deviation}
\item[2] \footnotesize{Frequencies in boldface are related to the period spacing, frequencies in italics denote the pressure mode candidates}
\item[3] \footnotesize{SNR calculated in the residual periodogram over a 6\,d$^{-1}$ interval}
\item[4] \footnotesize{Normalised BIC}
\item[5] \footnotesize{$T$ denotes the total time span}
\end{list}             
}                      
\end{appendix}

\end{document}